\documentclass{aastex631}

\usepackage[T1]{fontenc}
\usepackage[utf8]{inputenc}
\usepackage[british]{babel}
\usepackage{graphicx}
\usepackage{amsmath}
\usepackage{mathastext}    
\usepackage[caption=false]{subfig}
\usepackage{xcolor}
\usepackage{makecell}
\usepackage{tabularx}      
\usepackage{multirow}
\usepackage{booktabs}      
\usepackage{fancyhdr}   
\usepackage{pbox}

\newcommand{\be}{\begin{equation}}
\newcommand{\ee}{\end{equation}}

\newcommand{\ba}{\begin{eqnarray}}
\newcommand{\ea}{\end{eqnarray}}
\newcommand{\om}{\omega}
\newcommand{\Alfven}{Alfv\'{e}n }
\newcommand{\Alfvenic}{Alfv\'{e}nic }

\newcommand{\E}{{\bf E}}
\newcommand{\B}{{\bf B}}
\newcommand{\J}{{\bf J}}
\renewcommand{\v}{{\bf v}}
\renewcommand{\k}{{\bf k}}

\newcommand\eg{{\it{e.g.\ }}}

\newcommand{\NS}{neutron star}
\newcommand{\NSs}{{neutron stars}}
\newcommand{\EM}{electromagnetic}
\newcommand{\BH}{{black hole}}

\newcommand{\ms}{magnetosphere}

\renewcommand{\vec}[1]{\mathbf{#1}}

\newcommand{\pd}[2]{\frac{\partial #1}{\partial #2}}

\shortauthors{Sharma et al.}

\begin{document}

\title{Production of Jets before Neutron Star Mergers}

\author{Praveen Sharma}
\affiliation{Department of Physics \& Optical Engineering, Rose-Hulman Institute of Technology, 5500 Wabash Ave, Terre Haute, IN 47803, USA}

\author{Slava G. Turyshev}
\affiliation{Jet Propulsion Laboratory, California Institute of Technology, 4800 Oak Grove Drive, Pasadena, CA 91109, USA}

\author{Maxim V. Barkov}
\affiliation{Institute of Astronomy, Russian Academy of Sciences, Moscow, 119017, Russian Federation}

\author{Maxim Lyutikov}
\affiliation{Department of Physics, Purdue University, 525 Northwestern Avenue, West Lafayette, IN 47907, USA}

\begin{abstract}

We demonstrate that magnetospheric interactions between merging neutron stars (NSs) generate dual-jetted current outflows, analogous to the Alfv\'{e}n wings observed during planetary interactions in the Solar System. Using 3D relativistic MHD simulations, we model the interaction as a conducting sphere moving through a highly magnetized plasma of the companion's magnetosphere. Unusually, the interaction operates in a regime that is relativistic yet sub-Alfv\'{e}nic. Electromagnetic draping amplifies magnetic fields in a narrow layer near the stellar surface, leading to the generation of electric currents along the local magnetic field. 
The generation of beamed outflows enhances the instantaneous power of the pulsar-like radio and high-energy emission, produces spin/orbital modulations, and is likely to lead to observable precursor emission preceding the main gravitational wave event. 

\end{abstract}




\section{Introduction}
\label{intro_section}

Neutron star mergers represent key targets for Multi-messenger astronomy, serving as sources of gravitational waves and potential progenitors of short gamma-ray bursts (GRBs). Detecting electromagnetic precursors to such mergers would provide valuable early warnings and insights into merger physics. The most promising mechanism for generating such precursor emission is the electromagnetic interaction between merging compact objects through inductive electric fields created by relative motion within their common magnetosphere \citep{2001MNRAS.322..695H,2012ApJ...757L...3L,2012ApJ...746...48M,2020ApJ...893L...6M,2024arXiv241216280M,2024arXiv240216504L}.

Since merger timescales are typically long \citep{1996A&A...312..670P}, by the time of the merger, the neutron stars are likely to have spun down and become electromagnetically ``dead'' -- not actively producing pairs in the classical pulsar manner \citep{GJ}. Even fast merger channels are long enough for typical pulsars to switch off \citep{Andrews_2020}. However, as orbital inspiral progresses, the magnetospheres can be reactivated through their relative orbital motion \citep{2019MNRAS.483.2766L,2024arXiv240216504L}.

Both single magnetized (1M-DNS) and double-magnetized (2M-DNS) cases are possible \citep{2019MNRAS.483.2766L}. Our work focuses on the 1M-DNS case, where an unmagnetized, highly conducting neutron star moves through the magnetic field of its companion. This scenario shares qualitative similarities with planetary magnetosphere interactions \citep{1969ApJ...156...59G}, but exhibits crucial differences due to the extreme conductivity of neutron stars and the possibility of relativistic velocities during late-stage inspiral. 

Magnetic interaction of merging \NSs\ is expected to proceed in an unusual (by space physics) regime: (i) plasma is expected to be highly magnetized, with $\sigma = (\om_B/\om_p)^2  \geq 1$ (and,  correspondingly $\beta \ll 1$; (ii) interaction velocities may be mildly relativistic $\sim c$, as near merger the Keplerian velocity is close to the speed of light. This unusual regime is the goal of the present study. (For example, the interaction may be in a regime of relativistic but sub-\Alfvenic.)

The key points of the present work are:  (i) motion of a conducting star through \ms\ of a companion creates an EMF, and can possibly lead to 
The generation of observable emission; (ii) resulting current structure resembles \Alfven wings observed planetary interactions; the resulting emission pattern is then expected to be anisotropic, leading to higher peak fluxes and orbital or spin modulations. 

The rest of the paper is organized as follows: \S\ \ref{theory} describes non-relativistic case (planetary \Alfven wings),  theoretical background and expectations for the relativistic regimes; \S\ \ref{simulations} details the numerical setup; \S\ \ref{non_rel_results} presents simulation results for non-relativistic flows spanning sub-\Alfvenic to super-\Alfvenic regimes; \S\ \ref{rel_results} presents results for mildly relativistic flows and compares them with classical predictions; and \S\ \ref{discussion} provides discussion and conclusions regarding the implications for neutron star merger precursor emission.

\section{Theoretical Background}  
\label{theory}

\subsection{\Alfven wings in planetary interactions}
\label{sec:planetary_wings}

The interaction of conducting bodies with magnetized plasma is well established in space plasma environments and displays markedly different behavior depending on both conductivity and flow regime. A particular feature of interaction is the formation of so-called \Alfven wings.

\Alfven wings are routinely realized in planetary and satellite environments whenever a conducting (or effectively conducting) obstacle is embedded in a sub-\Alfvenic magnetized flow ($\mathcal{M}_A<1$). In this regime, the response to the obstacle is not mediated by a detached bow shock but by stationary \Alfvenic characteristics, which allow stresses and currents to propagate both upstream and downstream along the ambient magnetic field \citep{1962P&SS....9..321G,1965PhRvL..14..171D,1980JGR....85.1171N,2013A&A...552A.119S}. In the obstacle frame, the two \Alfven wings are aligned with the characteristic directions
\begin{equation}
\mathbf{u}_{\pm} \equiv \mathbf{v}_0 \pm v_A \hat{\mathbf{b}},
\label{eq:aw_characteristics}
\end{equation}
where $\hat{\mathbf{b}} \equiv \mathbf{B}_0/|\mathbf{B}_0|$ and $v_A$ is the \Alfven speed. For the commonly used geometry $\mathbf{v}_0 \perp \mathbf{B}_0$, the wing inclination relative to $\mathbf{B}_0$ satisfies $\tan\theta_A \simeq \mathcal{M}_A$, where $\mathcal{M}_A$ is upstream \Alfven Mach number.

In the solar system, \Alfven wing structures are observed when a conducting obstacle is embedded in a sub-\Alfvenic flow. For example, Earth’s magnetosphere has been observed forming \Alfven wing connected current systems in the solar wind \citep{2025GeoRL..5211931G}. Similar processes occur at the Moon, although its weak conductivity and lack of an intrinsic magnetic field lead to plasma wake formation rather than stable wings \citep{2012JGRA..117.9217C, 2016JGRA..12110698Z}. The most direct analogues for strongly conducting bodies are Jupiter’s moons. Io’s volcanically sustained ionosphere and Ganymede’s intrinsic magnetic field each generate coherent \Alfven wings and associated field-aligned current systems \citep{2017P&SS..137...40V, 1999JGR...10428671N, 2002JGRA..107.1490K}, which are observable through auroral and radio signatures \citep{2014A&A...569A..86M, 2024EPSC...17..726C}. Extensions of Alfvén wing theory to relativistic flows have been explored in the context of pulsar winds and orbiting bodies by \cite{2011A&A...532A..21M}, demonstrating that strong currents can be driven even when the body is immersed in a relativistic plasma flow. These examples clearly demonstrate that conducting obstacles in magnetized plasma flows produce electromagnetic structures that efficiently channel energy and can be detected observationally.

Quantitatively, the planetary \Alfven-wing interaction can be expressed as an effective electrodynamic circuit in which the moving obstacle acts as a unipolar generator. For an obstacle of radius $R_p$ embedded in a magnetized flow with $\mathbf{v}_0 \perp \mathbf{B}_0$, the motional field is $E_0 \simeq v_0 B_0$, yielding an induced electromotive force $\Delta\Phi \simeq 2R_p E_0 \simeq 2R_p v_0 B_0$ \citep{1965PhRvL..14..171D,1969ApJ...156...59G,1980JGR....85.1171N}. The associated field-aligned current is limited by the \Alfven-wave conductance $\Sigma_A \simeq (\mu_0 v_A)^{-1}$ (equivalently, impedance $Z_A \simeq \mu_0 v_A$), so in the high-conductance limit one expects $I \sim 4R_p E_0 \Sigma_A$ and a total electromagnetic power extraction $P_{\rm AW} \sim I\,\Delta\Phi \sim 8R_p^2 E_0^2 \Sigma_A$ transported along the two wings. For the Io--Jupiter interaction, inserting representative values ($R_p \simeq 1.82\times10^6\,{\rm m}$, $v_0 \sim 5\times10^4$--$7\times10^4\,{\rm m\,s^{-1}}$, $B_0 \sim 10^{-6}$--$2\times10^{-6}\,{\rm T}$, $v_A \sim 10^5$--$3\times10^5\,{\rm m\,s^{-1}}$) gives $\Delta\Phi \sim {\rm few}\times10^5\,{\rm V}$, $I \sim {\rm few}\times10^6\,{\rm A}$, and $P_{\rm AW}\sim 10^{12}\,{\rm W}$, consistent with in-situ magnetic perturbations and remote auroral footprint constraints \citep{1996Sci...273..337K,1996Sci...274..404C,2002Natur.415..997C,2005P&SS...53..395C}.

These planetary observations motivate two points directly relevant to the present neutron-star problem. First, they show that \Alfven wings can remain coherent over distances vastly exceeding the obstacle size, so long as $\mathcal{M}_A<1$ and dissipation does not destroy the field-aligned current channels \citep{1980JGR....85.1171N,2002Natur.415..997C}. Second, they demonstrate that the most readily observable signatures are frequently indirect: auroral footprints, lead angles, and radio emission reflect the Poynting flux and current carried by the wings rather than the near-body fields alone. Our simulations are designed to isolate the underlying MHD wing morphology and current topology, enabling a controlled comparison between the classical planetary regime and the neutron-star limit where the conductivity is effectively infinite and relativistic effects become important.

While in the above discussion, we used SI circuit notation ($\mu_0, \Omega$) for comparison with space-physics literature, all simulation variables and plasma parameters elsewhere are in CGS or code units.

\subsection{The unipolar inductor model}

Neutron stars represent the extreme limit of this physics. Their matter composition renders them effectively perfect conductors ($\sigma \to \infty$), enforcing a boundary condition that expels magnetic flux from the stellar interior. As a result, magnetic field lines drape tightly around the surface, forming thin current layers whose structure depends on the flow velocity regime. These layers primarily amplify magnetic fields through compression and can develop enhanced current concentrations, though the conditions for efficient particle acceleration and radiation remain to be explored with kinetic-scale simulations.


In a relativistic regime, the basic estimate of electromagnetic power from a relativistic unipolar inductor  \citep{2002luml.conf..381B} is
\begin{equation}
L \sim \frac{c}{4\pi} (\Delta \Phi)^2,
\label{L1}
\end{equation}
where $ \Delta \Phi \sim v B D$ is the EMF drop  
for a conductor of size $D$ moving at speed $v = \beta c$ in a magnetic field $B$. 

For a singly-magnetized case (1M-DNS) \citep{2019MNRAS.483.2766L}, the power scales as:
\begin{equation}
L_{1M-DNS}  \sim \frac{ B_{{NS}}^2 G M_{{NS}} R_{{NS}}^8}{c r^7} \approx 3 \times { 10^{43} } \;{t^{-7/4}}\,B_{13}^2\, {\rm \, erg \, s^{-1}} 
\label{L}
\end{equation}

The power in the double-magnetized (2M-DNS) case, $L_2$, is larger by a factor of $\approx (r/R_{NS})^2$. This enhancement arises because the magnetospheric interaction in the 2M-DNS scenario occurs on the scale of the orbit, whereas in the 1M-DNS case, it is confined to the scale of the companion NS. 

The estimates in Eq. (\ref{L}), normalized to a relatively large surface magnetic field of $10^{13}$ G, highlight a potential issue: the overall expected power is not particularly large. If emitted isotropically, it will be missed by high-energy all-sky X-ray monitors.

We demonstrate: (i) the emission is likely to be beamed; and 
(ii) one might expect the generation of coherent radio emission, {\it a la} pulsar-style. 
If a significant fraction of the energy is emitted in coherent radio, the expected peak flux density is given in Jansky \citep{2001MNRAS.322..695H,2019MNRAS.483.2766L,2023MNRAS.519.3923C}. 
The radio signal will also experience a dispersive delay of
\be
\Delta t \simeq 14~{sec} 
\left(\frac{\nu}{1~{GHz}}\right)^{-2} 
\left(\frac{d}{200~{pc}}\right),
\ee
where $\nu$ is the observing frequency, $d$ is the source distance in parsecs, and $\Delta t$ is the dispersive delay \citep{2010Ap&SS.330...13P,2023PASA...40...50T}.

\section{Numerical Simulations}
\label{simulations}


We conduct three-dimensional magnetohydrodynamics (MHD) simulations using the \texttt{PLUTO} code version~4.3 \citep{2007ApJS..170..228M} to investigate magnetic draping around solid obstacles in both classical non-relativistic and relativistic regimes. The code solves the full set of ideal (R)MHD equations using high-resolution shock-capturing schemes optimized for curvilinear coordinate systems. Our focus remains on magnetohydrodynamic processes, neglecting external and self-gravity effects, as electromagnetic interactions dominate the dynamics in the regimes of interest.

\subsection{Code Setup}

We model the neutron star as a dense, non-rotating, perfectly conducting sphere of radius $R_{\star}$ embedded in an ambient magnetized plasma. The star is assumed to be unmagnetized and highly conducting, and interacts with the magnetic field of a companion star. We work in the star’s rest frame so that the neutron star remains fixed in space. Far from the sphere ($r \gg R_{\star}$), the plasma flows with velocity $\mathbf{v} = -v_0 \hat{\mathbf{x}}$ from right (upstream) to left (downstream).


Non-relativistic simulations use a Courant-Friedrichs-Lewy (CFL) number of 0.4 and the Harten-Lax-van Leer (HLL) Riemann solver. 
The divergence constraint $\nabla\!\cdot\vec{B}=0$ was enforced using 
\textsc{pluto}'s eight-wave formulation \citep{1994arsm.rept.....P, 1999JCoPh.154..284P}. We monitored the dimensionless volume-averaged normalized divergence error  $\Delta x \langle|\nabla\!\cdot\vec{B}|\rangle / \langle|\vec{B}|\rangle$. This metric remained below $10^{-3}$, indicating good divergence control.


Relativistic runs adopt CFL=0.25, the HLLD solver \citep{2009MNRAS.393.1141M}, and an adiabatic index $\gamma = 4/3$. The domain uses a stretched grid with $N_x = N_y = N_z = 170$, giving a minimum cell size $\Delta x \sim 0.04\,R_\star$. Divergence is controlled through hyperbolic cleaning \citep{2002JCoPh.175..645D}. The divergence constraint was handled through the hyperbolic divergence cleaning method \citep{2002JCoPh.175..645D}. The normalized divergence error remained below $10^{-3}$ with the characteristic cell.


The stretched grid concentrates resolution near the stellar surface. For the fiducial relativistic runs (N=170), $\Delta x_{min} \approx 0.034\,R_\star$, marginally resolving the predicted draping layer thickness $\delta_r \sim (v_0/c)^2 R_\star \sim 0.04 R_\star$. Convergence studies (Appendix~\ref{numerical_convergence}) confirm that integrated quantities, such as wing current and magnetic amplification, are robust across resolutions $N=113$–$190$.

All runs employ \textsc{pluto} with second-order Runge-Kutta (RK2) time-stepping. 
A detailed description of the numerical setup, boundary conditions, along with the discussion of divergence cleaning and convergence, is discussed in detail in the Appendices \ref{simulation_setup}, \ref{pluto_code}.

\subsection{Interaction parameters}

To characterize the interaction of magnetized plasmas with obstacles, we now discuss key dimensionless parameters that govern the flow. The primary parameter across different runs is the upstream \Alfvenic Mach number
$\mathcal{M}_A = \frac{v_0}{v_{A0}}$ where $v_{A0} = \sqrt{B_0^2/\rho_0}$ is the \Alfven speed in the ambient medium far away from the central obstacle.

Another important parameter is the plasma beta $\beta = \frac{p_{thermal}}{p_{mag}} $ which measures the relative importance of thermal plasma pressure to magnetic pressure. Since neutron star magnetospheres are typically magnetically dominated, we focus on low $\beta$ regimes ($0.01 \le \beta \le 0.02$) in our non-relativistic runs. Our relativistic runs explore slightly higher values ($\beta \approx 0.16$--$0.80$), chosen to ensure numerical stability in the relativistic MHD regime.


For relativistic flows, these parameters must be evaluated using comoving-frame quantities. Primed variables denote the plasma rest frame. We introduce the magnetization parameter
\begin{equation}
\sigma' \equiv  \frac{2p'_{mag}}{\rho'_0 c^2}
\end{equation}
which compares magnetic to rest-mass energy density: $\sigma' \ll 1$ corresponds to matter-dominated flows, while $\sigma' \gg 1$ corresponds to magnetically dominated flows.


The characteristic wave speeds in relativistic MHD (see Eqs.~38–46 in 
\citep{1993PhRvE..47.4354G}; also \citealp{2012PhRvE..85b6401L}) evaluated in the co-moving frame are given by
\begin{equation}
\begin{aligned}
v'_A &= c\sqrt{\frac{\sigma'}{h' + \sigma'}},\\
v'_s &= \sqrt{\frac{\gamma p}{\rho' h'}},\\
v'_f 
&= \sqrt{{v'_A}^2 + {v'_s}^2 - \frac{{v'_A}^2 {v'_s}^2}{c^2}}
\end{aligned}
\end{equation}
where $h' \equiv \frac{\epsilon' + p}{\rho' c^2} $ is the specific enthalpy, with $\epsilon'$ the internal energy density and $\rho'$ the rest-mass density in the plasma rest frame. For an ideal-gas equation of state, $\epsilon' = \rho' c^2 + \frac{p}{(\gamma-1)}$ \citep{2023A&A...679A..49M}, with $\gamma $ being the adiabatic index.

The relativistic \Alfven Mach number is defined using 
\begin{equation}
\mathcal{M}_A = \frac{\Gamma v_0}{\Gamma_A v'_A}, \quad
\Gamma_A = \left(1 - \frac{{v'_A}^2}{c^2}\right)^{-1/2}
\label{eq:rmhd_mach}
\end{equation}
which can also be written as
\begin{equation}
\mathcal{M}_A = \frac{\Gamma v_0}{c}\sqrt{\frac{h'}{\sigma'}}.
\label{eq:rmhd_mach_simple}
\end{equation}

with analogous definitions for $\mathcal{M}_s$ and $\mathcal{M}_f$.

A simplified non-relativistic analogy $\mathcal{M}_A \approx v_0/v'_A$ is sometimes used in the literature, but it neglects the Lorentz transformation properties of both the bulk flow and the characteristic wave speeds. The definition in Eq.~\ref{eq:rmhd_mach} is fully relativistic and frame-invariant (cf. Eq.~37 of \cite{2015MNRAS.452.3010Z}; \cite{2023MNRAS.524...90C}).


\subsection{Simulation Runs}
\label{simulation_runs}

Using the dimensionless parameters defined above, we construct a set of representative models spanning sub-\Alfvenic and super-\Alfvenic regimes in both classical and relativistic limits. The principal runs discussed in this work are summarized in Table~\ref{tab:runs_params}.

\textbf{Non-relativistic runs (A-series):} The fiducial sub-\Alfvenic run A1 ($\beta = 0.01$, $\mathcal{M}_A = 0.08$) represents the strongly magnetized plasma conditions characteristic of certain planetary magnetospheres and of regions immediately surrounding neutron stars. 
Runs A2 ($\mathcal{M}_A = 0.63$) and A3 ($\mathcal{M}_A = 2.19$) probe the moderately sub-\Alfvenic and trans-\Alfvenic regimes.

\textbf{Relativistic runs (B-series):} The relativistic runs span the sub-\Alfvenic to trans-\Alfvenic regimes. The fiducial run B1 ($\Gamma_0 = 1.05, \mathcal{M}_A \approx 0.21, \beta = 0.40$) explores the mildly relativistic sub-\Alfvenic flow. Run B2 ($\Gamma_0 = 1.15,\mathcal{M}_A \approx 0.65, \beta = 0.40$) represents the moderately sub-\Alfvenic regime. Run B3 ($\mathcal{M}_A \approx 1.91, \beta = 0.80$) approaches trans-\Alfvenic conditions. Simulations at substantially lower $\beta$ (values similar to the non-relativistic runs) become unstable in the relativistic module, owing to the increased stiffness of the RMHD system and associated difficulties in primitive-variable recovery.

\begin{table}[ht]
\centering
\subfloat[Non-relativistic (A-series)]{%
\begin{tabular}{l  c c c}
\toprule
Run & $\beta$ & $\mathcal{M}_A$ & $\mathcal{M}_f$ \\
\midrule
A1  & 0.01 & 0.08 & 0.073 \\
A2  & 0.01 & 0.63 & 0.61 \\
A3  & 0.02 & 2.19 & 1.96 \\
\bottomrule
\end{tabular}
}
\subfloat[Relativistic (B-series)]{%
\small
\begin{tabular}{l c c c c c}
\toprule
Run & $v_0/c$ ($\Gamma_0$) & $\beta$ & $\mathcal{M}_A$ & $\mathcal{M}_f$ \\
\midrule
B1 & 0.30 (1.05) & 0.16 & 0.21 &  0.17 \\
B2 & 0.50 (1.15) & 0.40 & 0.65 &  0.47 \\
B3 & 0.50 (1.15) & 0.80 & 1.91 & 1.42 \\
\bottomrule
\end{tabular}
}
\caption{Simulation parameters for the selected runs discussed in this work. Columns list the plasma $\beta$, and \Alfvenic ($\mathcal{M}_A$) and fast magnetosonic ($\mathcal{M}_f$) Mach numbers. Relativistic runs include the Lorentz factor $\Gamma_0$, with velocities expressed in terms of the speed of light, with the corresponding Lorentz factors shown in brackets.}
\label{tab:runs_params}
\end{table}

The characteristic timescale of the interaction is the \Alfven crossing time,
\begin{equation}
\tau_A \equiv \frac{R_{\star}}{v_A},
\end{equation}
and all simulations are evolved for several $\tau_A$ to ensure the development of steady-state structures.

\section{Non-relativistic Flow}
\label{non_rel_results}

Although our primary interest is in the relativistic MHD runs, we first analyze a set of non-relativistic simulations to establish a baseline for interpreting the relativistic results. By systematically varying the \Alfven Mach number ($\mathcal{M}_A$), we characterize how the global morphology transitions from magnetically controlled to shock-dominated flow.  

We begin with the classical sub-\Alfvenic limit and then consider the trans-\Alfvenic regime, before turning to the relativistic cases in \S\ref{rel_results}.

\subsection{Sub-\Alfvenic Regime}
\label{results_nonrel_sub_alfvenic}

\begin{figure}[htbp]
\centering
\subfloat{%
\includegraphics[height=0.19\textwidth]{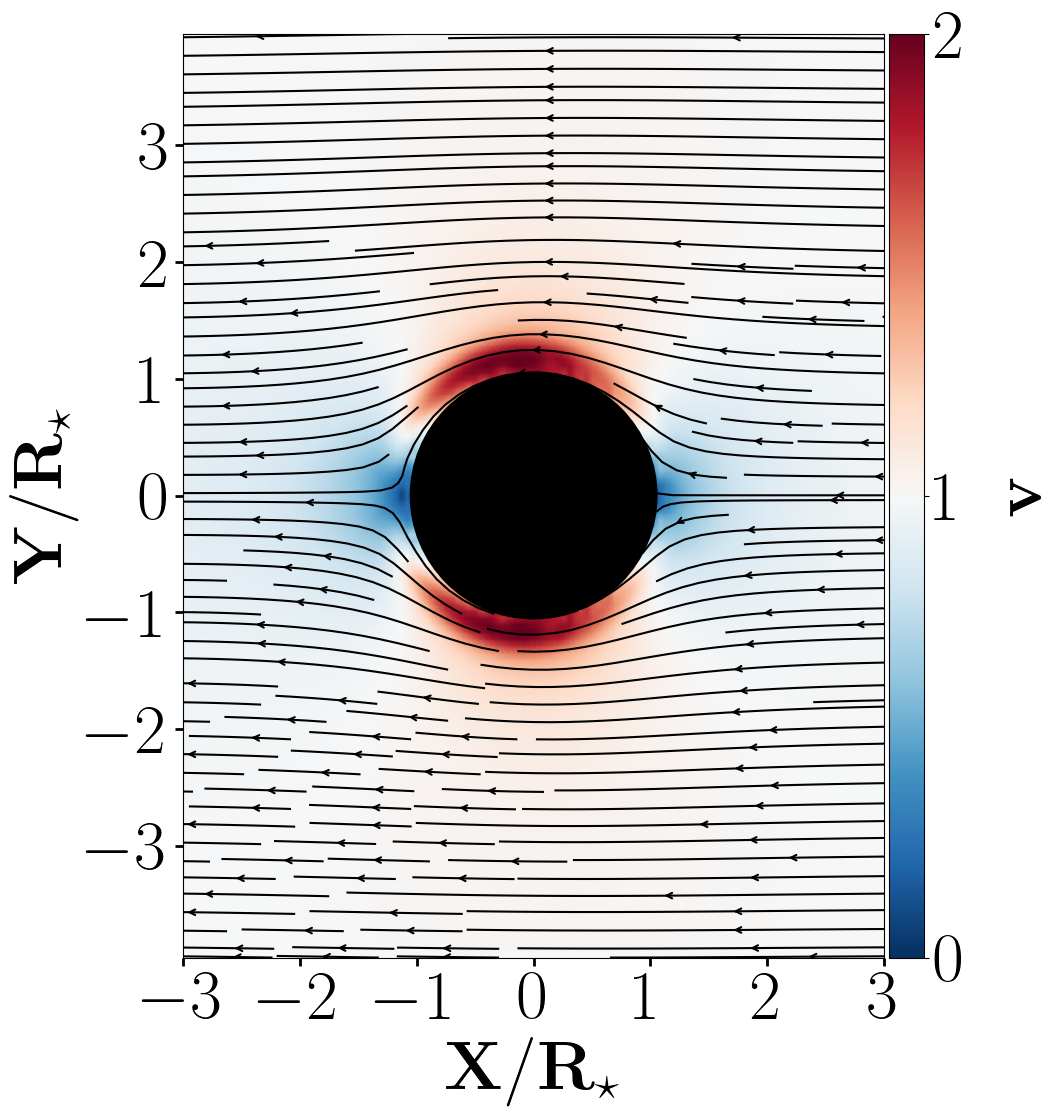}
}
\subfloat
{%
\includegraphics[height=0.19\textwidth]{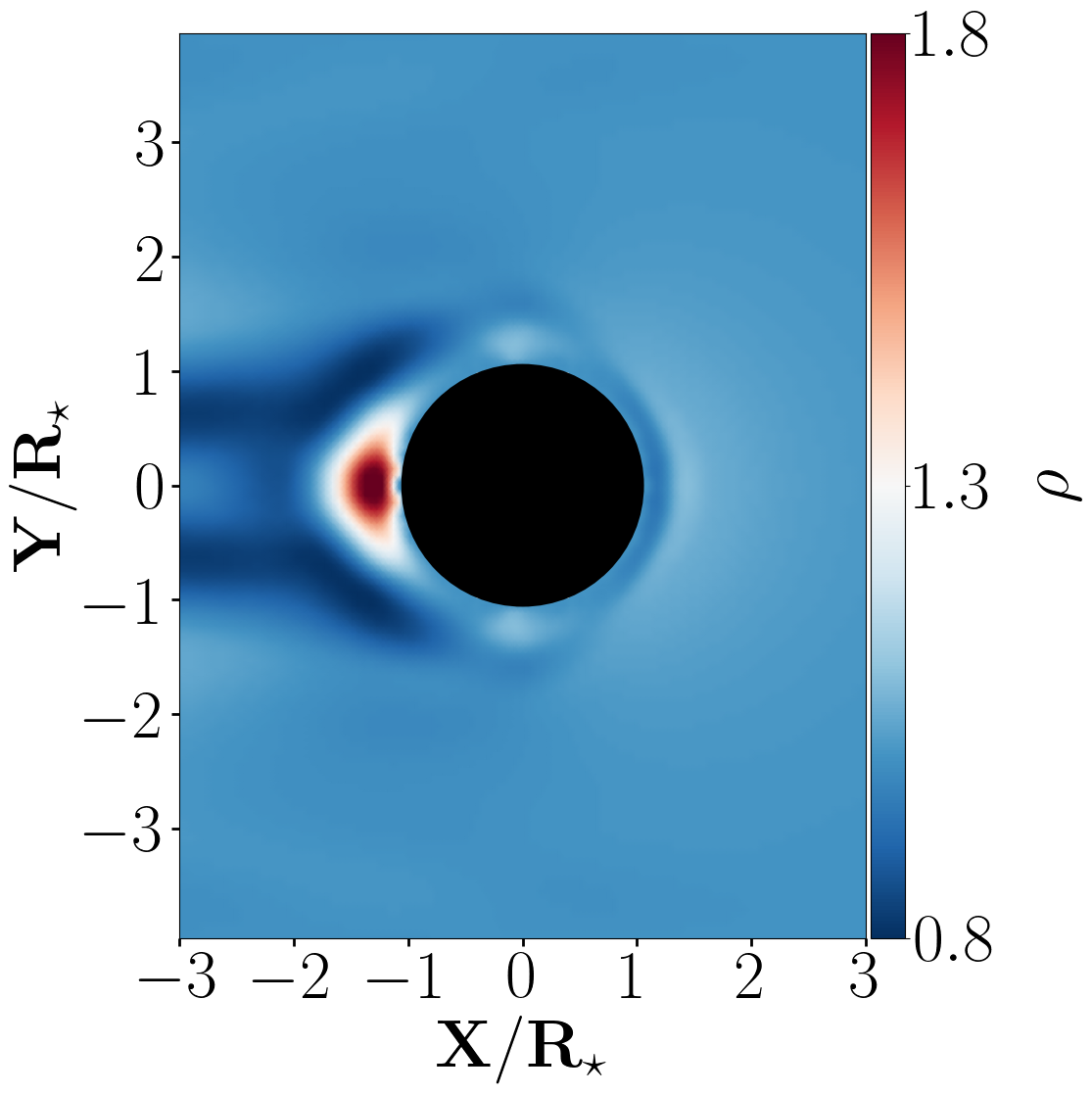}
}
\subfloat{%
\includegraphics[height=0.19\textwidth]{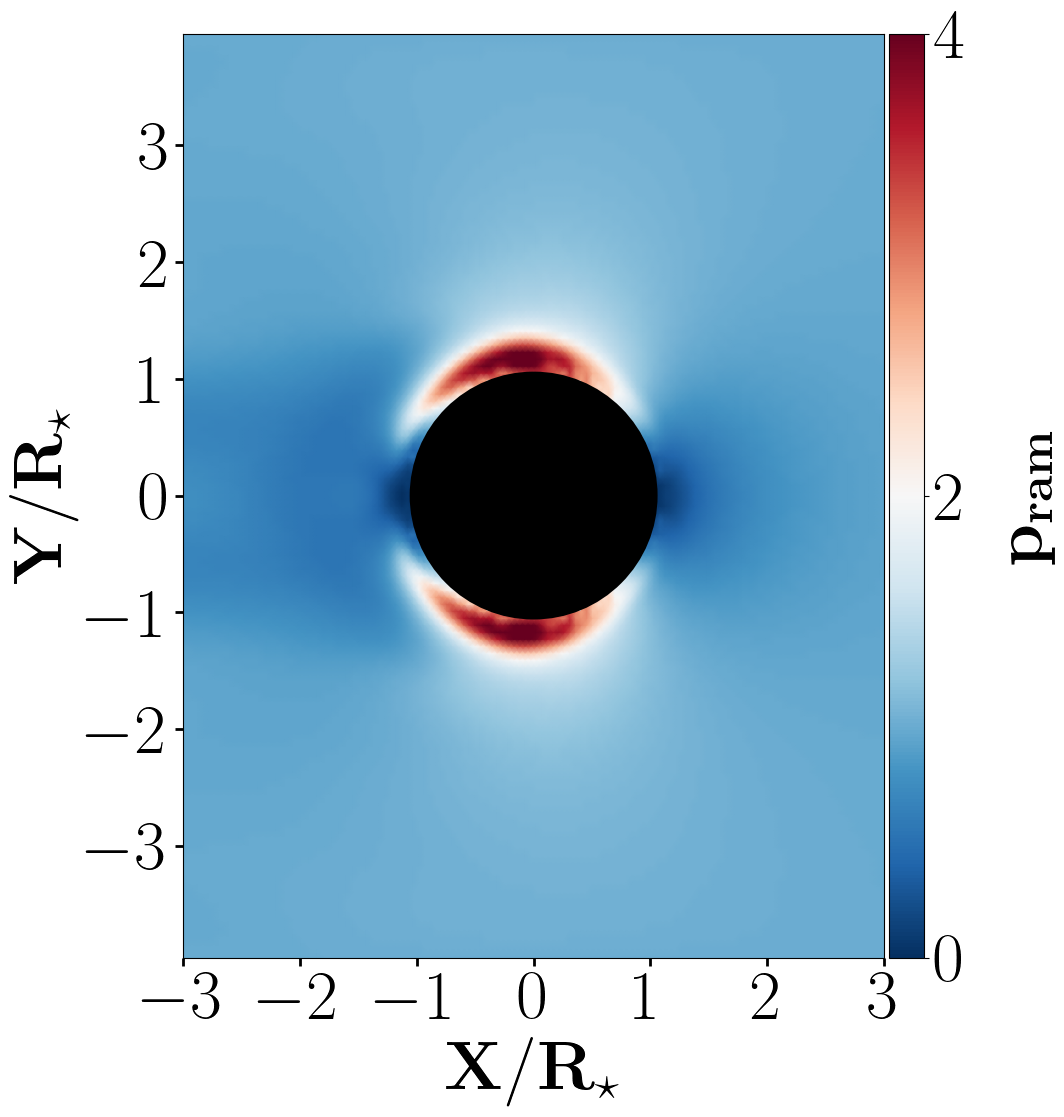}
}
\subfloat{%
\includegraphics[height=0.19\textwidth]{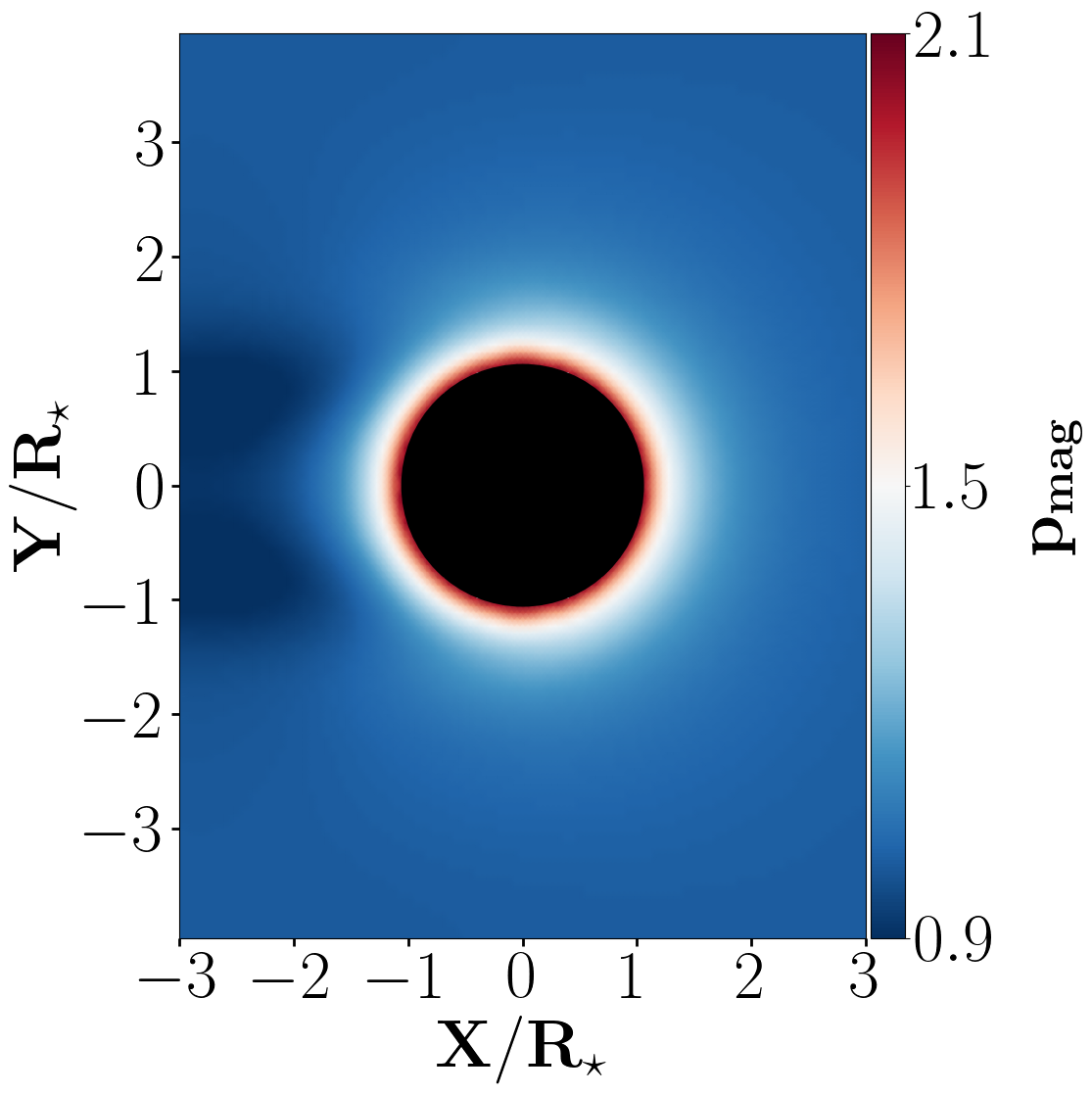}
}
\subfloat{%
\includegraphics[height=0.19\textwidth]{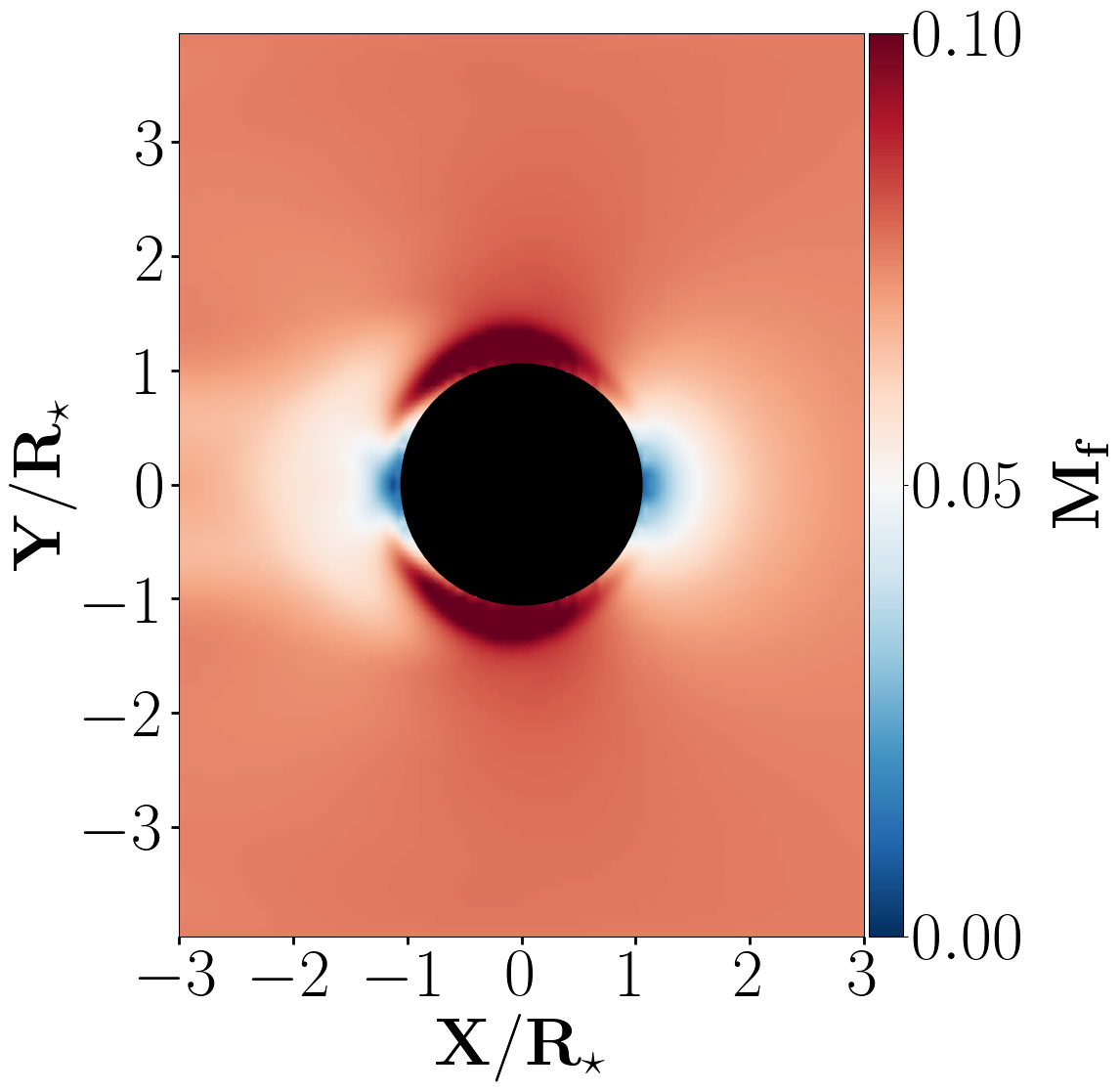}
}\\
\setcounter{subfigure}{0}

\subfloat[][Velocity $v$]{%
    \includegraphics[height=0.19\textwidth]{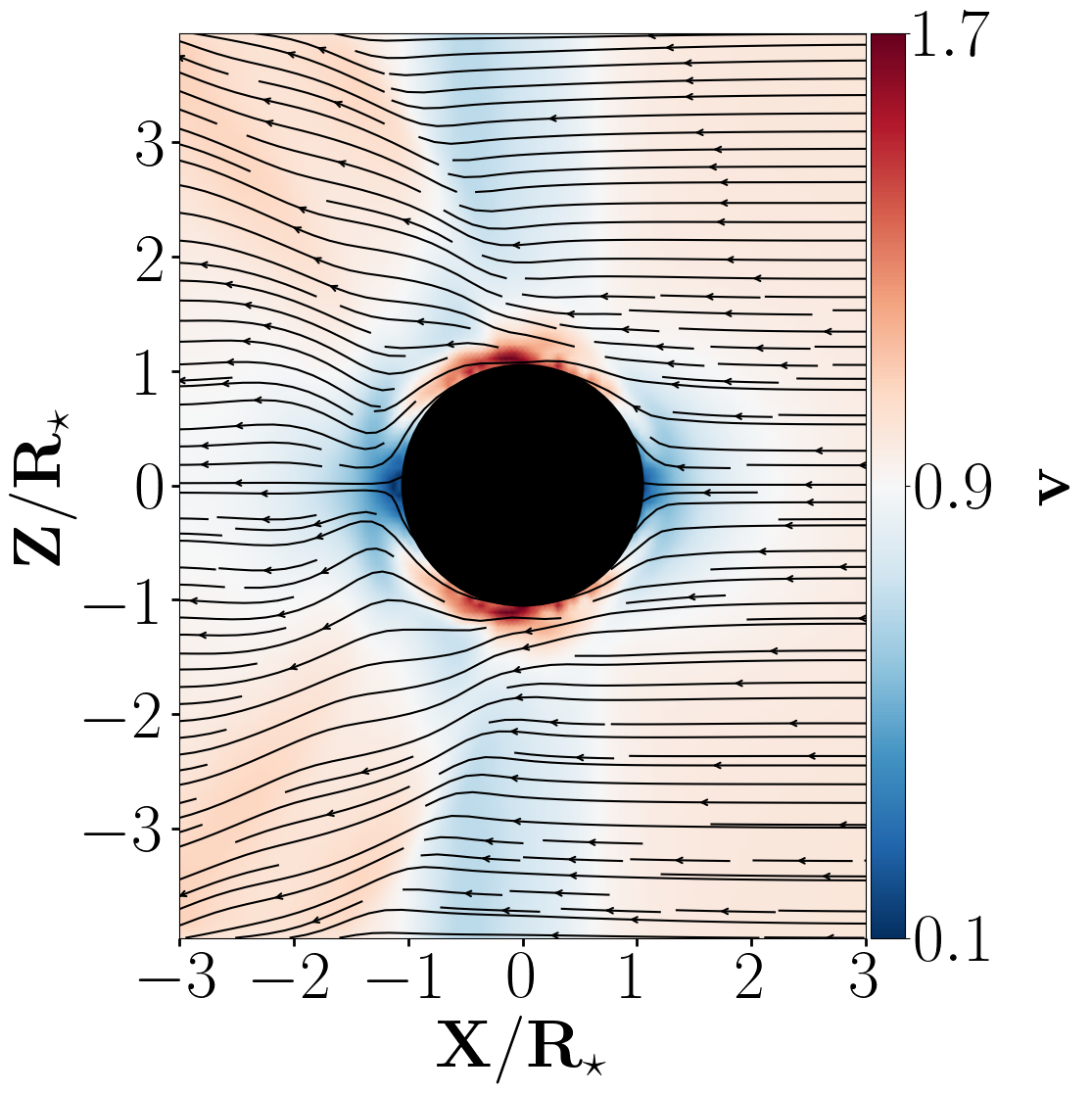}
}
\subfloat[][Density $\rho$]{%
\includegraphics[height=0.19\textwidth]{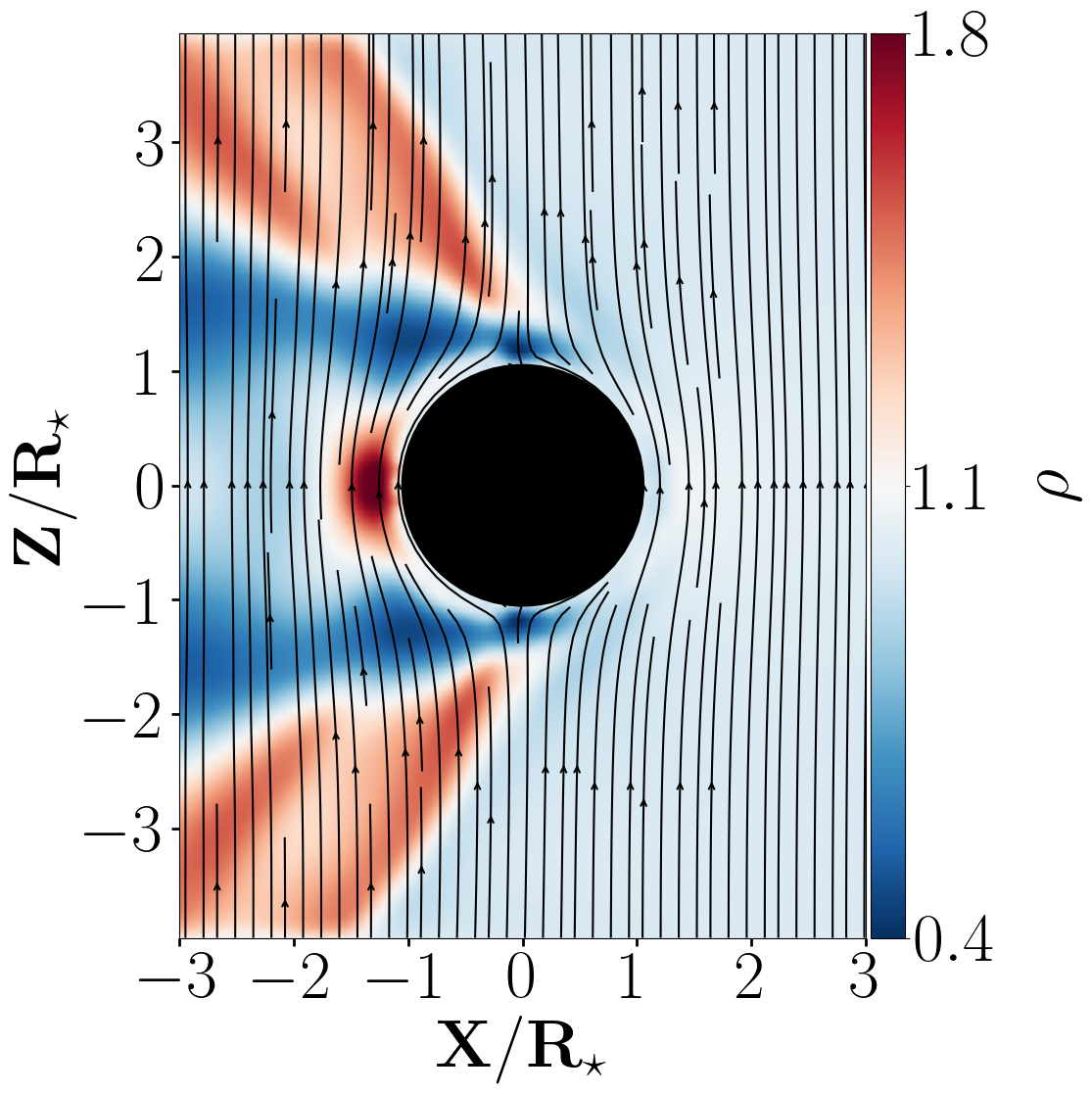}
}
\subfloat[][$p_{ram}$]{%
\includegraphics[height=0.19\textwidth]{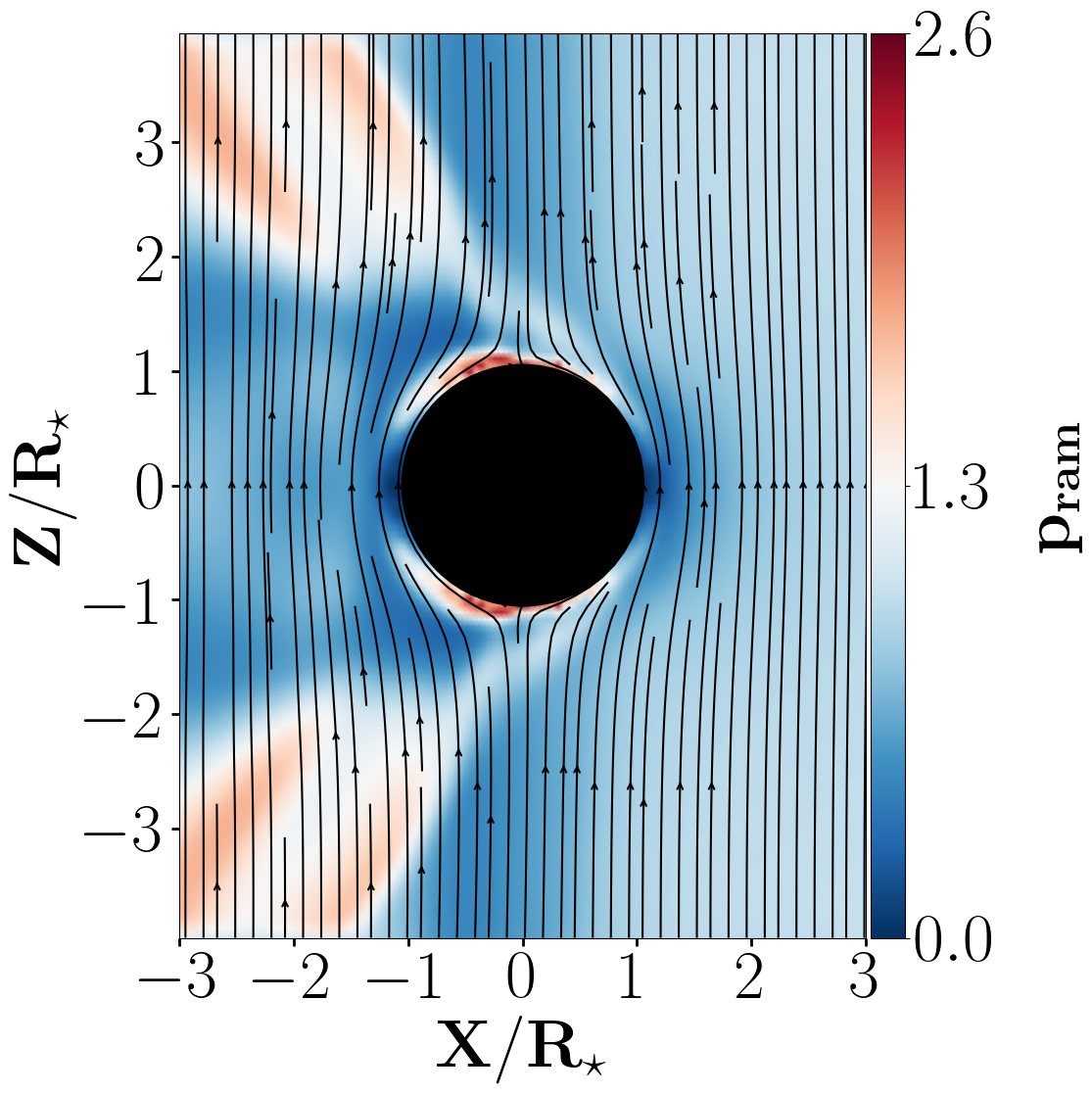}
}
\subfloat[][$p_{mag}$]{%
\includegraphics[height=0.19\textwidth]{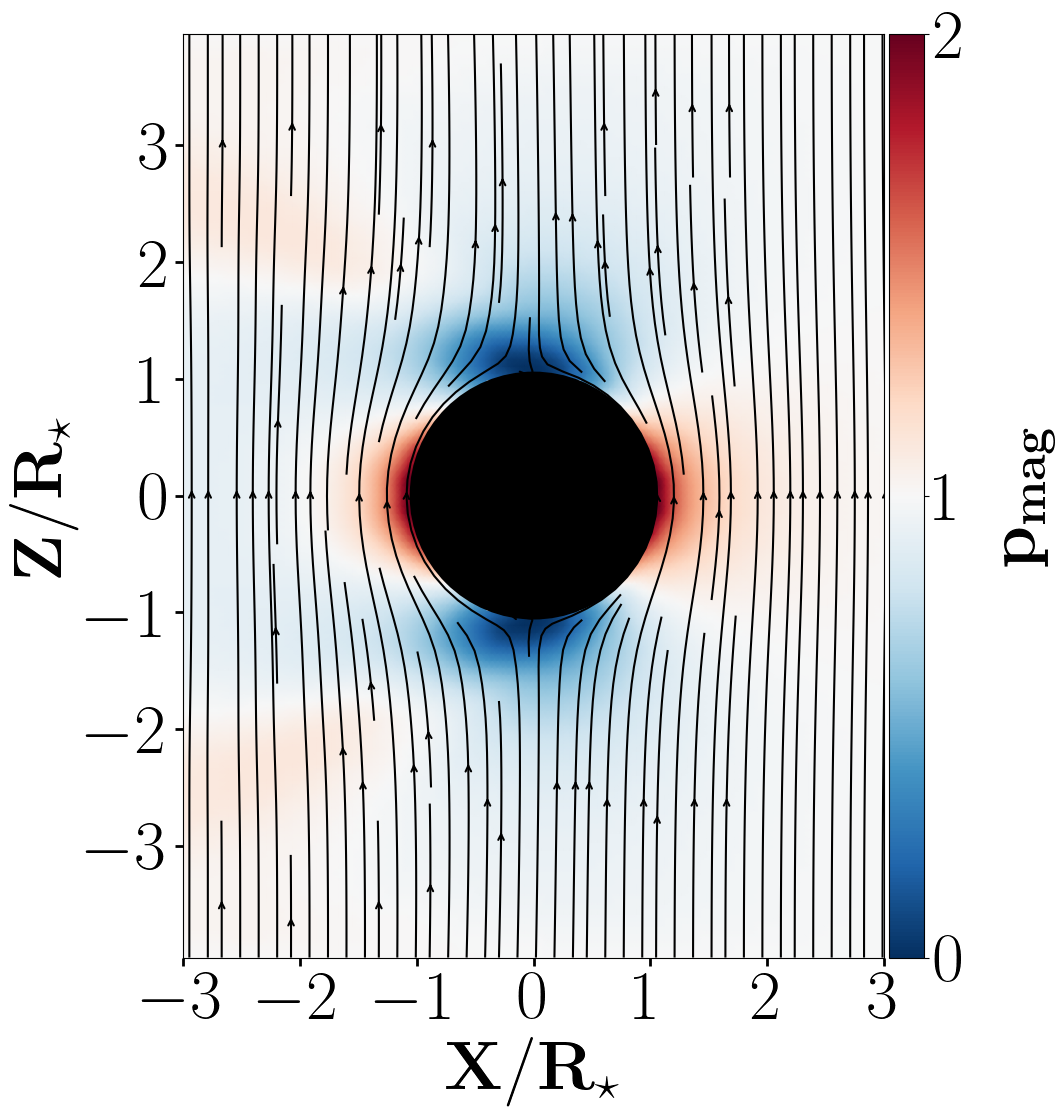}
}
\subfloat[][Fast Mach $\mathcal{M}_{f}$]{%
\includegraphics[height=0.19\textwidth]{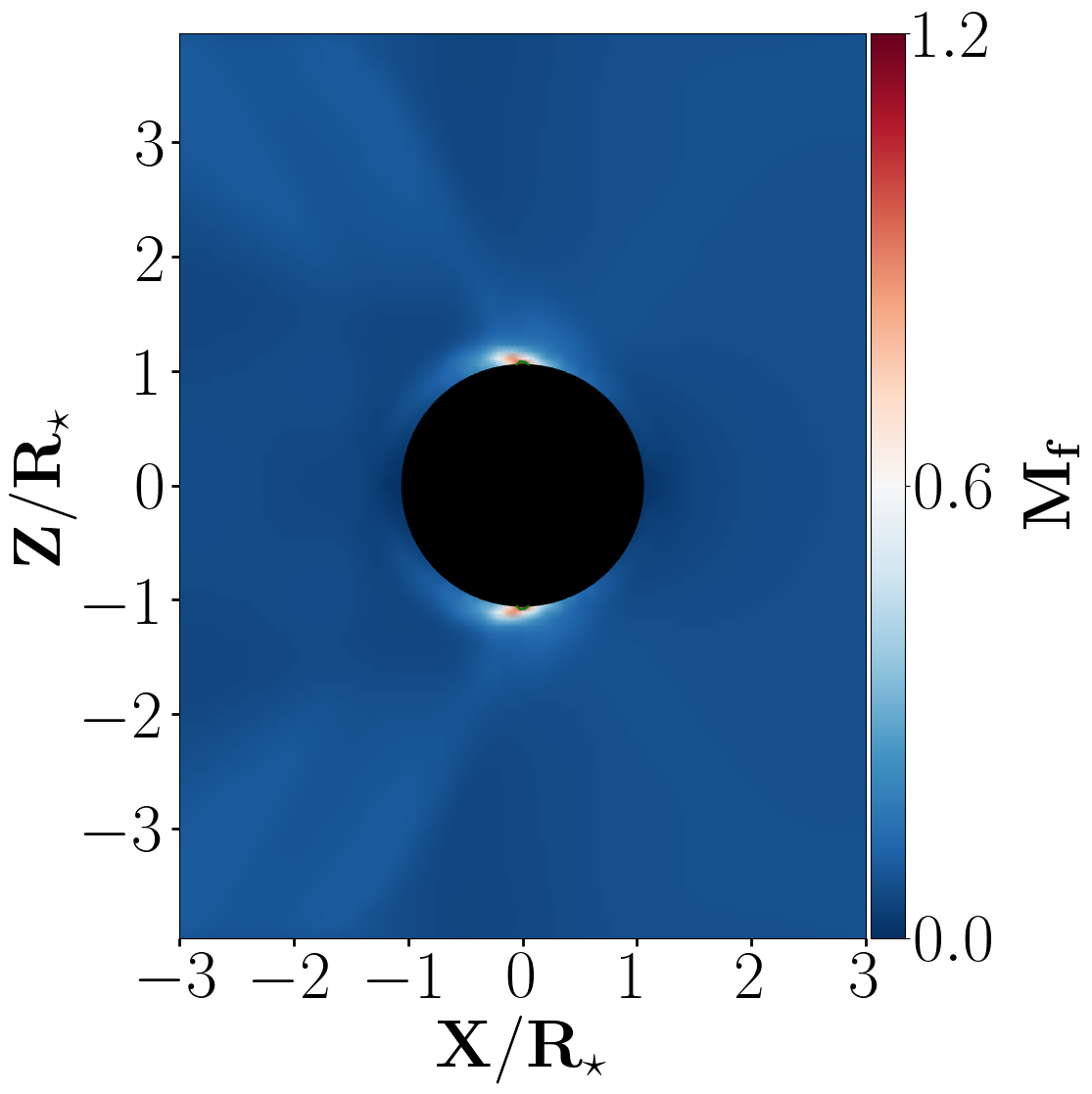}
}
\caption{Non-relativistic Sub-\Alfvenic run A1.  Top row: $xy$-plane slices of flow velocity ($v$), density ($\rho$), ram pressure ($p_{\mathrm{ram}}$), and fast magnetosonic Mach number ($\mathcal{M}_{f}$).  
Bottom row: corresponding $xz$-plane slices.  
All quantities, except $\mathcal{M}_{f}$, are normalized by the upstream values. Velocity Streamlines and magnetic field lines are overlaid where applicable. The green contour in fast Mach plots marks the boundary $\mathcal{M}_{f}=1$.}
\label{nonrelativistic_colorplots_A1_2S}
\end{figure}

We first examine the fiducial strongly magnetized and strongly sub-\Alfvenic case (Run A1), which provides a reference for magnetically dominated interactions.
Fig.~\ref{nonrelativistic_colorplots_A1_2S}(a) shows 2D slices of the normalized flow velocity across two physically relevant planes. The flow remains largely symmetric about the upstream axis and closely follows the imposed background state far from the star. As the plasma approaches the conducting sphere, it decelerates and is smoothly redirected along magnetic field lines, producing coherent \Alfven\ wings at an angle $\theta_A \simeq \tan^{-1} \mathcal{M}_A \approx 5^\circ$. These appear as extended regions of reduced velocity and gentle deflection in the meridional plane, while perturbations perpendicular to the flow remain small.

The density and ram-pressure distributions reflect this magnetic channeling ( Fig.~\ref{nonrelativistic_colorplots_A1_2S}b–c). In the equatorial ($xy$) plane, the upstream plasma remains mostly undisturbed, except for an underdense ring very close to the star. Immediately downstream, a dense lobe forms behind the star, followed by a sharp transition to an extended low-density wake. In the $xz$ plane, a similar dense lobe appears immediately behind the star, but the downstream structure extends farther, forming a broad V-shaped under-dense region. High-density regions surrounding this wedge form wing-like structures, with these features corresponding to \Alfven wings and their associated compression zones. Regions of high density coinciding with low ram pressure correspond to stagnation near the conducting surface, where the velocity approaches zero.

\begin{figure}[!htb]
\centering
\includegraphics[width=0.45\textwidth]{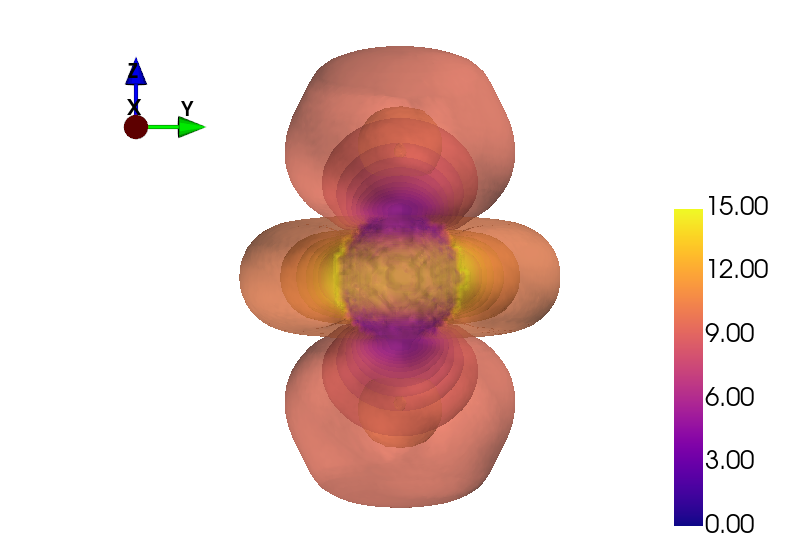}
\caption{Non-relativistic Sub-\Alfvenic run A1. 3D contour plot of the magnetic pressure ${B^2}/{8\pi}$ at the steady state viewed along the x-axis.}
\label{nonrelativistic_pmag_3dcontours_A1_2S}
\end {figure}

Magnetic pressure maps further highlight this behavior. We first present the 3D isocontour plots of the magnetic pressure distribution in Fig.~\ref{nonrelativistic_pmag_3dcontours_A1_2S}, followed by 2D slices in Fig.~\ref{nonrelativistic_colorplots_A1_2S}(d). In the $xy$ plane, we observe a thin ring-like structure with enhanced magnetic pressure surrounding the obstacle. In the $xz$ plane, magnetic pressure reaches local minima near the poles, while roughly symmetric peaks appear in the equatorial regions on either side of the star. The downstream wake shows strong magnetic pressure depletion. This organized topology is consistent with a sub-\Alfvenic flow unable to penetrate the stellar magnetosphere. The fast magnetosonic Mach number remains globally sub-fast ($\mathcal{M}_f \lesssim 1$), with only small localized super-fast patches near the flanks and poles. Overall, this run exhibits the expected picture of a magnetically controlled interaction characterized by coherent wings, modest compression, and smooth flow deflection.

\begin{center}
\textbf{Current Systems and Dissipative Regions}
\end{center}

\begin{figure}[!htb]
\centering
\subfloat[\label{fig:alfven_wing_jmag}]{%
\includegraphics[width=0.49\textwidth]{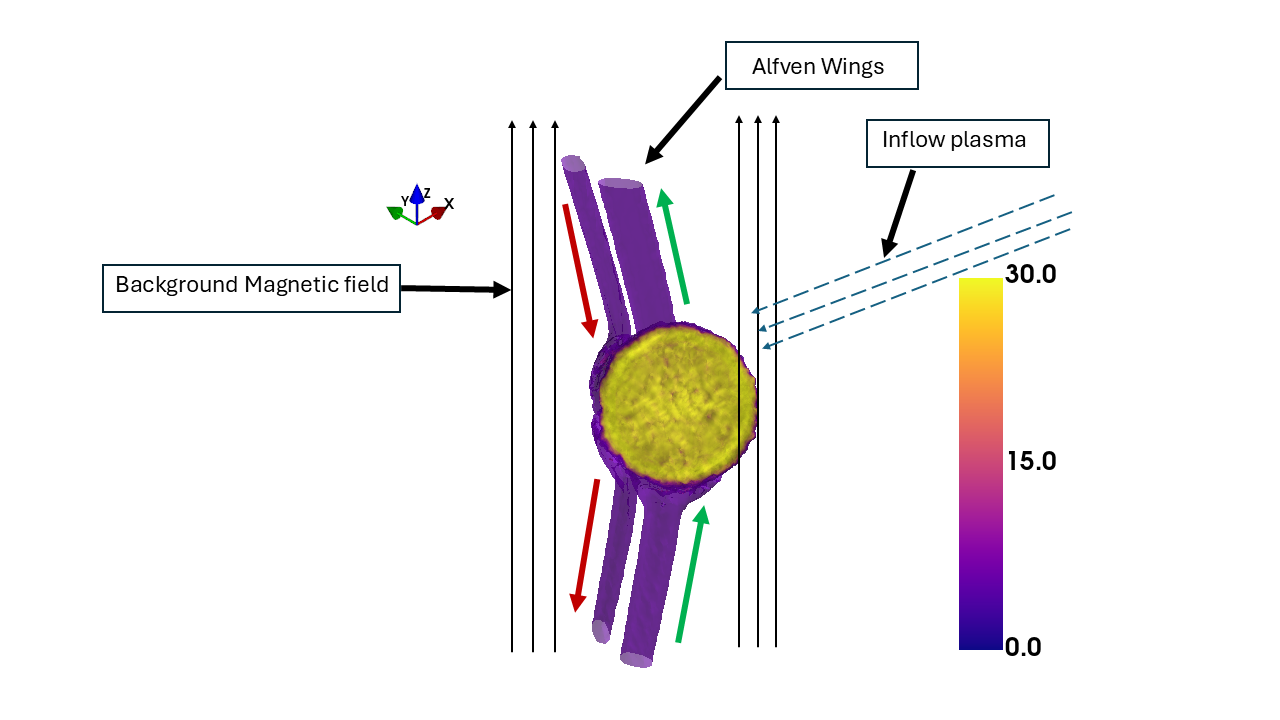}
}
\hfill
\subfloat[\label{fig:alfven_wing_streamlines}]{%
\includegraphics[width=0.49\textwidth]{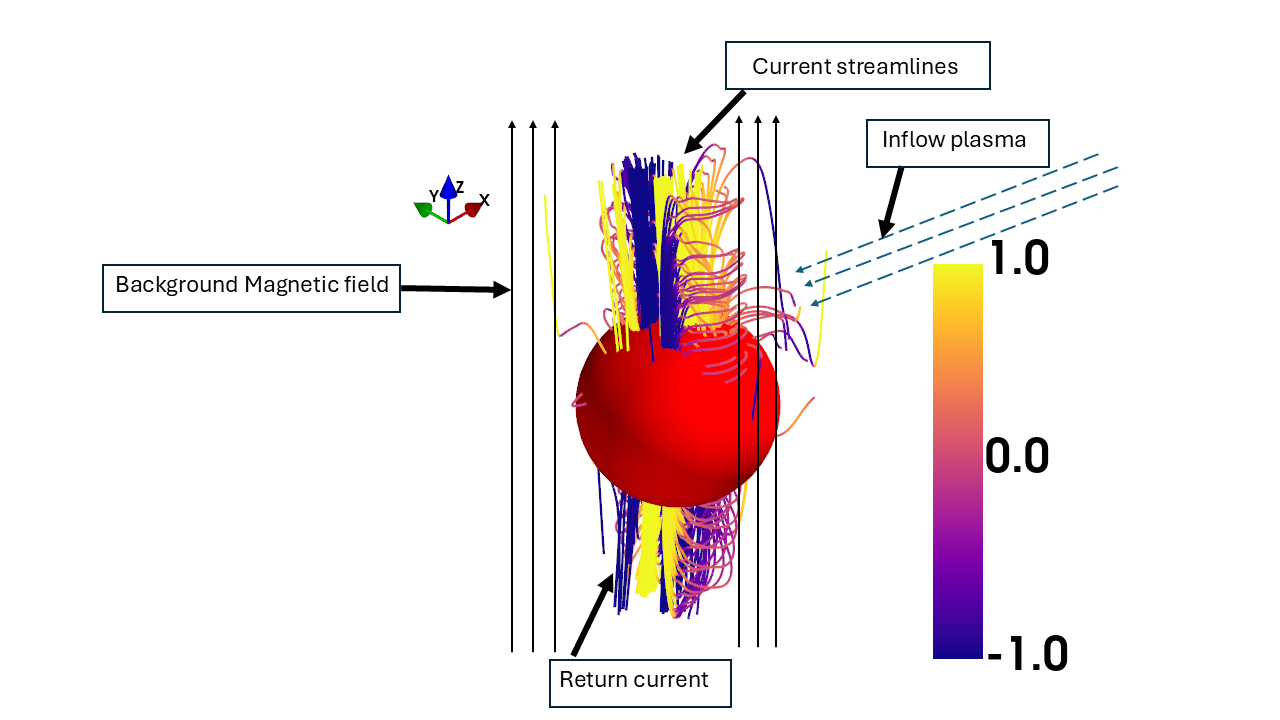}
}
\caption{\Alfven wing structure in illustrative run A1c.
\textbf{(a)} Isosurface of current density magnitude $|\mathbf{J}|$ showing bipolar field-aligned current channel (purple) extending from the neutron star surface. The green and red arrows on the wings indicate upward (+z) and downward (-z) directed currents, respectively, forming the bipolar \Alfven wing current system.
\textbf{(b)} Current streamlines colored by $J_z$ component, revealing the field-aligned nature of the currents.
Both panels show arrows indicating ambient magnetic field $\mathbf{B}_0$ (black) and flow velocity $\mathbf{v}_0$ (dark teal). }
\label{alfven_wing_current_annotated_runA1c}
\end{figure}

The sub-\Alfvenic interaction establishes a set of coupled current systems linking the stellar surface to the surrounding flow. Surface currents form along the conducting boundary to enforce flux exclusion and cancel tangential electric fields, while field-aligned currents are diverted into a pair of symmetric \Alfven wings that carry electromagnetic energy away from the obstacle. Fig.~\ref{alfven_wing_current_annotated_runA1c} illustrates this topology using the sub-\Alfvenic run A1c ($\mathcal{M}_A=0.26$), where the larger opening angle ($\theta_A \simeq \tan^{-1}(\mathcal{M}_A) \approx 15^\circ$) makes the wing structure visually prominent. We are still in the deeply sub-\Alfvenic regime, so the overall physics remains consistent with run A1. The background magnetic field $\mathbf{B}_0$ is shown by vertical black arrows, whereas the inflow plasma velocity is represented by the dark teal dashed arrows. 

Iso-surfaces of $|\mathbf{J}|$ reveal two collimated current channels extending from the stellar surface at the expected \Alfven angle. Streamlines colored by $J_z$ show that these currents are predominantly field-aligned and flow in opposite directions above and below the equatorial plane, forming the characteristic bipolar \Alfven wing system. Enhanced current density immediately surrounding the star marks the thin draping layer where magnetic flux is compressed around the conducting surface. The overall morphology is consistent with classical \Alfven wing interactions observed in planetary magnetospheres \citep{2020JGRA..12527485B, 2017P&SS..137...40V}.

\begin{figure}[!htb]
\centering
\includegraphics[width=0.45\textwidth]{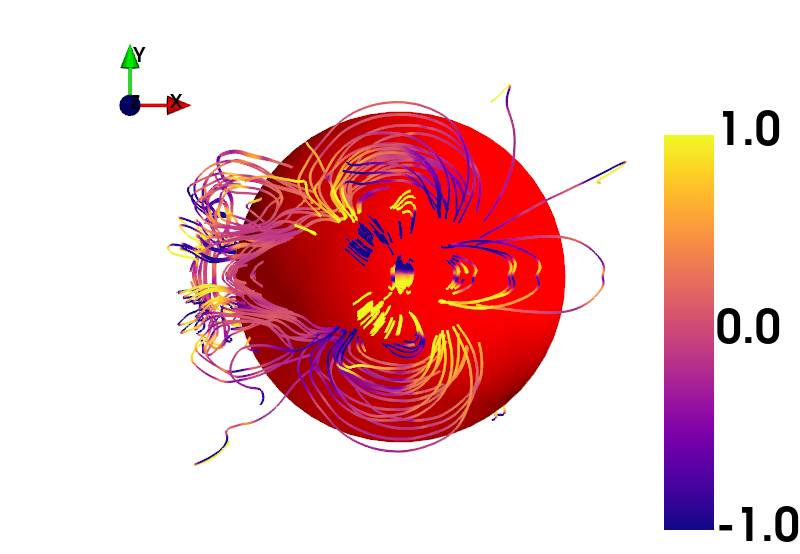}
\includegraphics[width=0.45\textwidth]{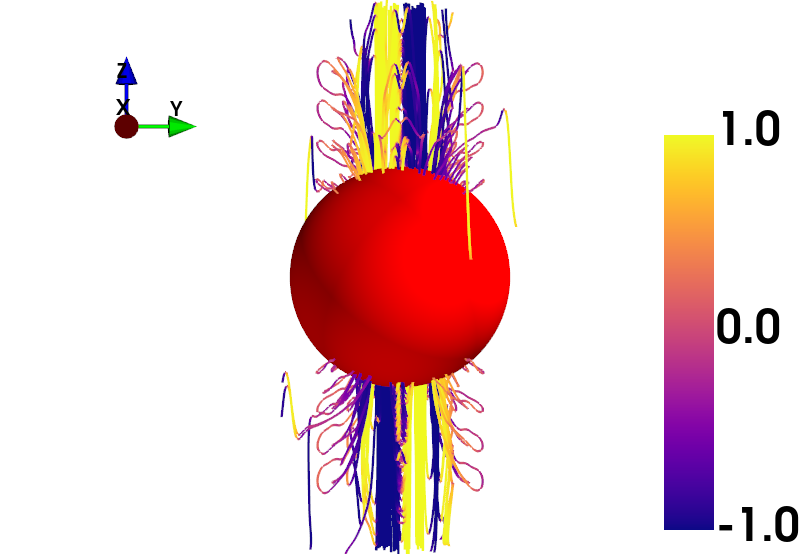}
\caption{Non-relativistic Sub-\Alfvenic run A1. We show 3D streamlines of the current colored by $J_z$ for flow past a perfectly conducting sphere.
\textit{Left:} View along $+z$ direction. The upstream hemisphere ($x>0$) hosts an antisymmetric pair of organized current sheets wrapping around the nose. Downstream ($x<0$), current streamlines show lateral circulation patterns that provide closure paths connecting surface currents to the \Alfven wing channels.
\textit{Right:} View along $+x$ direction. Bipolar, field-aligned $J_z$ structures (\Alfven wings) extend above and below the equator, with organized collimated current channels propagating downstream.}
\label{J_3Dstreamlines_A1_2S}
\end {figure}

Having established the \Alfven wing morphology with the more visually prominent case A1c, we now return to our highly sub-\Alfvenic fiducial run A1. The same current system persists, but with more tightly collimated wings ($\theta_A \approx 5^\circ$). Viewed along the $+z$ direction (left), an upstream--downstream asymmetry is evident. Upstream, currents form organized loops from the magnetic field bending around the star's surface. Downstream ($x<0$), the wake contains current channels that extend along field lines, representing sites of potential electromagnetic emission. In the lateral regions ($\pm y$ directions), current streamlines circulate in opposite senses above and below the equator (clockwise vs.\ counterclockwise), providing closure paths that connect the surface currents to the \Alfven wing channels.

We emphasize that these field-aligned current channels represent \Alfven-wing structures characteristic of sub-Alfvénic magnetospheric interactions, rather than mass-loaded relativistic jets. The ideal MHD framework employed here does not capture jet formation mechanisms such as particle acceleration or radiative cooling that would be required to model astrophysical jets in the traditional sense.

\begin{figure}[!htb]
\centering
\subfloat[$J_z(x = 0)$\label{fig:Jz_x0}]{%
\includegraphics[width=0.31\textwidth]{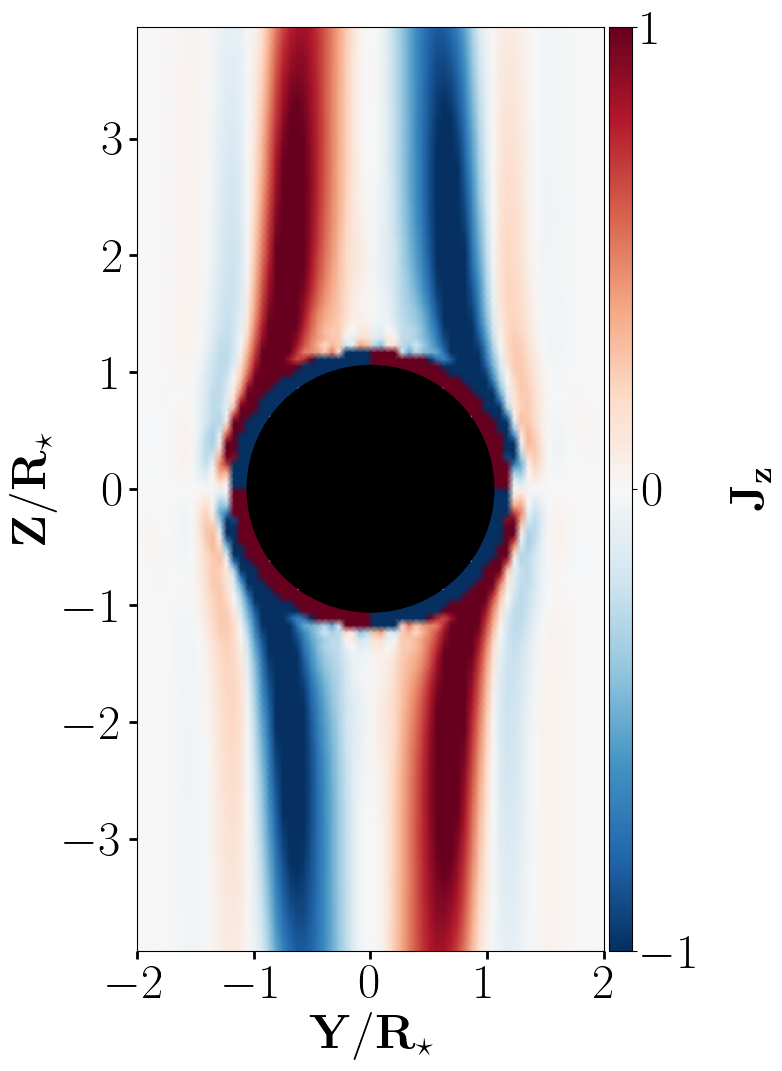}
}
\subfloat[$J_z(y = 0.5)$\label{fig:Jz_y0}]{%
\includegraphics[width=0.31\textwidth]{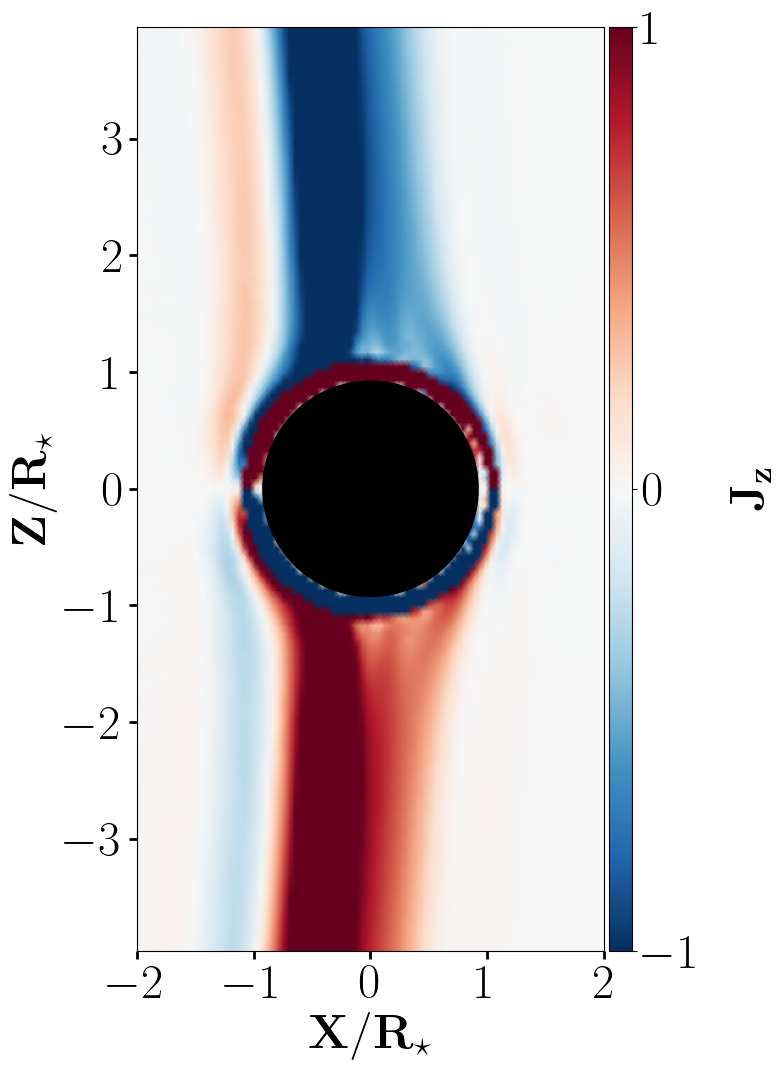}
}
\caption{Non-relativistic Sub-\Alfvenic run A1. Here we show 2D slices of parallel current $J_z$ along $x=0$, and  $y=0.5$. In the left panel, we observe currents with alternating directions, while an upstream-downstream asymmetry is clearly visible in the right image.}
\label{Jz_slices_A1_2S}
\end{figure}

Two-dimensional slices of the parallel current component (Fig.~\ref{Jz_slices_A1_2S}) further highlight this structure. The $x=0$ plane shows a clear bipolar pattern aligned with the background magnetic field and confined near the stellar surface by the draping layer, whereas in the $y=0.5$ plane, the current distribution exhibits an upstream-downstream asymmetry, upstream currents remain weak and diffuse, whereas downstream currents form well-organized channels within the wake.

\subsection{Moderate Sub-\Alfvenic Flow}
\label{results_nonrel_moderate_sub_alfvenic}

\begin{figure}[htbp]
\centering
\subfloat{%
\includegraphics[height=0.19\textwidth]{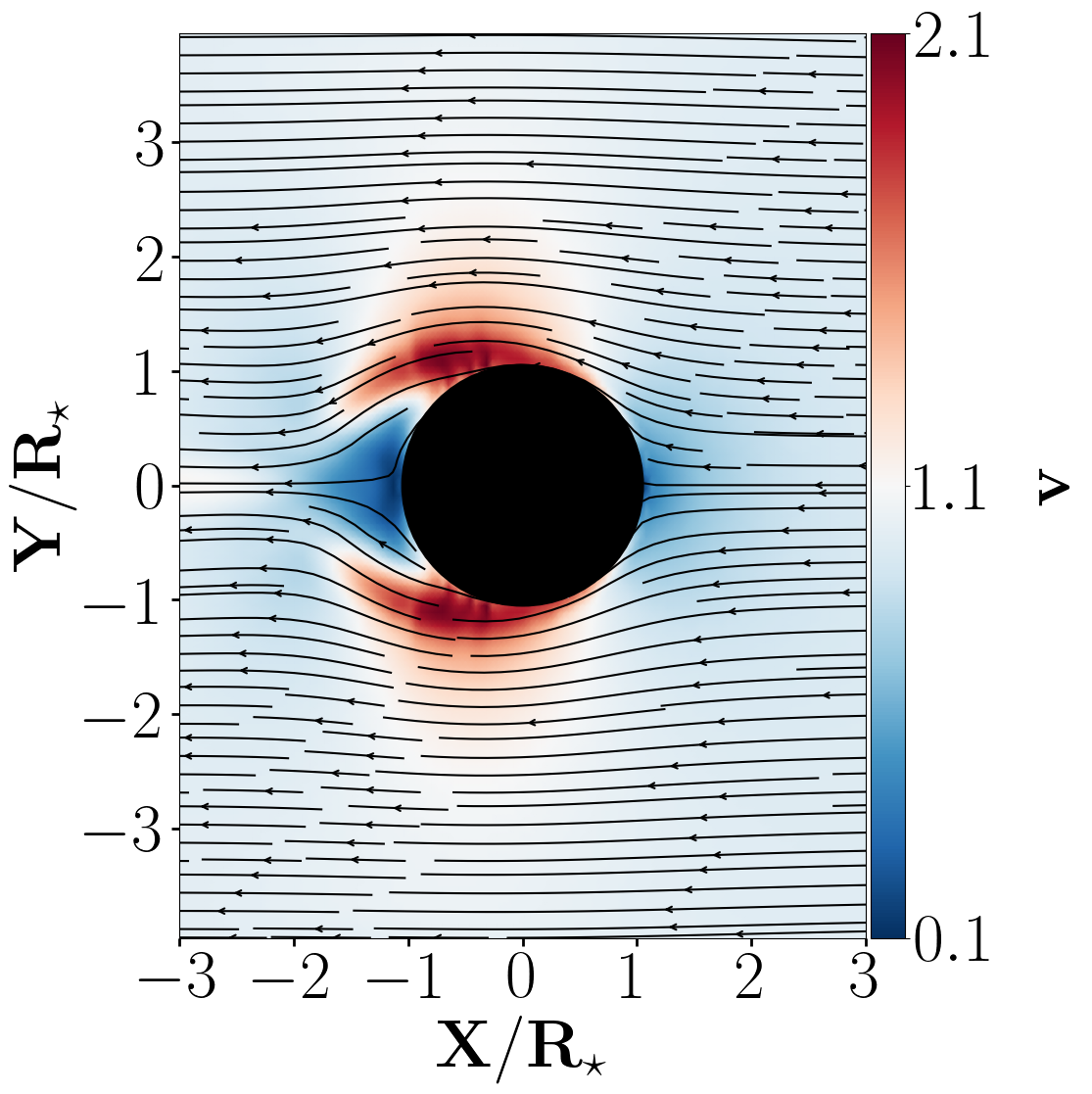}
}
\subfloat{%
\includegraphics[height=0.19\textwidth]{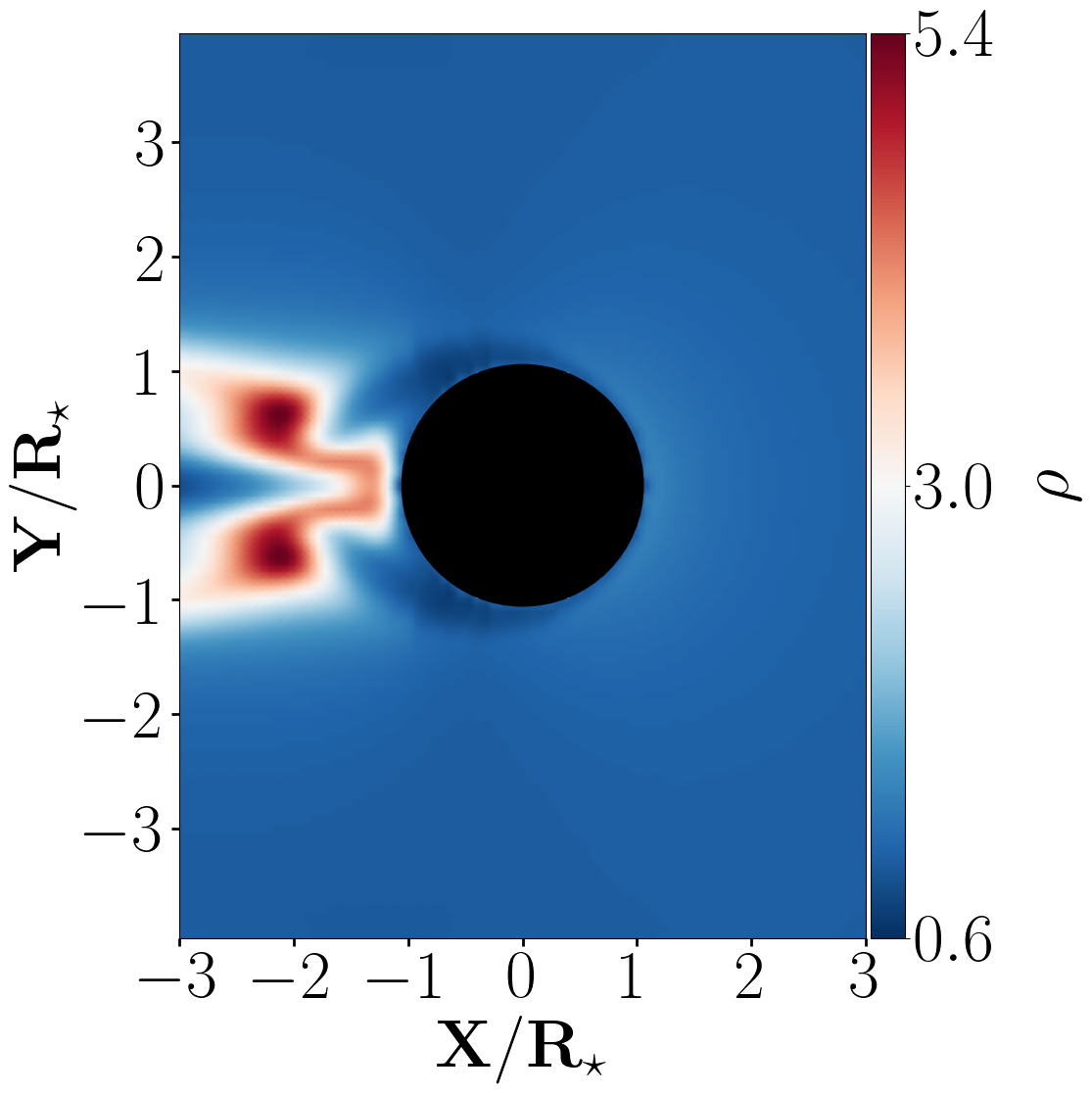}
}
\subfloat{%
\includegraphics[height=0.19\textwidth]{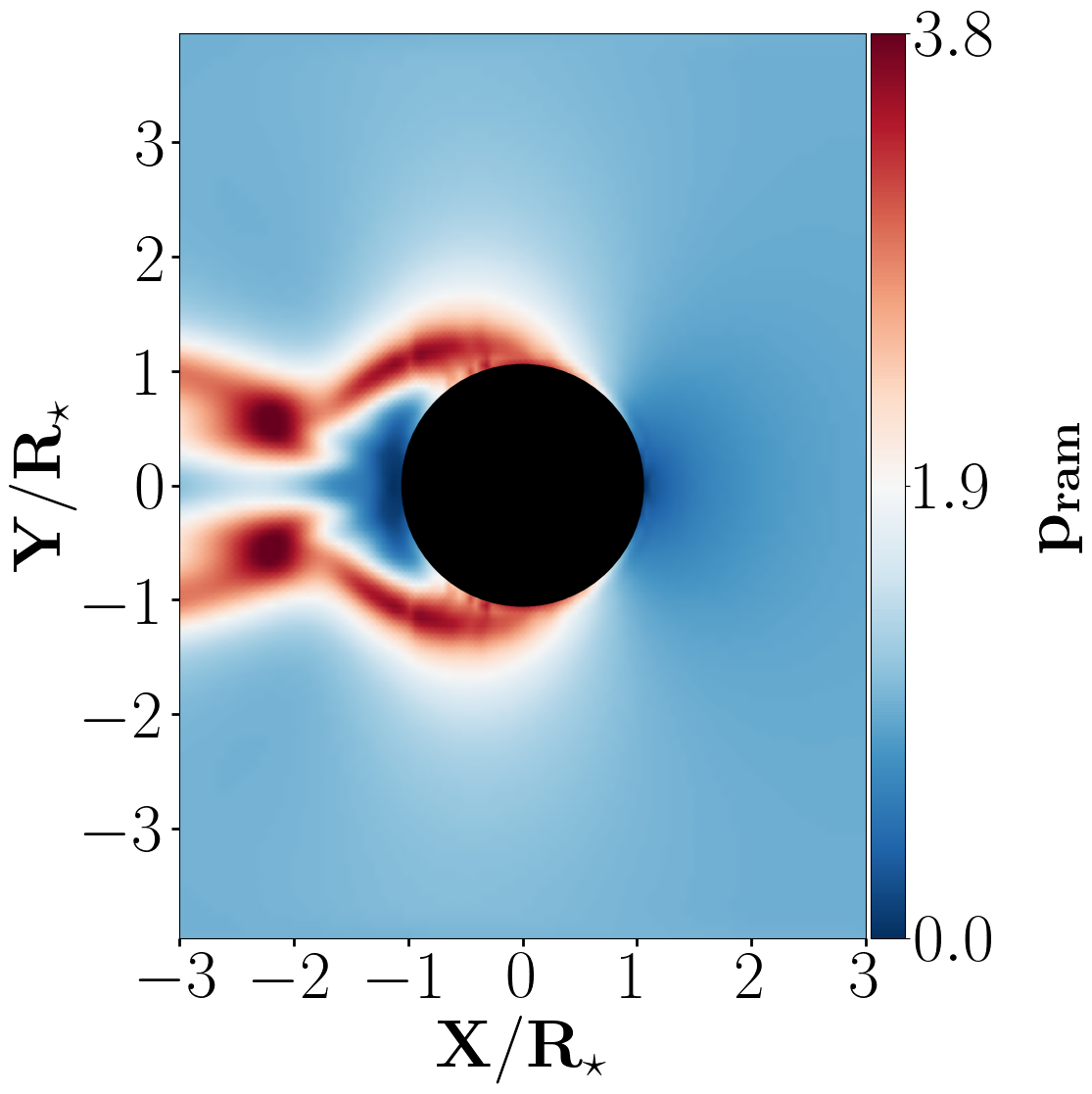}
}
\subfloat{%
\includegraphics[height=0.19\textwidth]{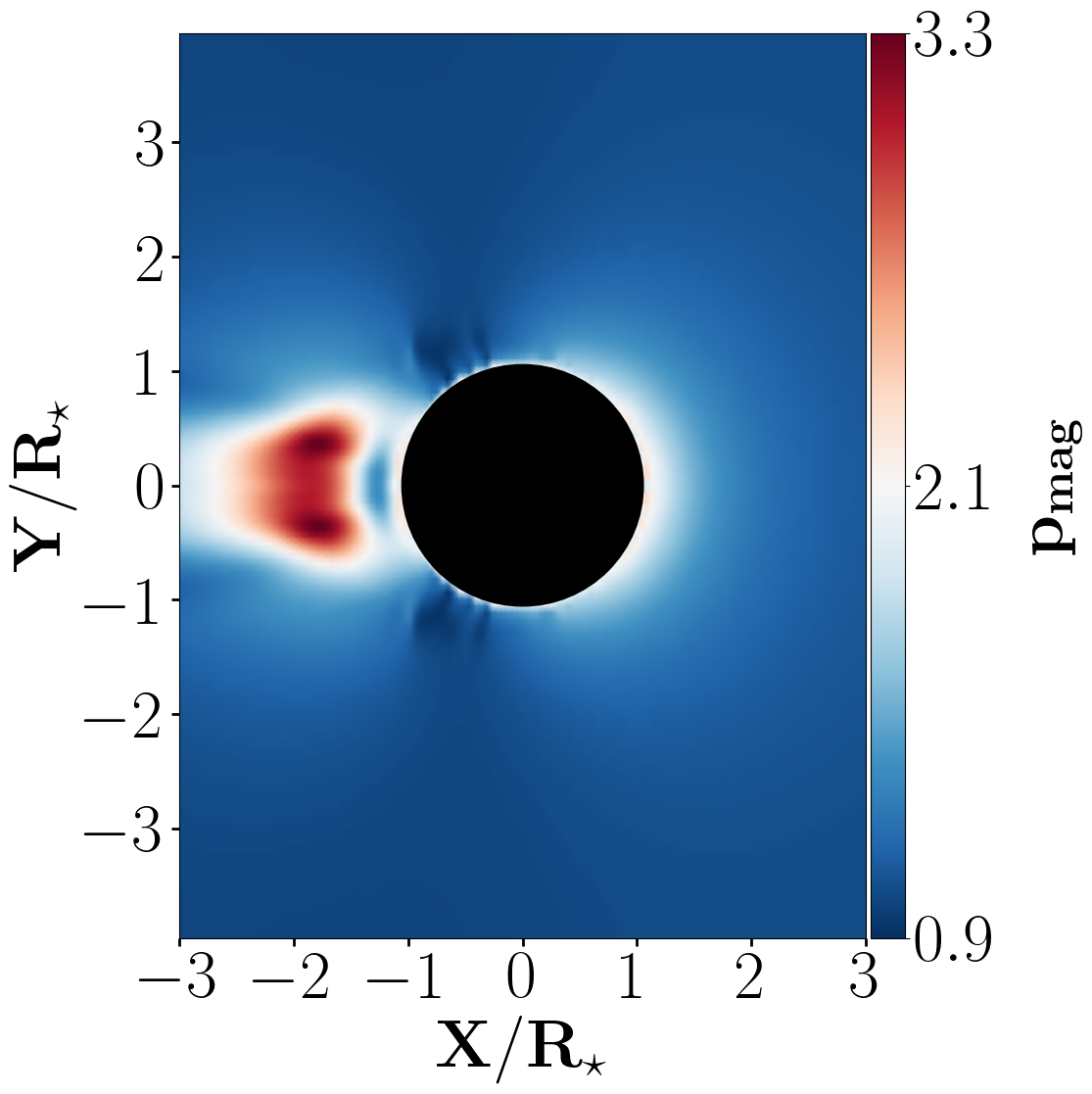}
}
\subfloat{%
\includegraphics[height=0.19\textwidth]{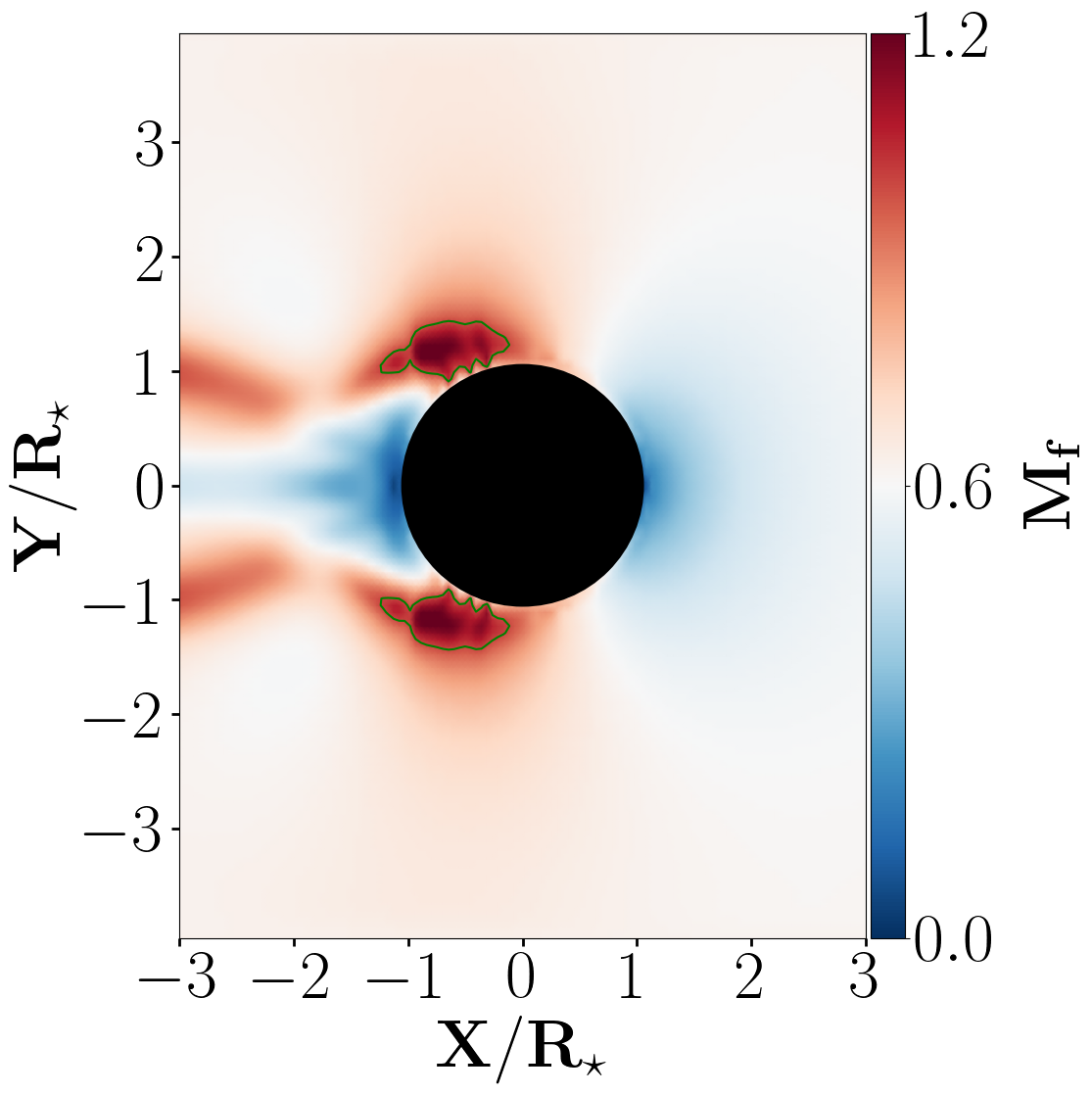}
}\\
\setcounter{subfigure}{0}
\subfloat[Velocity $v$]{%
\includegraphics[height=0.19\textwidth]{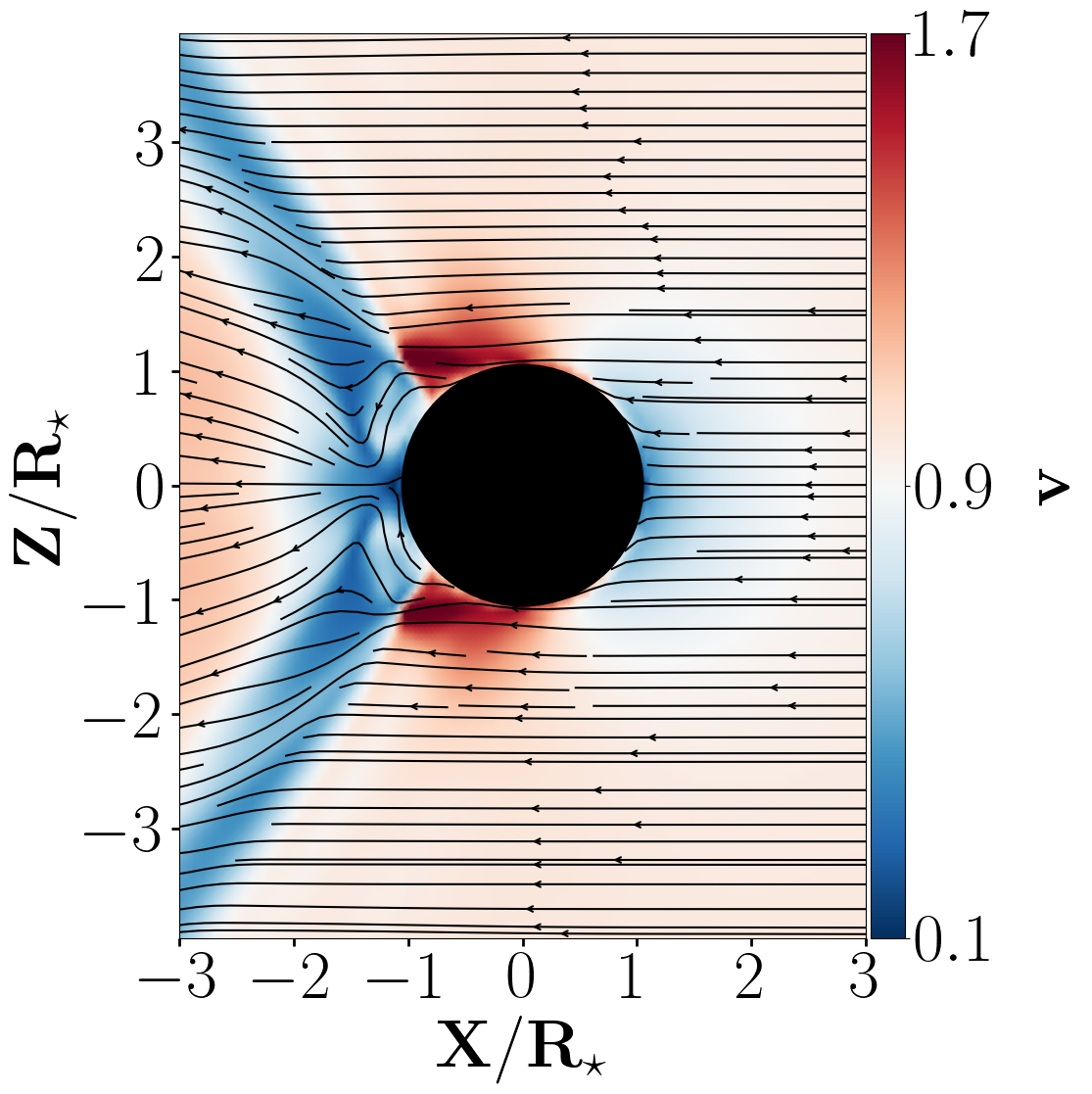}
}
\subfloat[Density $\rho$]{%
\includegraphics[height=0.19\textwidth]{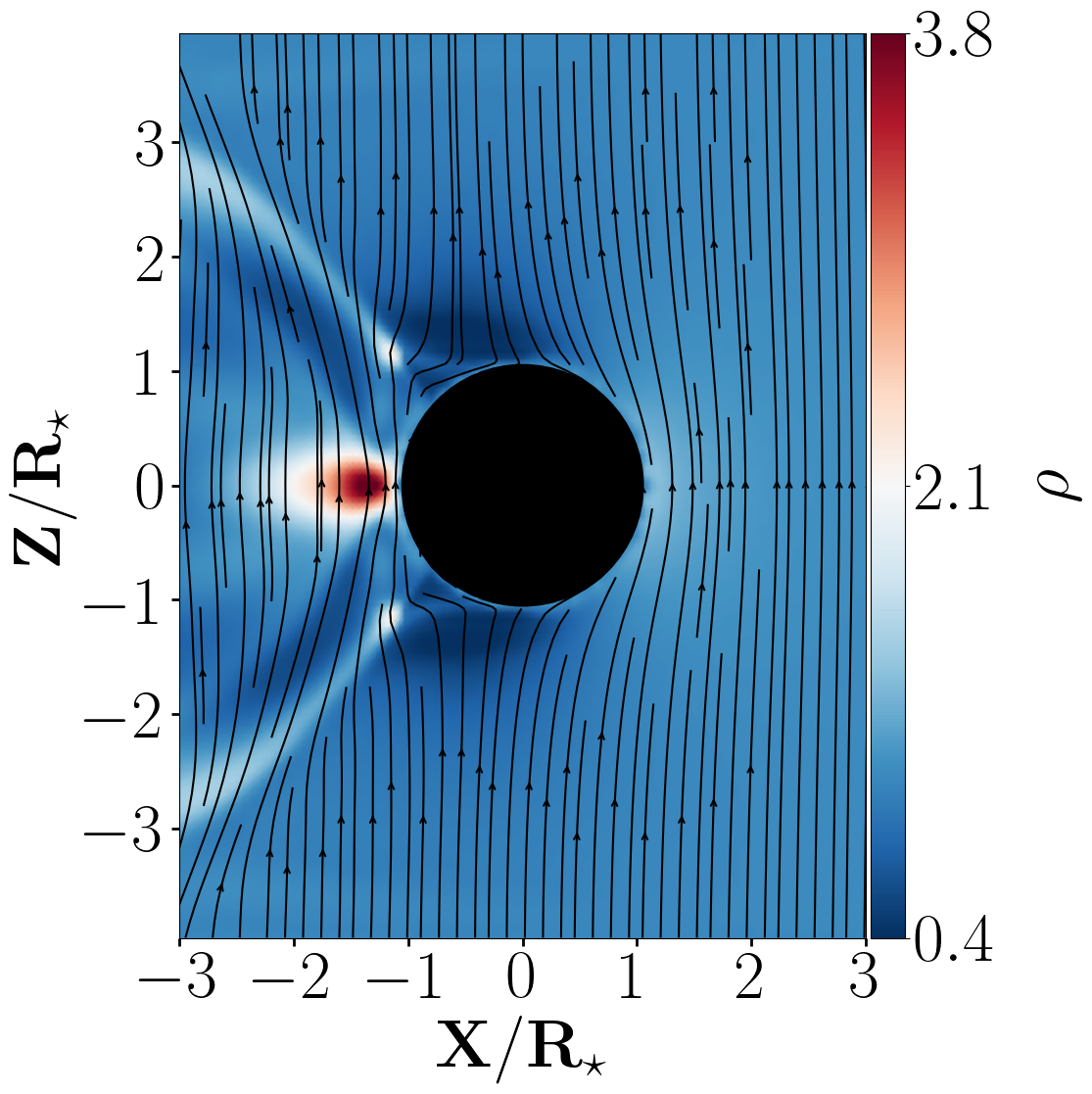}
}
\subfloat[$p_{ram}$]{%
\includegraphics[height=0.19\textwidth]{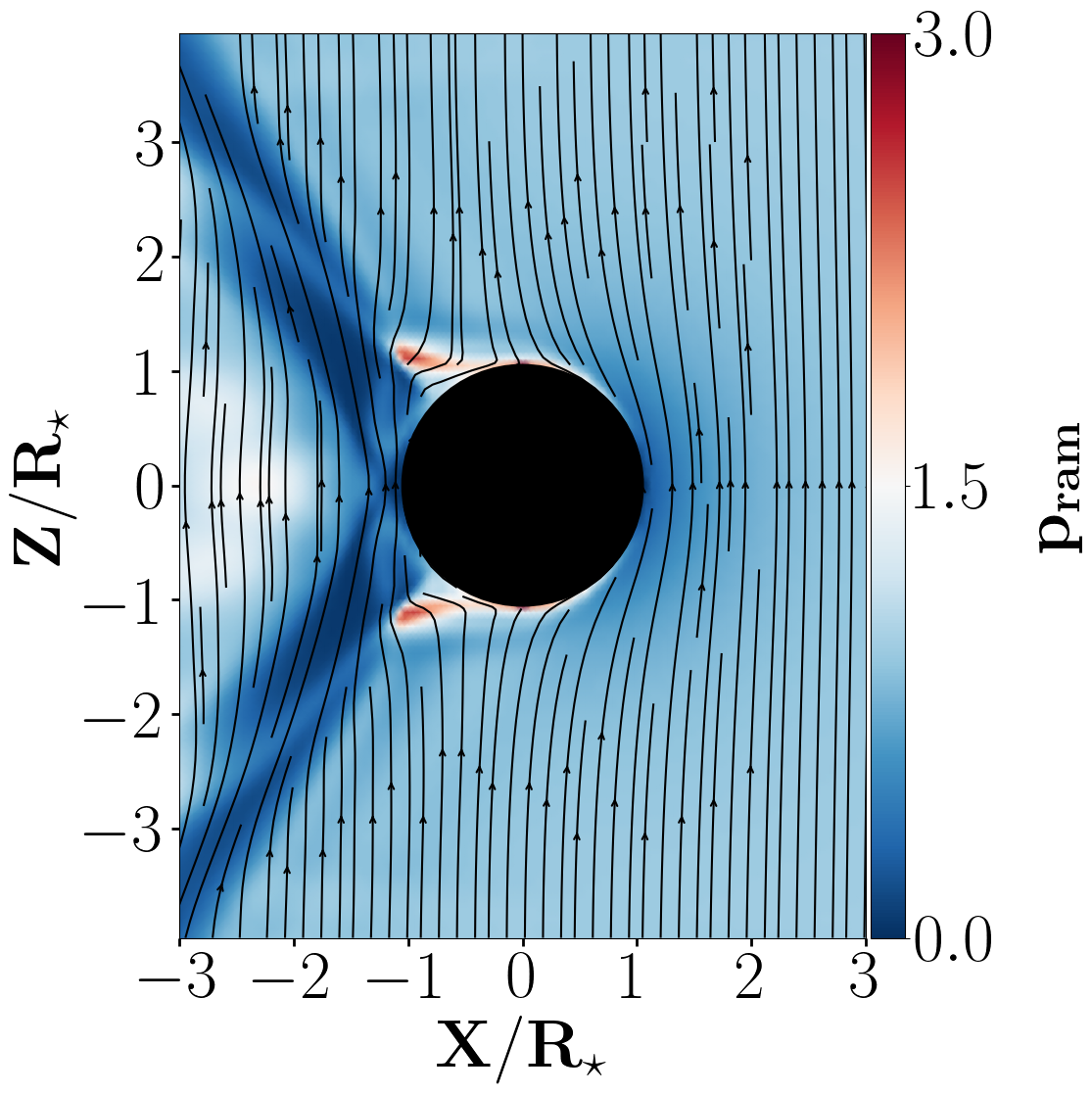}
}
\subfloat[$p_{mag}$]{%
\includegraphics[height=0.19\textwidth]{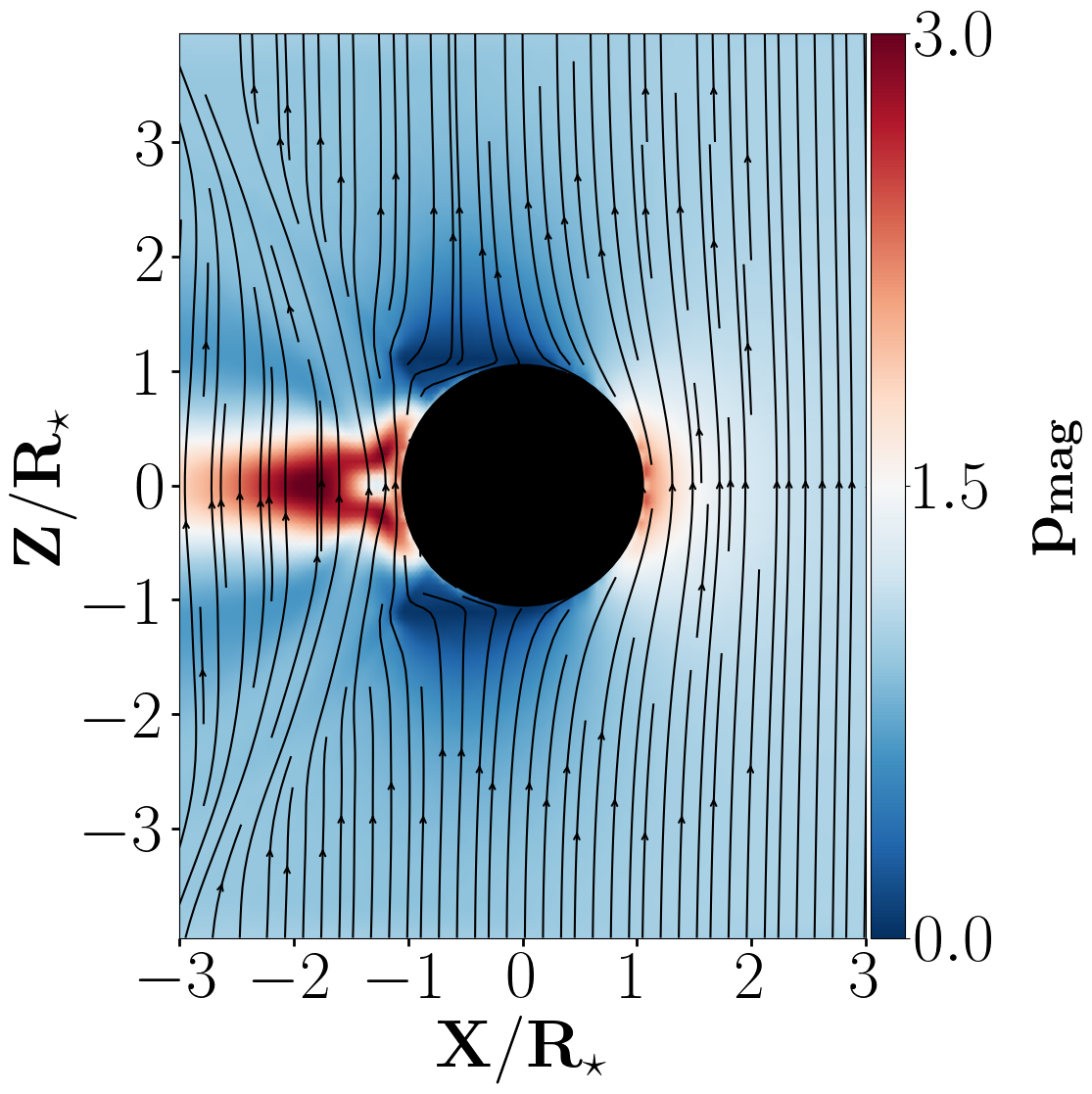}
}
\subfloat[Fast Mach $\mathcal{M}_{f}$]{%
\includegraphics[height=0.19\textwidth]{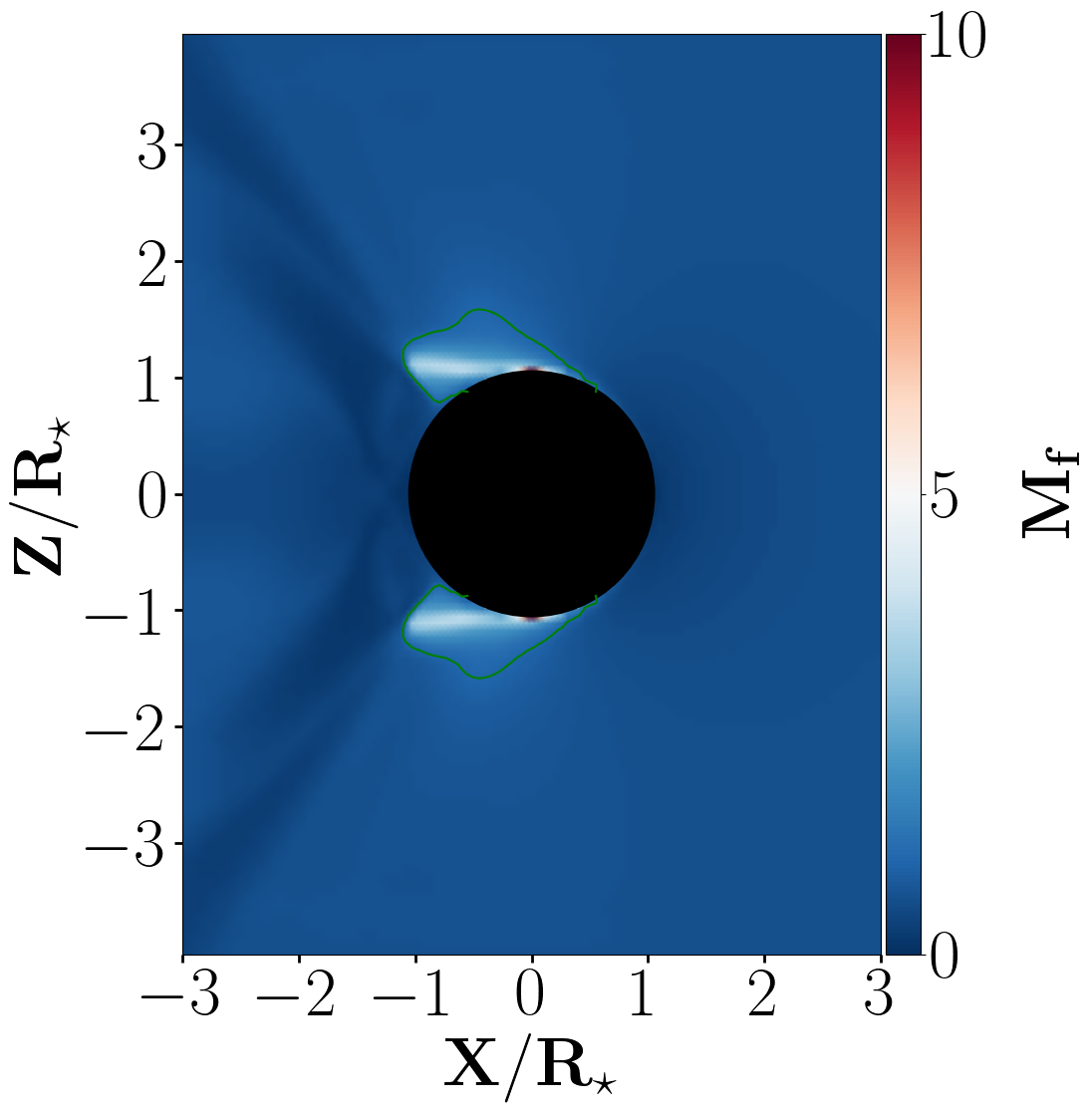}
}
\caption{Moderate sub-\Alfvenic run A2 ($\mathcal{M}_{A} \approx 2.2, \beta = 0.1$). The details are the same as in Fig.~\ref{nonrelativistic_colorplots_A1_2S}.}
\label{nonrelativistic_colorplots_A2_2S}
\end{figure}

We next consider a moderately sub-\Alfvenic case (run A2; $\mathcal{M}_A \approx 0.63$, $\beta \approx 0.01$), which lies in the transitional regime between coherent \Alfven wings and more compressive, asymmetric flow. In contrast to the smooth, magnetically guided structure of the strongly sub-\Alfvenic case, the interaction here becomes more compressive and asymmetric. The velocity field (Fig.~\ref{nonrelativistic_colorplots_A2_2S}a) shows stronger deflection around the obstacle and the development of an extended downstream wake where the flow nearly stagnates. Correspondingly, the density and ram-pressure maps (Fig.~\ref{nonrelativistic_colorplots_A2_2S}b–c) exhibit substantial compression immediately behind the star, with peak values reaching $\rho/\rho_0 \sim 5$ and enhanced ram pressure along the equatorial flanks. These high-pressure lobes trace regions where the converging flow accumulates before being redirected into the wake.

Magnetic pressure enhancements become more localized and concentrated than in the strongly sub-\Alfvenic run (Fig.~\ref{nonrelativistic_colorplots_A2_2S}d), indicating stronger field compression near the obstacle. Consistent with this picture, the fast magnetosonic Mach number (Fig.~\ref{nonrelativistic_colorplots_A2_2S}e) depicts extended super-fast regions and a wide downstream wedge. At the same time, sub-fast flow is now confined to regions near the stellar surface and portions of the upstream flow.

\begin{figure}[!htb]
\centering
\includegraphics[width=0.45\textwidth]{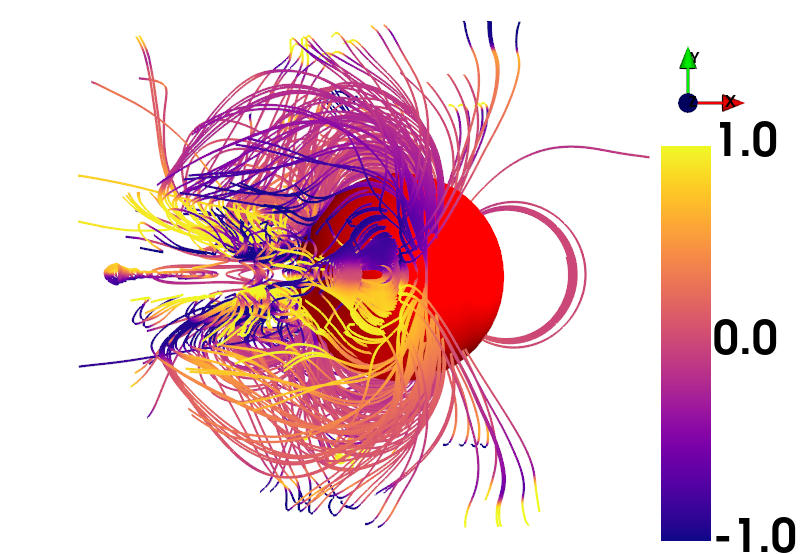}
\includegraphics[width=0.45\textwidth]{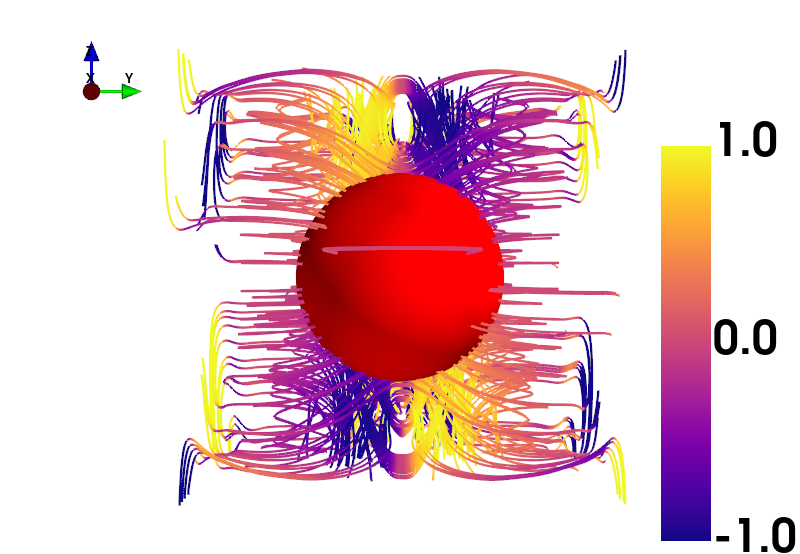}
\caption{Moderate sub-\Alfvenic run A2. 3D streamlines plot of current density at the steady state as viewed across $+z$ axis (left panel) and $+x$ axis (right panel)} 
\label{J_3Dstreamlines_A2_2S}
\end {figure}

Fig.~\ref{J_3Dstreamlines_A2_2S} shows the three-dimensional current streamlines for this run. Although the current structure is more complex than in the strongly sub-\Alfvenic regime (run A1), organized field-aligned patterns remain visible. Viewed along the $+x$ direction (left), current streamlines wrap around the obstacle with increased concentration in the upstream and lateral regions, while the downstream wake contains denser and more bundled current channels. Viewed along the $+z$ direction (right), a bipolar current system persists above and below the equator, indicating that the \Alfven wing geometry survives as the flow approaches the trans-\Alfvenic regime. This coexistence of compressed, tangled currents with residual wing-like structure is consistent with the transitional nature of this regime.

Finally, we quantify how the field-aligned current strength depends on the upstream flow and magnetic field by measuring the integrated current in the \Alfven\ wings across several non-relativistic, sub-\Alfvenic\ runs. 

\begin{table}[ht]
\centering
\begin{tabular}{cccc|cccc}
\toprule
$v_0$ & $B_0$ & $M_A$ & $I_{wings}$ & $v_0$ & $B_0$ & $M_A$ & $I_{wings}$ \\
\midrule
0.10 & 7.09 & 0.014 & 0.1031 & 0.80 & 15.85 & 0.05 & 0.6445 \\
0.20 & 1.59 & 0.126 & 0.1569 & 1.20 & 15.85 & 0.076 & 0.8890 \\
0.40 & 5.01 & 0.08 & 0.3208 & 4.20 & 15.85 & 0.26 & 3.0736 \\
0.40 & 15.85 & 0.025 & 0.3880 & 10.0 & 15.85 & 0.63 & 9.683 \\
\bottomrule
\end{tabular}
\caption{Measured wing current $I_{wings}$ through $z=1.5$ for various non-relativistic sub-\Alfvenic runs.}
\label{tab:nonrel_sub_alfvenic_currents}
\end{table}

To quantify this behavior, we computed the flux of field–aligned current density through transverse $xy$ slices at fixed $z=z_0$. We first calculated the field-aligned current density $\mathbf{J}_{\parallel}= (\mathbf{J} \cdot \mathbf{\hat b})\mathbf{\hat b}$, where $\mathbf{\hat b}$ is the local magnetic field unit vector, and then calculated the current via $\iint \mathbf{J}_{\parallel} \cdot d\mathbf{A}$, giving us

\begin{equation}
\begin{aligned}
I_{+}(z_0)= \iint_{y>0} J_{\parallel,z}(x,y,z_0)\; dA \\
I_{-}(z_0)=\iint_{y<0} J_{\parallel,z}(x,y,z_0)\; dA 
\label{I_parallel_z0}
\end{aligned}
\end{equation}

Due to the anti-symmetry of the current system, this magnitude is roughly the same in the other half ($y < 0$), so the total current carried by the two \Alfven wings is $I_{wings}(z_0)=|I_{+}(z_0)|+|I_{-}(z_0)|$.
The wings are cleanly separated by the sign of $y$, reflecting the expected bipolar structure. For this calculation, we evaluated the current at $z_0 = 1.5 R_\star$, a distance sufficient to capture the wing structure outside the stellar surface.

\begin{figure}[!htb]
\centering
\includegraphics[width=0.75\textwidth]{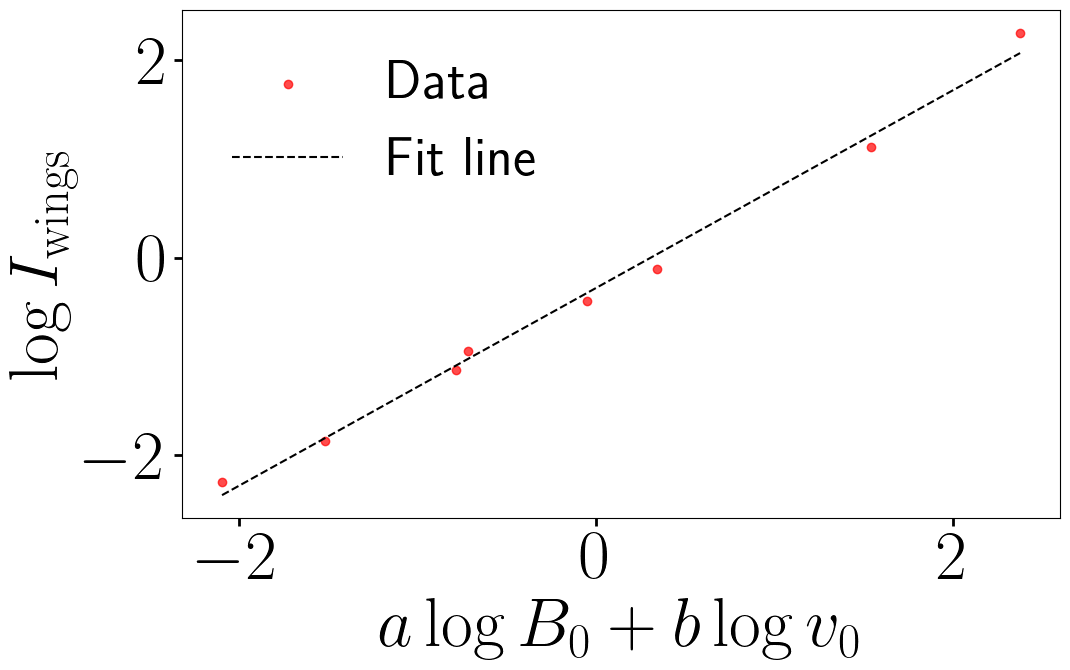}
\caption{Scaling of the integrated field-aligned current through the $z_0=1.5 R_\star$ transverse plane ($I_{wings}$) as a function of the inflow velocity $v_0$ and background magnetic field strength $B_0$. 
A log-log regression over selected 7 sub-Alfvénic runs (Table~\ref{tab:nonrel_sub_alfvenic_currents}) gives
$I_{wings} \propto v^{0.96 \pm 0.05 } B_0^{0.06 \pm 0.08}$ with $R^2 \approx 0.99$. The stronger sensitivity to flow speed suggests that faster flows drive substantially larger currents.}
\label{Iz_scaling_nonrel}
\end{figure}

As shown in Fig.~\ref{Iz_scaling_nonrel}, the total field-aligned current shows a power-law dependence:
\begin{equation}
\begin{aligned}
I_{wings} &\propto v_0^{0.96} B_0^{0.06}  \propto \mathcal{M}_{A}^{0.96} B_0^{1.02}
\end{aligned}
\end{equation}
indicating that the current is primarily controlled by the flow speed, with only a weak dependence on $B_0$. Because $\mathcal{M}_A$ depends on $B_0$ through $v_A$ at fixed density, the apparent $\mathcal{M}_A$ scaling should be interpreted as an empirical fit over the explored parameter range rather than as independent scaling with $\mathcal{M}_A$ and $B_0$.

Similar scalings arise in analytic and semi-analytic models of planetary \Alfven\ wing interactions \citep{1980JGR....85.1171N, 2025MNRAS.539.3459D, 2011PhRvD..83f4001L}. We would like to mention that our definition of $I_{wings}$ refers specifically to the field-aligned current flux crossing a transverse $z=z_0$ plane. In contrast, traditional models generally consider currents integrated over the full surface of the obstacle. So variations from theory are expected. The strong sensitivity to $v_0$ suggests that faster flows can drive substantially larger currents. In astrophysical settings such as compact-object binaries, this trend may enhance electromagnetic energy transport during late inspiral, potentially strengthening any associated precursor emission.

\subsection{Super-\Alfvenic Flow}
\label{results_nonrel_super_alfvenic}

\begin{figure}[htbp]
\centering
\subfloat{%
\includegraphics[height=0.19\textwidth]{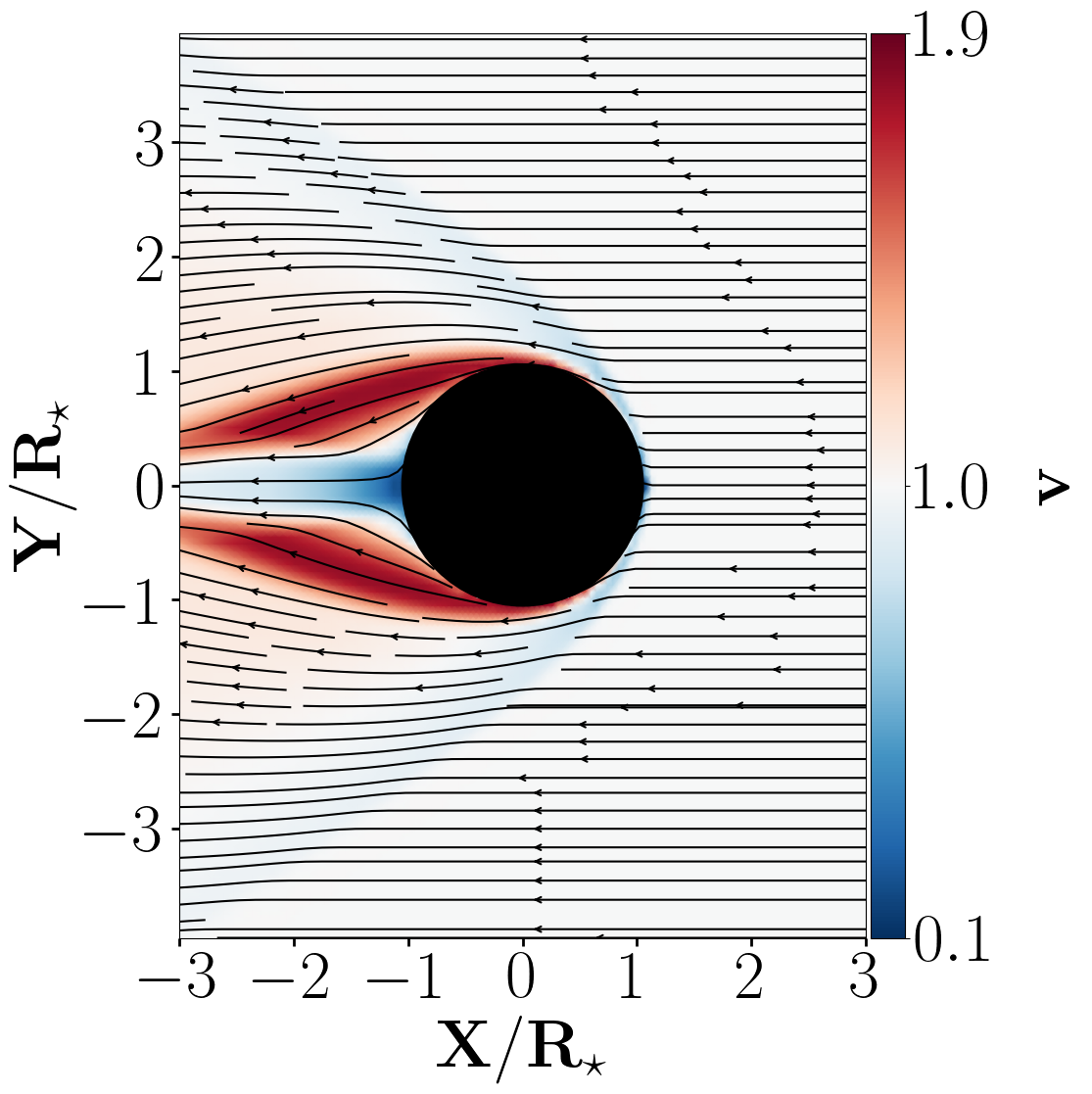}
}
\subfloat{%
\includegraphics[height=0.19\textwidth]{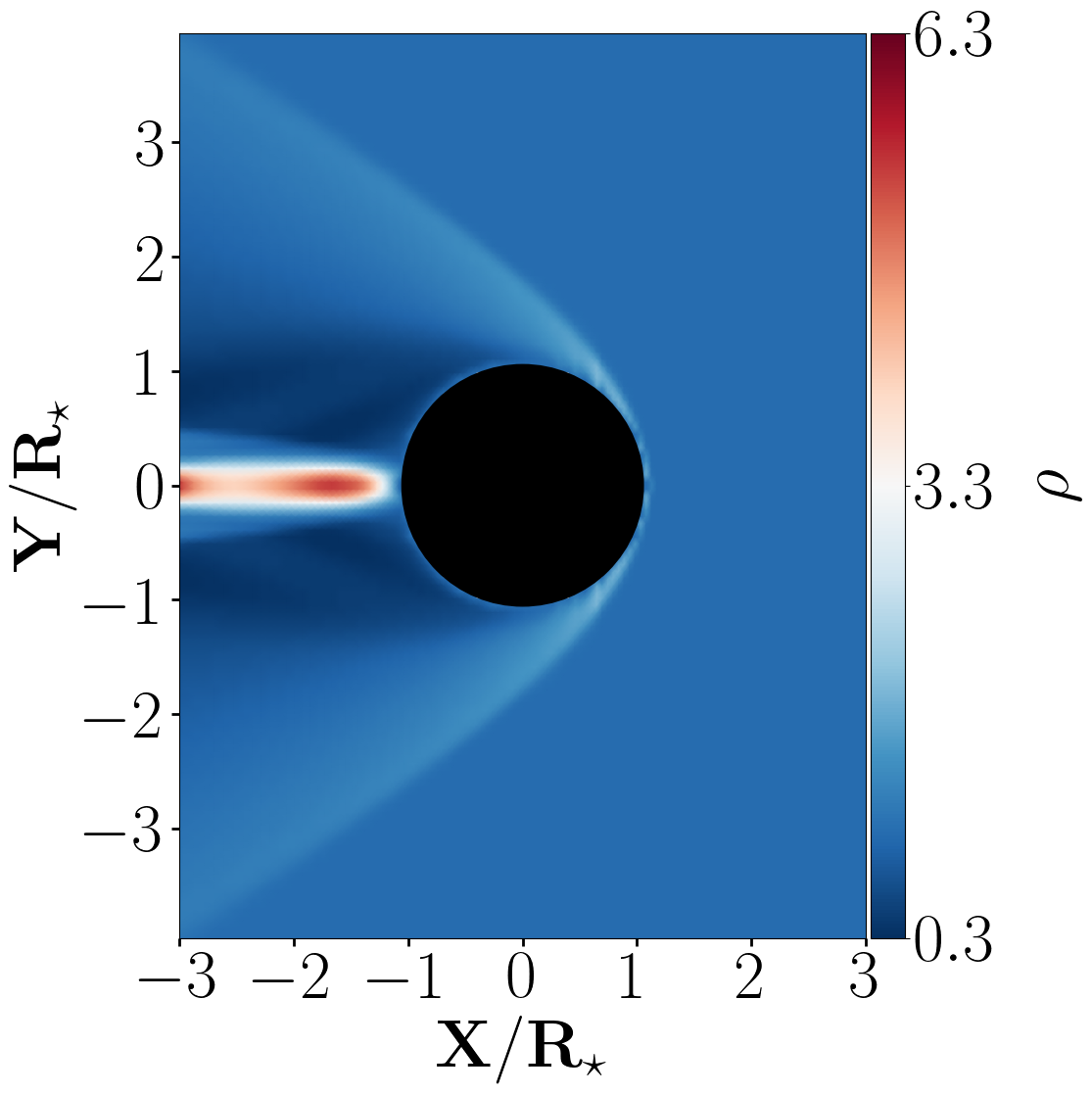}
}
\subfloat{%
\includegraphics[height=0.19\textwidth]{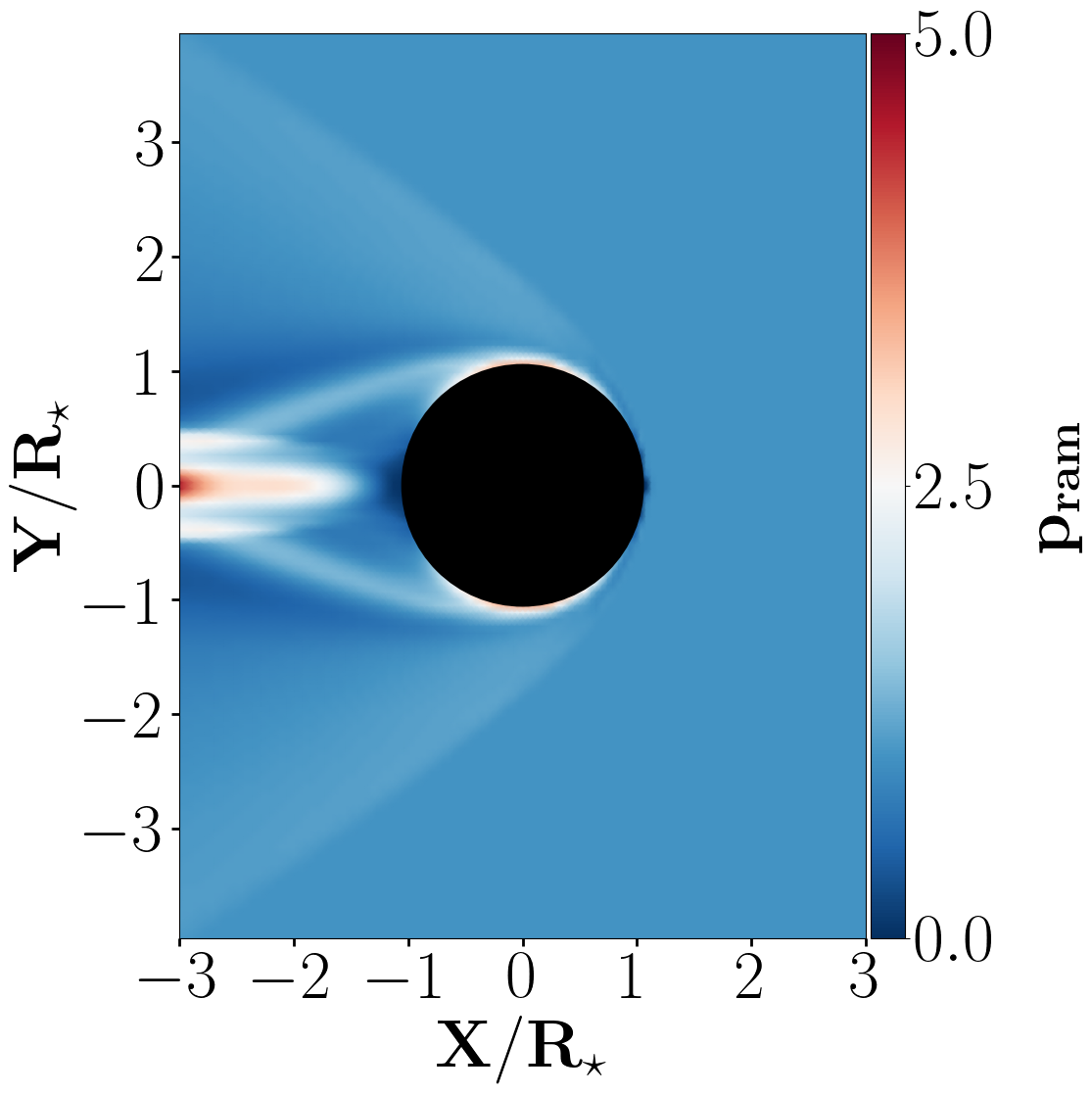}
}
\subfloat{%
\includegraphics[height=0.19\textwidth]{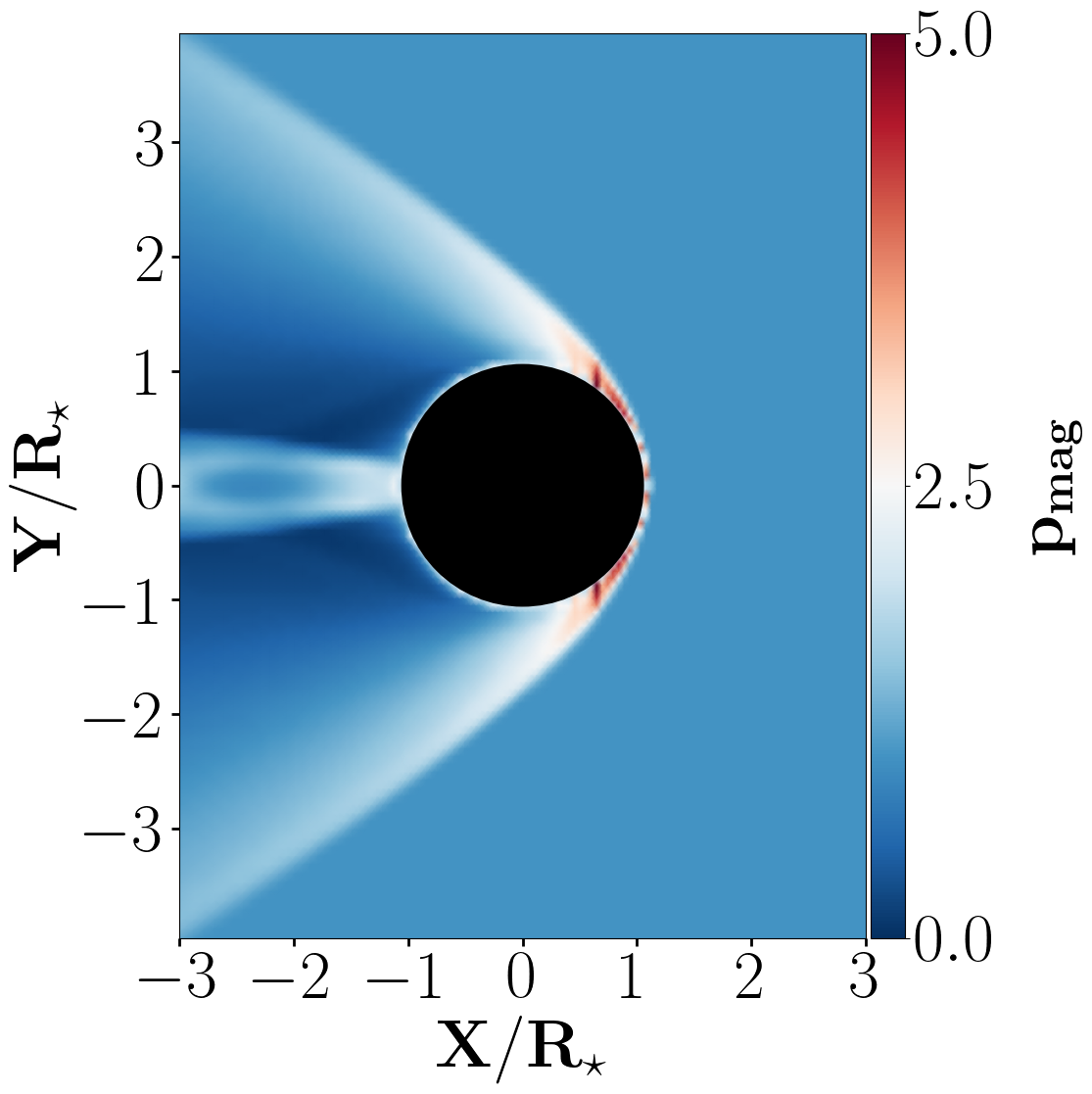}
}
\subfloat{%
\includegraphics[height=0.19\textwidth]{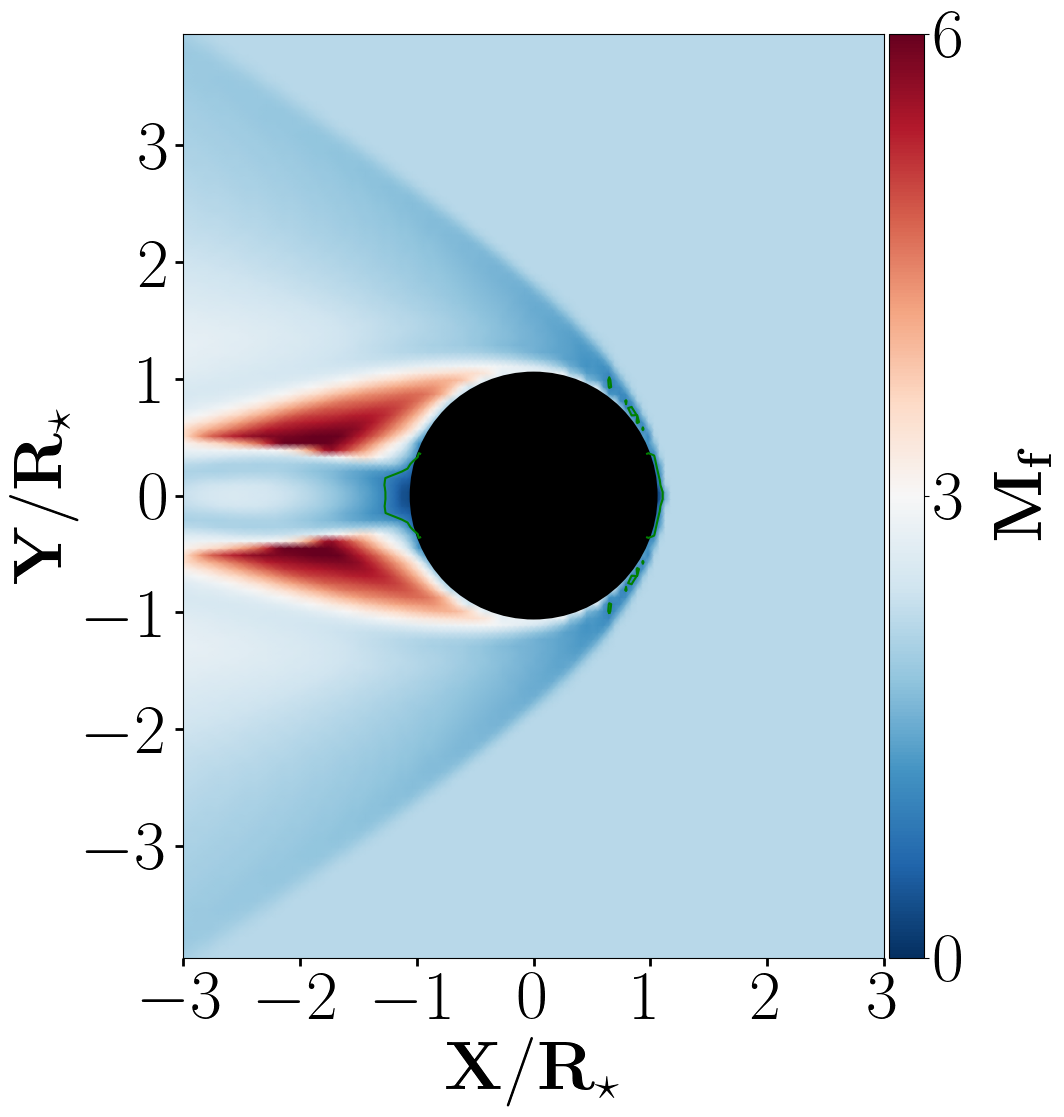}
}\\
\setcounter{subfigure}{0}

\subfloat[Velocity $v$]{%
\includegraphics[height=0.19\textwidth]{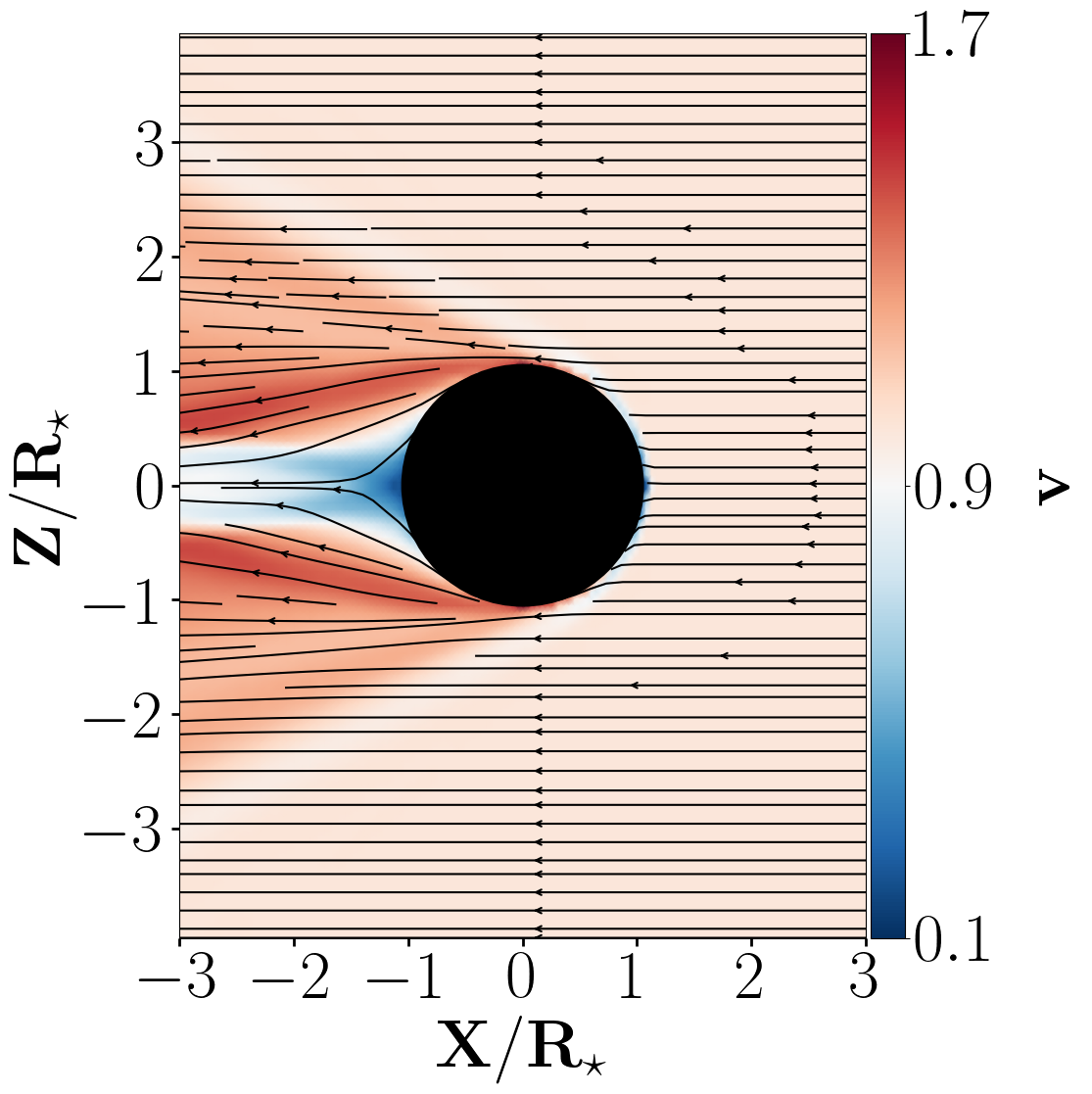}
}
\subfloat[Density $\rho$]{%
\includegraphics[height=0.19\textwidth]{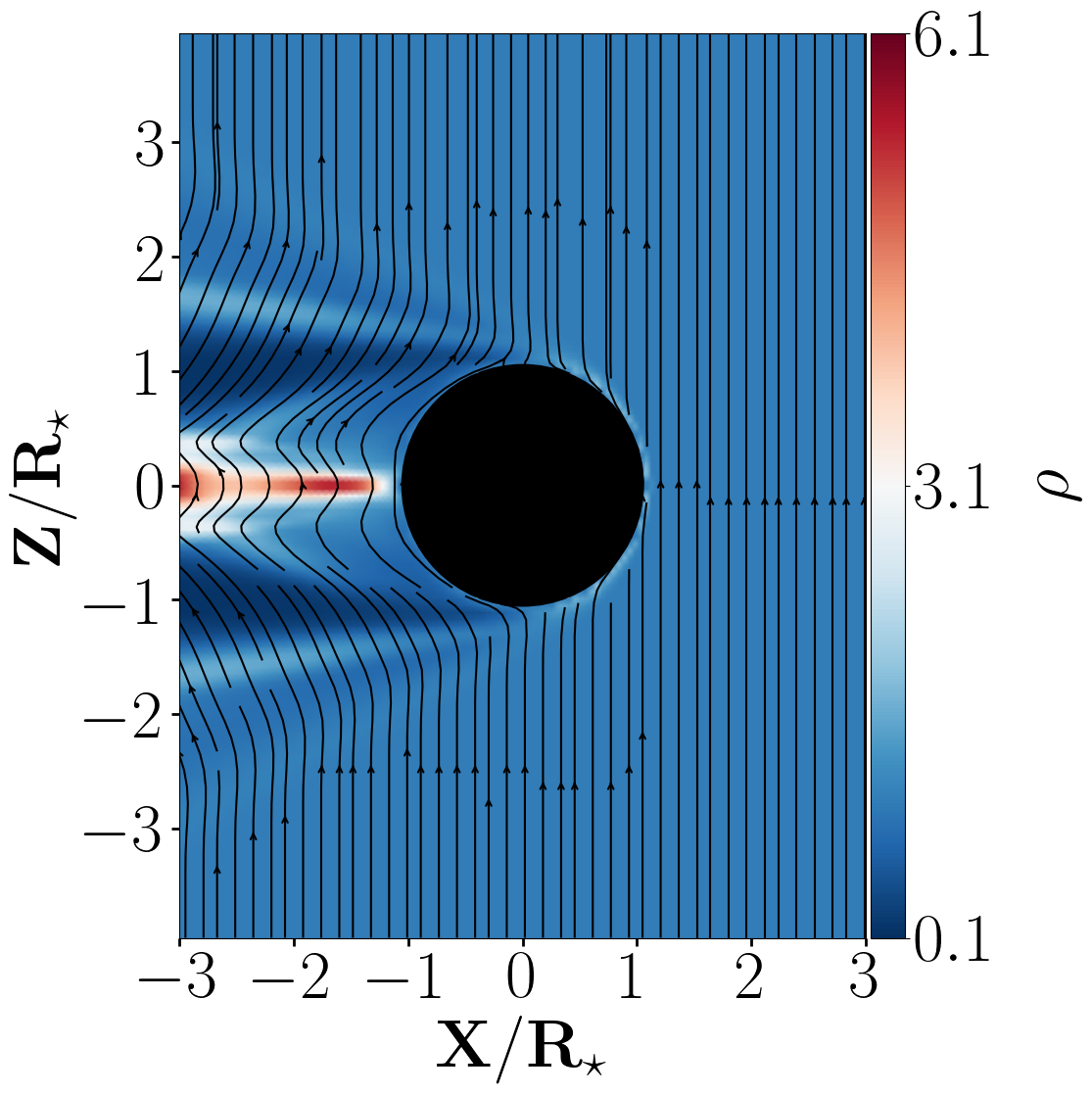}
}
\subfloat[$p_{ram}$]{%
\includegraphics[height=0.19\textwidth]{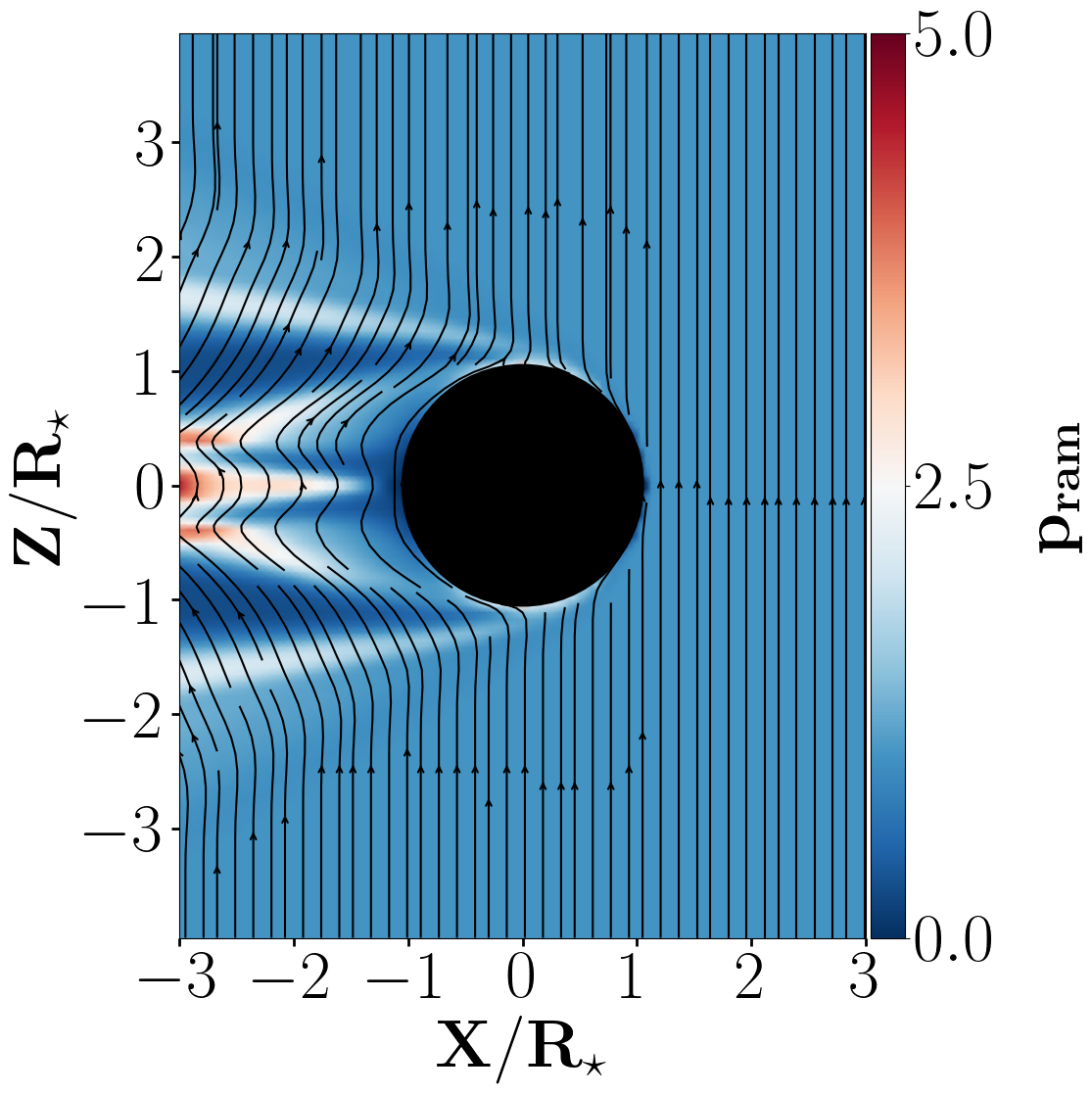}
}
\subfloat[$p_{mag}$]{%
\includegraphics[height=0.19\textwidth]{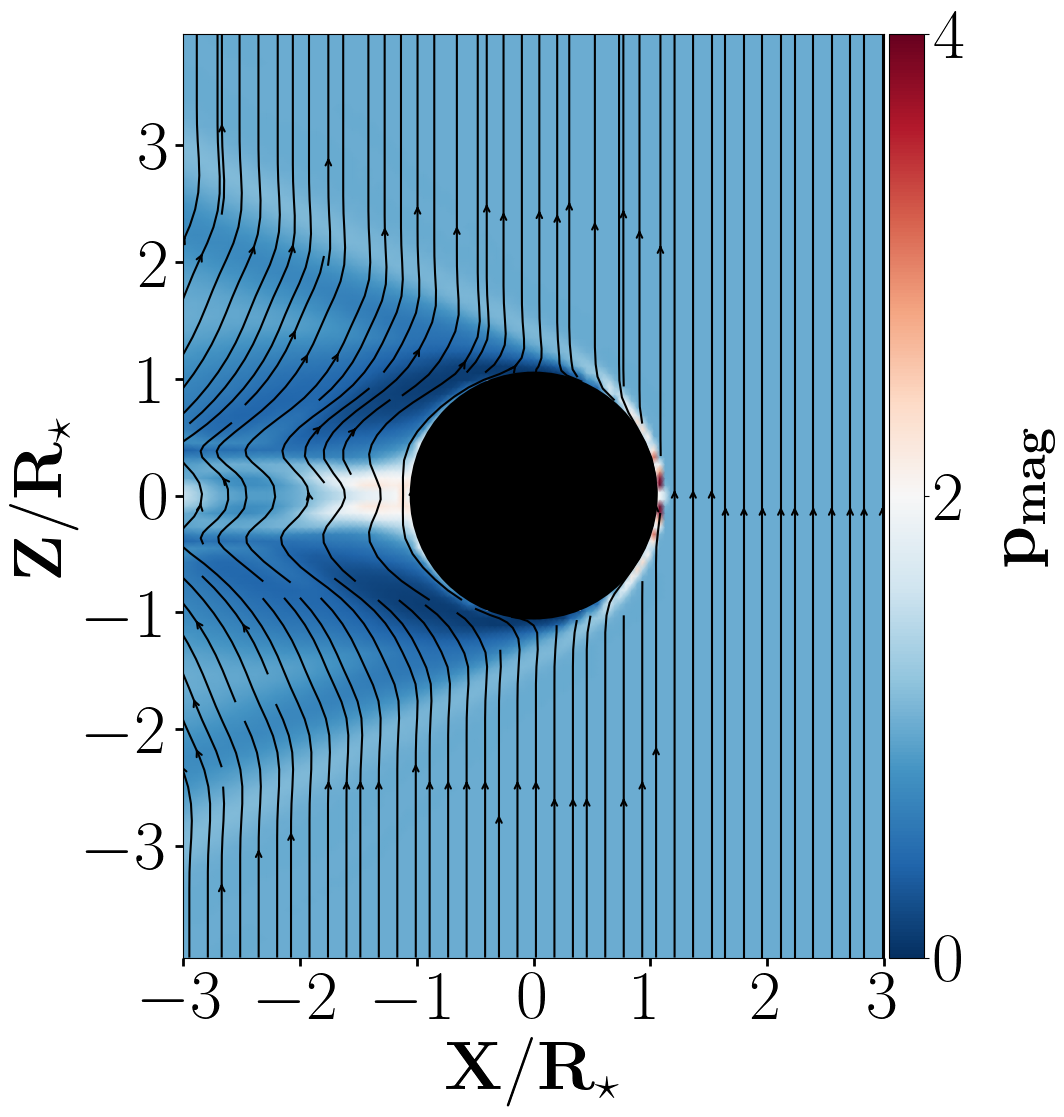}
}
\subfloat[$\mathcal{M}_{f}$]{%
\includegraphics[height=0.19\textwidth]{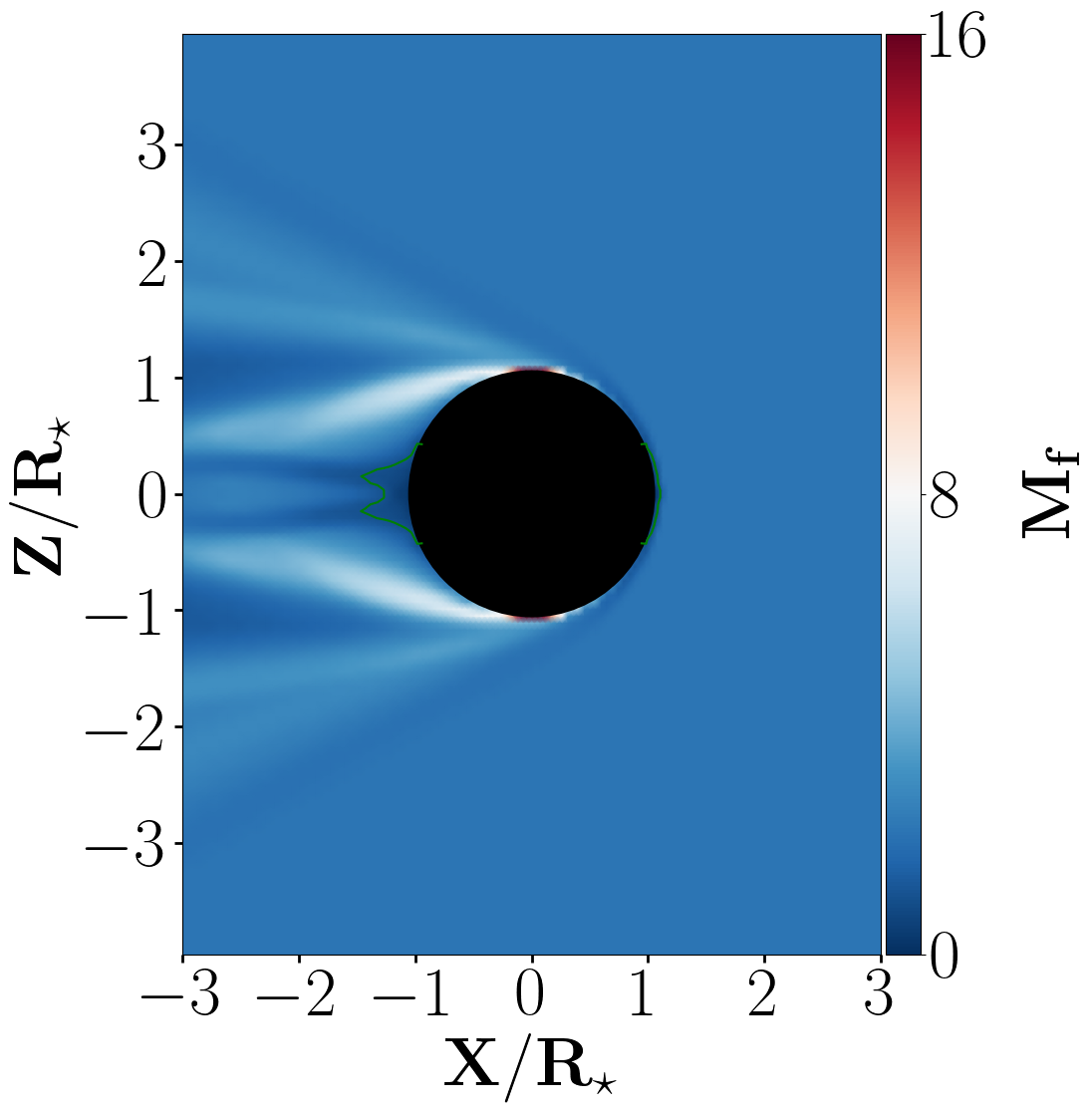}
}
\caption{Super-\Alfvenic run A3.  The details are the same as in Fig.~\ref{nonrelativistic_colorplots_A1_2S}. }
\label{nonrelativistic_colorplots_A3_2S}
\end{figure}

In the super-\Alfvenic\ case (run A3; $\mathcal{M}_A > 1$), the flow exhibits a hydrodynamic-dominated interaction. Magnetic effects are largely confined to a thin boundary layer around the stellar surface, while the bulk flow forms a sharp bow shock and narrow downstream wake (Fig.~\ref{nonrelativistic_colorplots_A3_2S}).  Density, ram-pressure, and fast magnetosonic Mach number distributions follow the expected flow-dominated pattern.

\begin{figure}[!htb]
\centering
\includegraphics[width=0.45\textwidth]{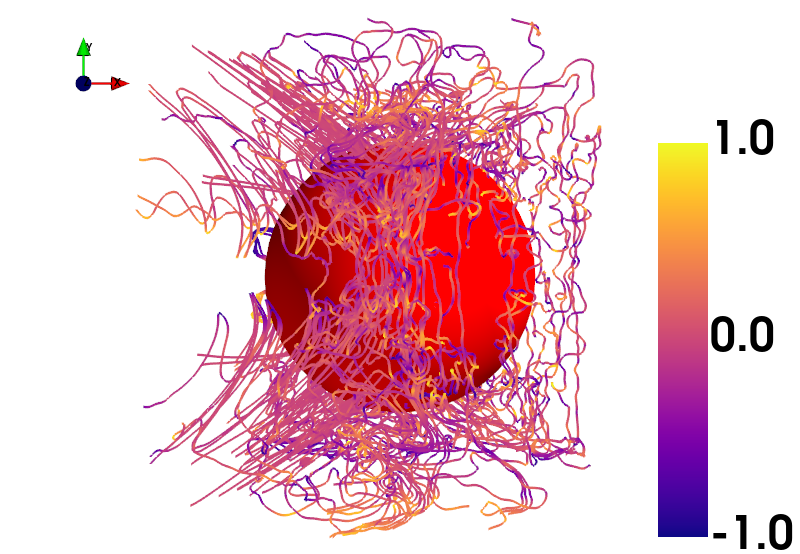}
\includegraphics[width=0.45\textwidth]{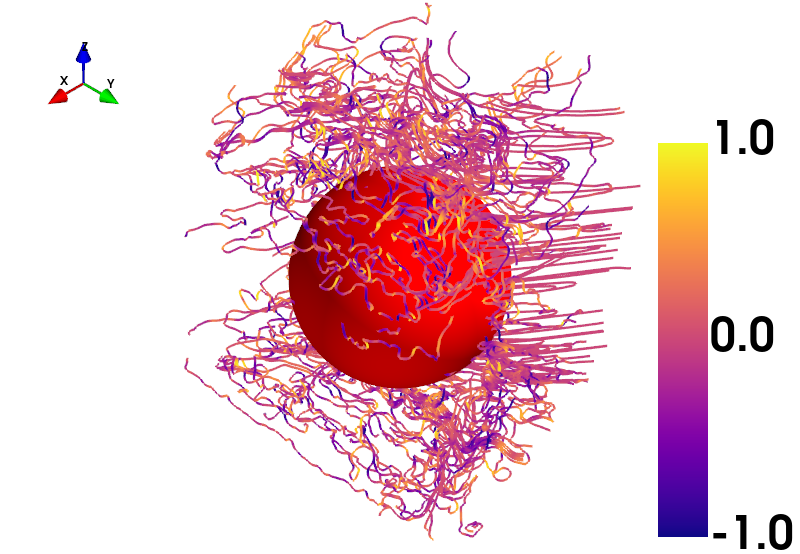}
\caption{Super-\Alfvenic run A3. 3D streamlines plot of current density at the steady state as viewed across various planes, with the viewing direction labeled in the plots.} 
\label{J_3Dstreamlines_A3_2S}
\end {figure}

The 3D current structure (Fig.~\ref{J_3Dstreamlines_A3_2S}) confirms a complete breakdown of the organized \Alfven-wing topology seen at lower $\mathcal{M}_A$. In both the $z=0$ equatorial plane (left) and the isometric view (right), current streamlines form tangled, irregular patterns with extensive looping. The upstream flow remains laminar far from the obstacle, but becomes disrupted immediately upon encountering the bow shock and compressed boundary layer. 

This regime represents a qualitative shift in the interaction physics. Whereas lower $\mathcal{M}_A$ runs channel electromagnetic energy along well-defined current systems, the super-\Alfvenic case confines magnetic structuring to the vicinity of the obstacle. This makes the regime a plausible source of strong electromagnetic emission, though the loss of coherence implies that any associated signatures should be more variable and broadband compared to the organized, wing-driven emission in the sub-\Alfvenic limit.

\section{Relativistic Flow}
\label{rel_results}

We now extend our investigation to relativistic velocities relevant to the inspiral phase of neutron star mergers. The combination of relativistic motion and perfect conductivity fundamentally alters the interaction physics compared to both the classical MHD case and planetary analogues. We analyze the system’s evolution across different plasma regimes by varying the \Alfven Mach number ($\mathcal{M}_A$) and plasma beta ($\beta$) parameters (see \S\ref{simulation_runs} for a complete list). Results for the sub-\Alfvenic and super-\Alfvenic regimes are presented separately below. All relevant quantities are evaluated in the lab frame.

\subsection{Relativistic Sub-\Alfvenic Flow}
\label{results_rel_sub_alfvenic}

Run B1 explores the mildly relativistic strongly sub-\Alfvenic regime with $v_0 = 0.30c$ (i.e., $\Gamma = 1.05$) and $\mathcal{M}_A = 0.21$ at plasma beta $\beta \approx 0.16$. This setup isolates relativistic effects while maintaining sub-\Alfvenic conditions, allowing direct comparison with non-relativistic results and providing a baseline for understanding magnetic field dominance in relativistic flows.

\begin{figure}[htbp]
\centering
\subfloat{%
\includegraphics[height=0.19\textwidth]{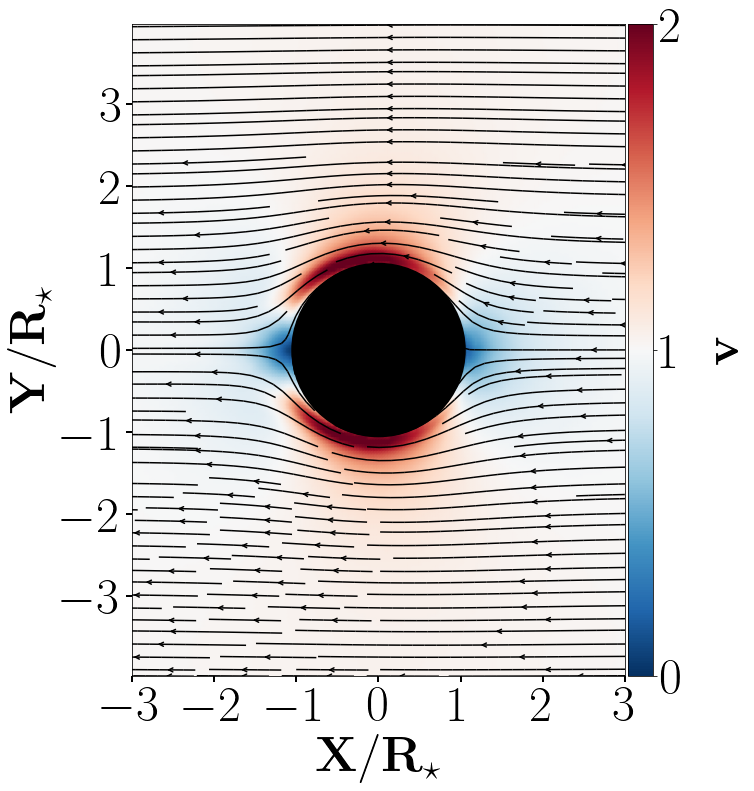}
}
\subfloat{%
\includegraphics[height=0.19\textwidth]{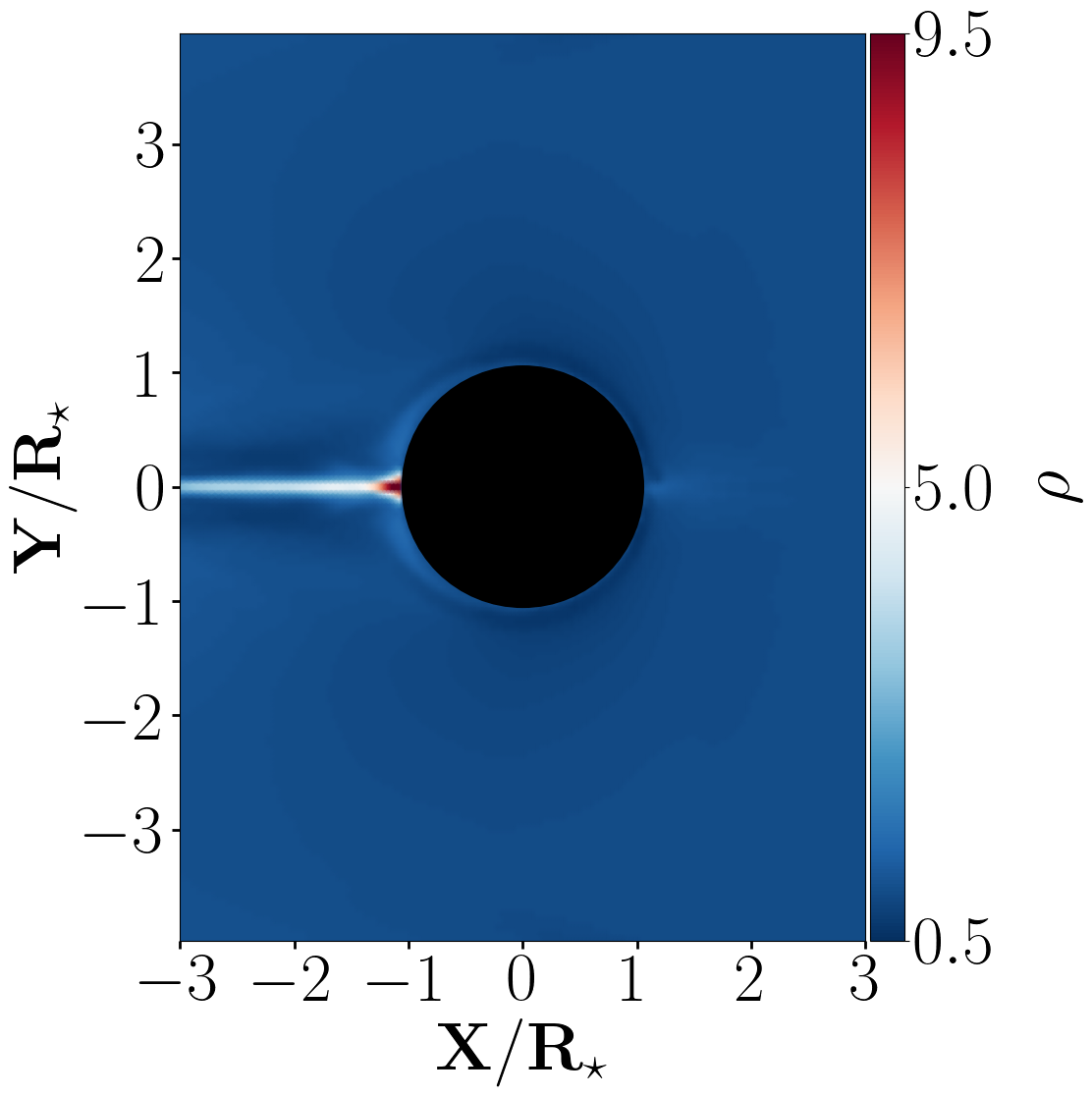}
}
\subfloat{%
\includegraphics[height=0.19\textwidth]{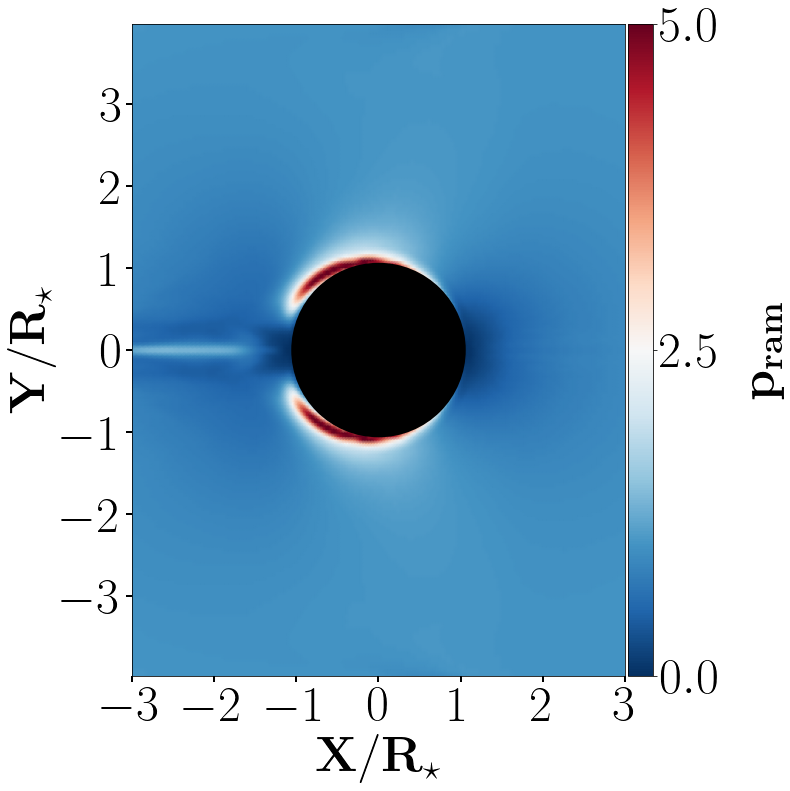}
}
\subfloat{%
\includegraphics[height=0.19\textwidth]{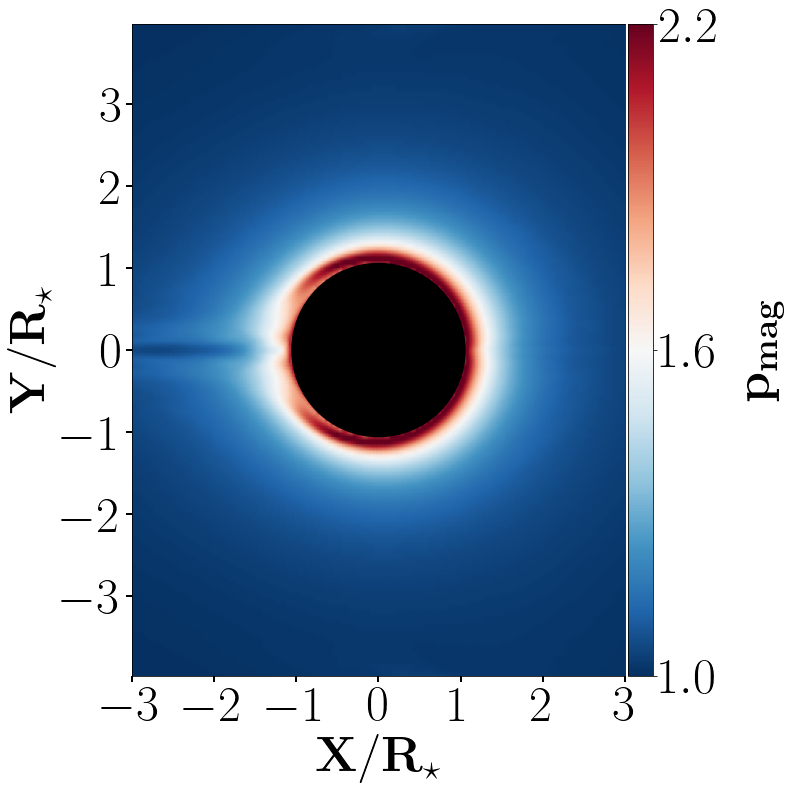}
}
\subfloat{%
\includegraphics[height=0.19\textwidth]{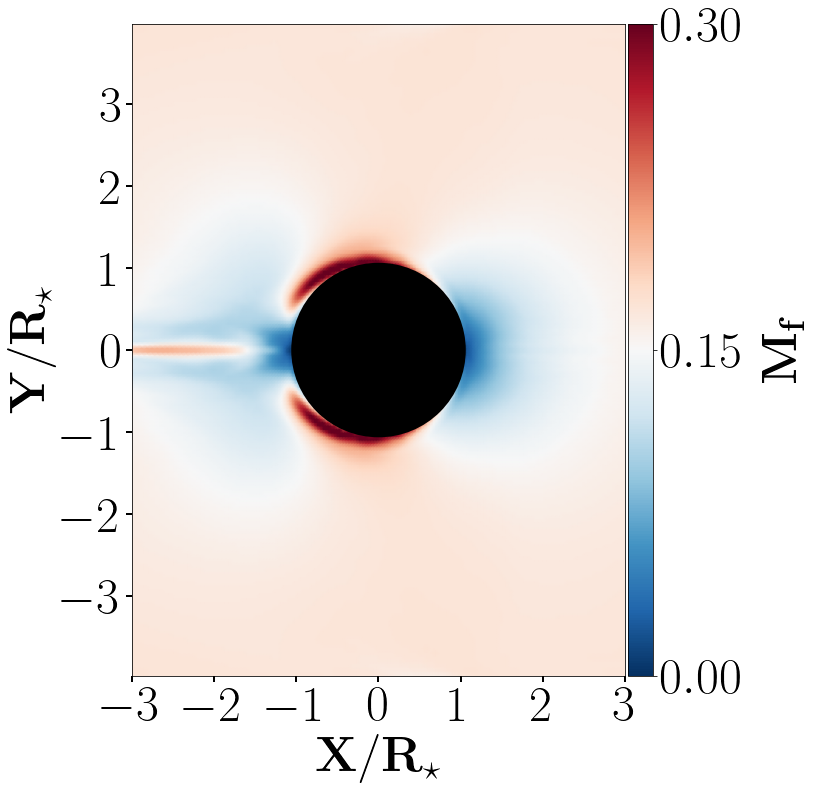}
}\\
\setcounter{subfigure}{0}

\subfloat[Velocity $v$]{%
\includegraphics[height=0.19\textwidth]{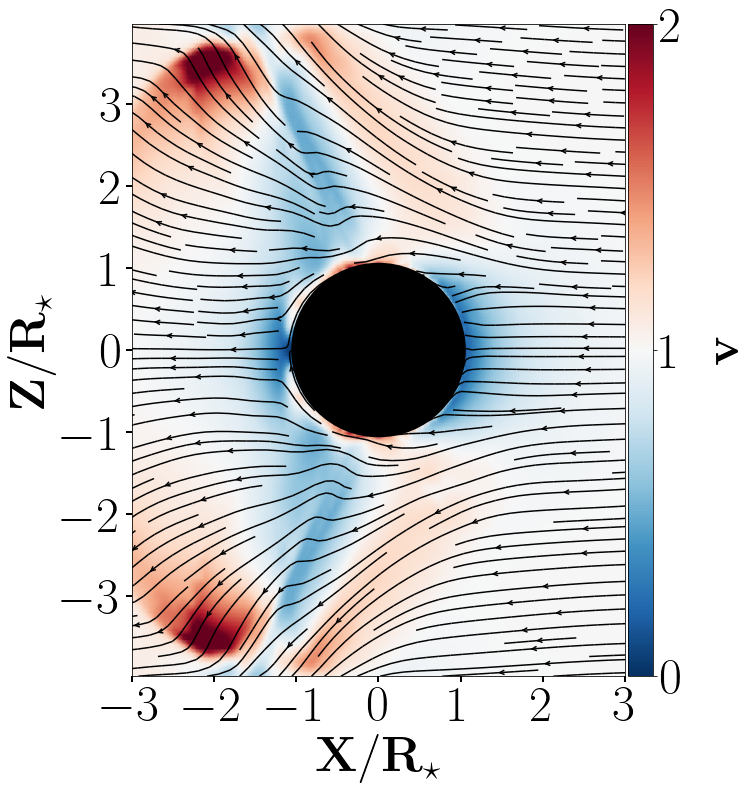}
}
\subfloat[Density $\rho$]{%
\includegraphics[height=0.19\textwidth]{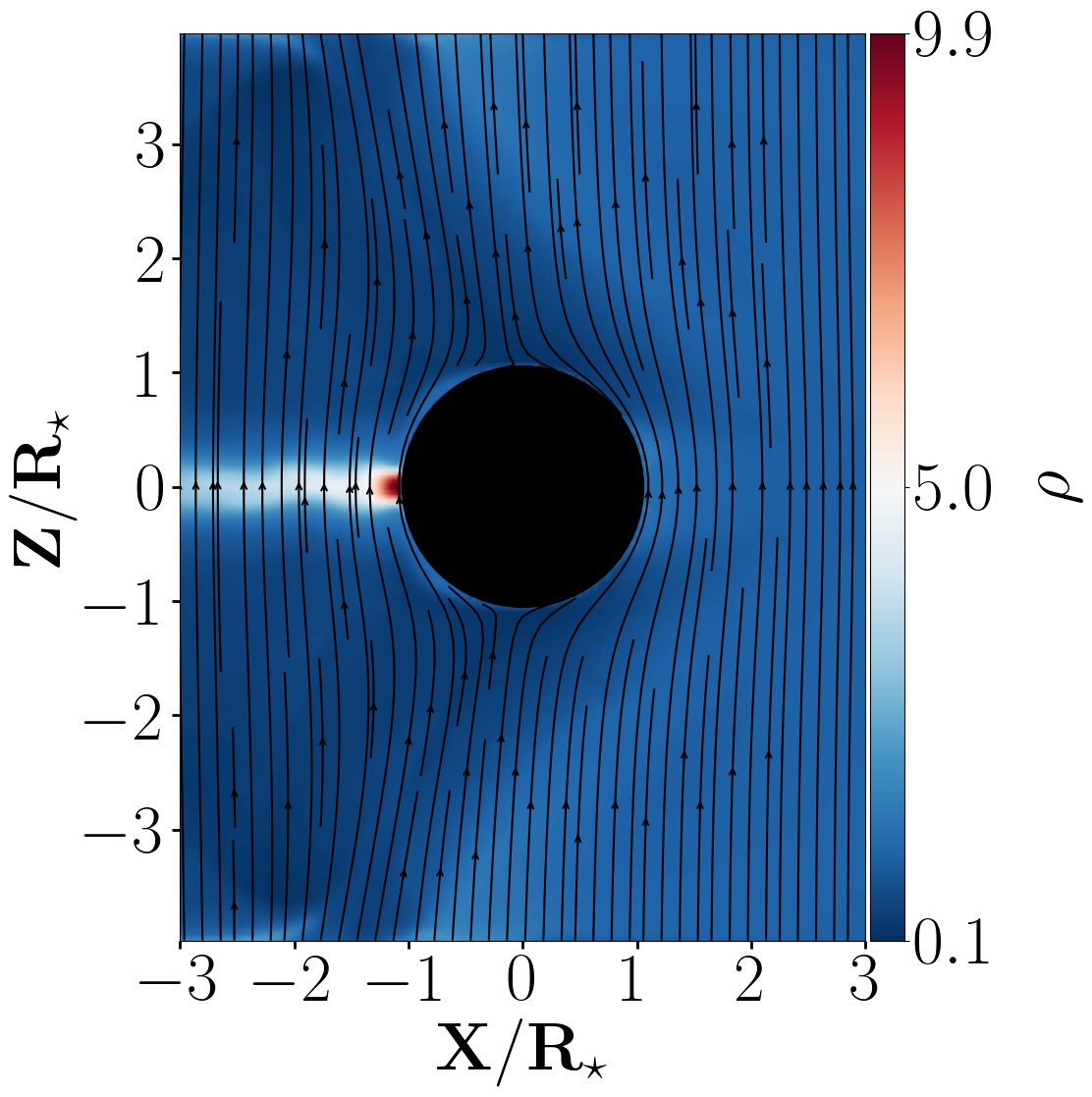}
}
\subfloat[$p_{ram}$]{%
\includegraphics[height=0.19\textwidth]{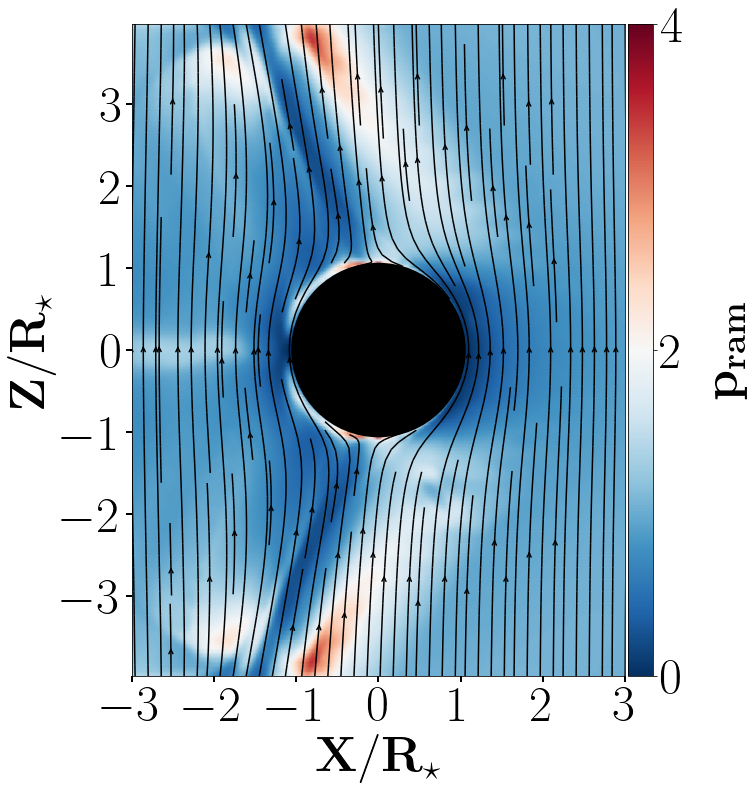}
}
\subfloat[$p_{mag}$]{%
\includegraphics[height=0.19\textwidth]{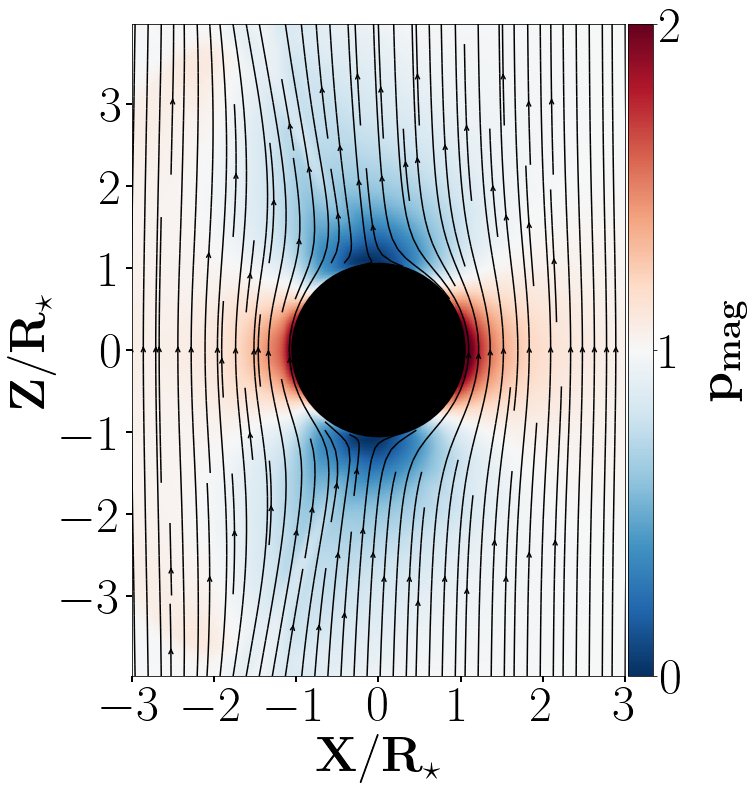}
}
\subfloat[Fast Mach $\mathcal{M}_{f}$]{%
\includegraphics[height=0.19\textwidth]{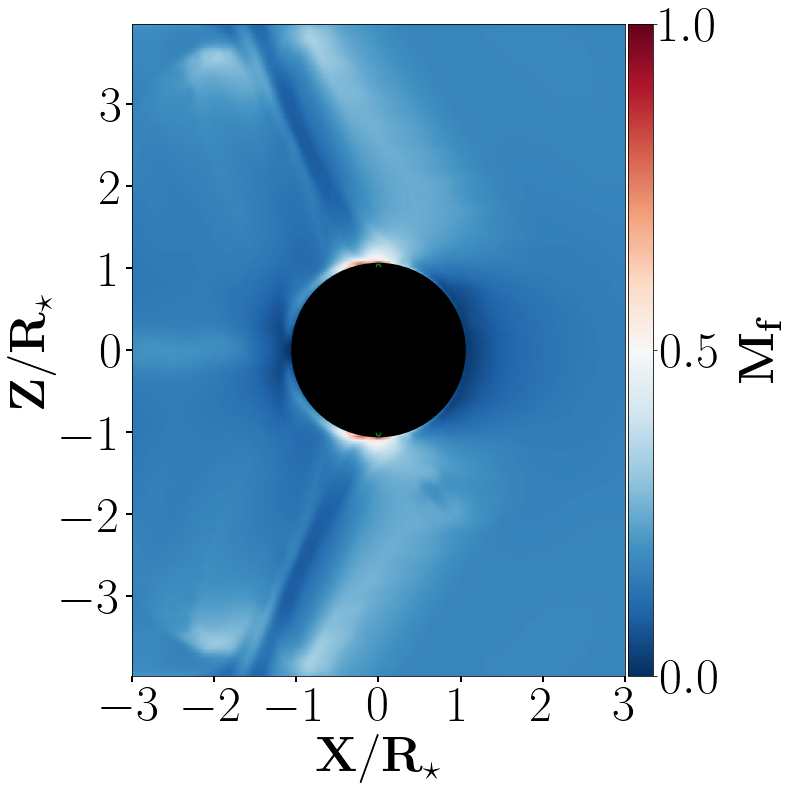}
}
\caption{Relativistic Sub-\Alfvenic run B1. The details are similar to those in Fig. \ref{nonrelativistic_colorplots_A1_2S}. All quantities are measured in the lab frame and normalized accordingly. }
\label{relativistic_colorplots_B1_1S}
\end{figure}

Fig.~\ref{relativistic_colorplots_B1_1S}(a) shows the flow velocity. In the equatorial plane, the flow undergoes smooth, symmetric deflection around the obstacle, with velocity enhancement near $y \approx \pm 1$ as plasma accelerates around the magnetosphere. In the $xz$ plane, the flow becomes asymmetric, with preferential deflection along oblique channels aligned with the background magnetic field. The plasma density shows a localized enhancement at $(x,z)\approx(-1,0)\, R_{\star}$, downstream of the obstacle. This enhancement extends along the equatorial plane ($z=0$) as a narrow density ridge that gradually decays downstream as the flow advects trapped material away from the interaction region. Far from the obstacle, the density remains close to ambient values. The physical origin of this on-axis wake compression, whether due to magnetic field line convergence in the equatorial current sheet, flow re-convergence behind the obstacle, or boundary condition effects, requires further investigation. The formation of a small downstream bubble was also observed by \citep{2025arXiv251114932C}.

Ram-pressure maps show enhanced compression along the equatorial flanks, forming a double-crescent pattern, whereas the downstream wake exhibits reduced ram pressure, consistent with diversion into the \Alfven wings. Magnetic pressure is modestly amplified ($p_{mag}/p_{mag,0} \sim 2-2.3$) in a thin shell surrounding the obstacle, forming a toroidal ring in the equatorial plane and stronger equatorial enhancement in meridional slices, while polar regions remain less affected.

The fast magnetosonic Mach number confirms that the interaction remains largely sub-unity throughout the domain. Overall, these results indicate that the magnetic field organizes the flow, producing coherent wing-like channels and localized compression, without any extended shock formation.

\begin{figure}[!htb]
\centering
\includegraphics[width=0.45\textwidth]{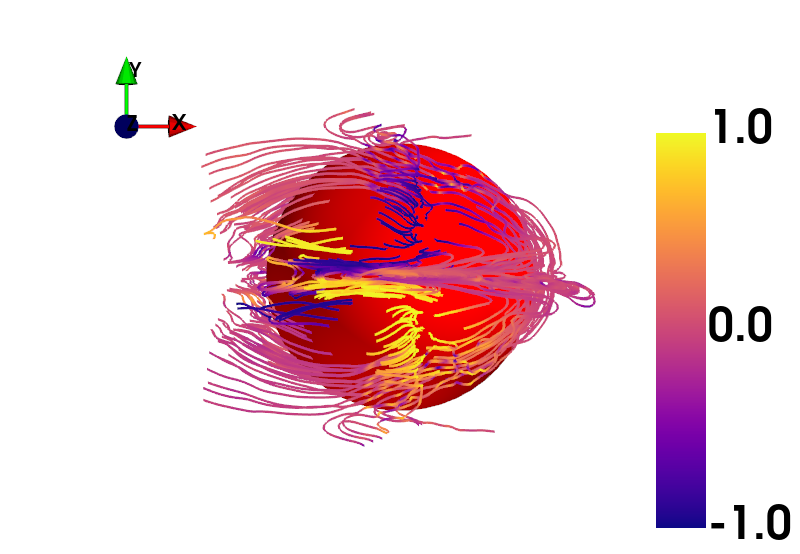}
\includegraphics[width=0.45\textwidth]{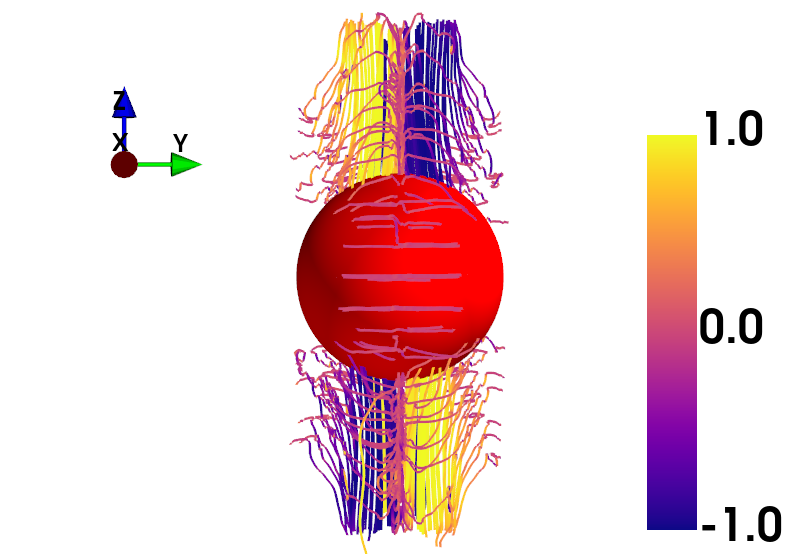}
\caption{Relativistic Sub-\Alfvenic run B1. We show 3D current density streamlines colored by $J_z$ for relativistic flow past a conducting sphere.
{Left:} $z=0$ equatorial view. {Right:} $x=0$ meridional view.}
\label{J_3dstreamlines_B1}
\end {figure}

Fig.~\ref{J_3dstreamlines_B1} then shows the current density. The system exhibits the characteristic \Alfven wing geometry, well-organized field-aligned channels form a bipolar structure above and below the equatorial plane, carrying oppositely directed currents along the background magnetic field. Near the obstacle, magnetic draping produces an elongated downstream current sheet, where oppositely directed components converge, and azimuthal currents flow in the equatorial plane. Color coding of the vertical current component ($J_z$) highlights the upward and downward flows feeding the wings, while the equatorial band shows predominantly azimuthal currents associated with the draping layer.
Minor turbulent features (twisted streamlines and small loops at the outer edges) suggest weak current-driven instabilities, but the overall structure remains coherent and organized, consistent with the magnetically-dominated regime where $B^2 \gg E^2$ throughout the domain.

\subsection{Moderate Sub-\Alfvenic Relativistic Flow}
\label{results_rel_moderate_sub_alfvenic}

Run B2 explores a moderately sub-\Alfvenic relativistic regime with $v_0 = 0.50c$ ($\Gamma = 1.15$), $\mathcal{M}_A \approx 0.65$, and $\beta = 0.40$. This case marks the transition from magnetically dominated to kinetically dominated flow, where ram pressure begins to exceed magnetic pressure and the flow becomes increasingly compressive.

Compared to the smooth, strongly sub-\Alfvenic flow (B1), the velocity field shows sharper gradients and pronounced deflection around the obstacle, with accelerated equatorial flanks and organized downstream channels. The density distribution reveals a patchy, partially compressed wake with a weaker downstream enhancement than in B1, indicating that magnetic confinement is still operative but less effective. Ram pressure peaks are substantially larger, concentrated along equatorial flanks and upstream regions, while the wake exhibits reduced values. Magnetic pressure remains amplified in an equatorial ring and polar lobes, but it is weaker than ram pressure, confirming kinetic dominance.

The fast magnetosonic Mach number indicates extended super-fast patches, following the trend seen earlier. Overall, this run shows a regime in which organized \Alfven-wing structures begin to fragment under the influence of strong ram pressure.

\begin{figure}[htbp]
\centering
\subfloat{%
\includegraphics[height=0.19\textwidth]{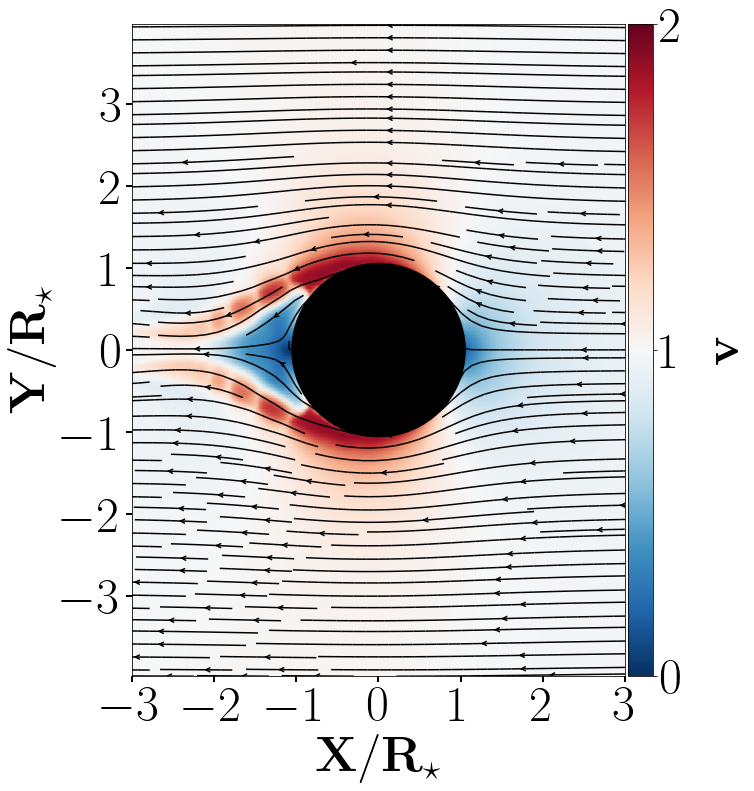}
}
\subfloat{%
\includegraphics[height=0.19\textwidth]{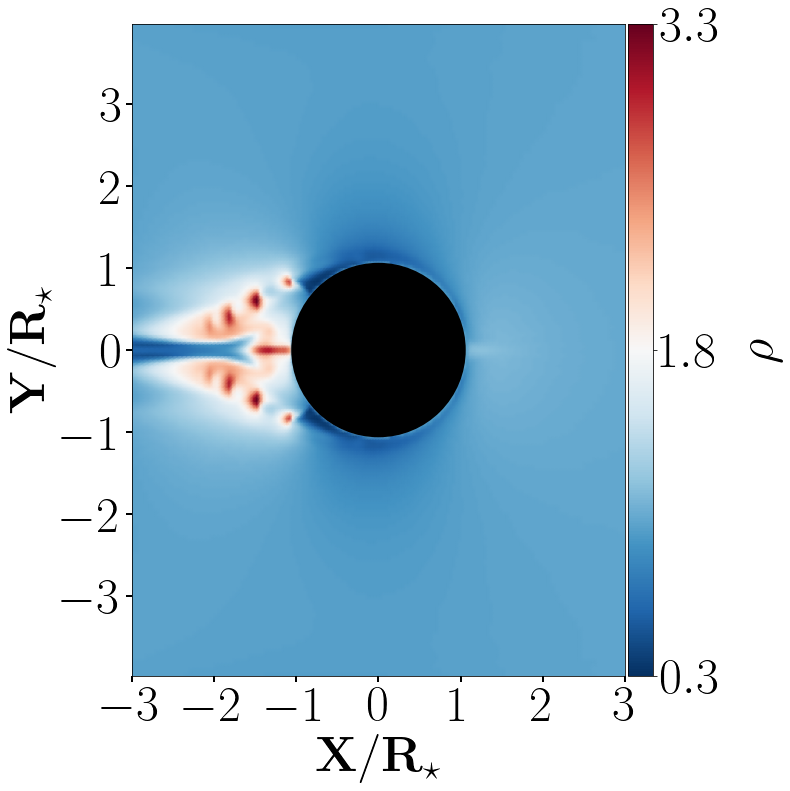}
}
\subfloat{%
\includegraphics[height=0.19\textwidth]{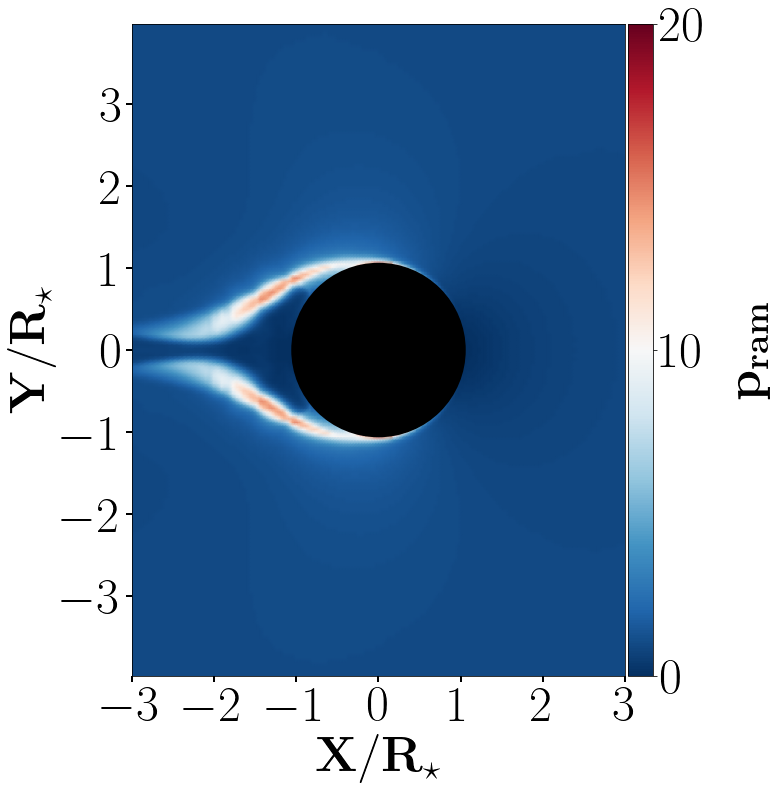}
}
\subfloat{%
\includegraphics[height=0.19\textwidth]{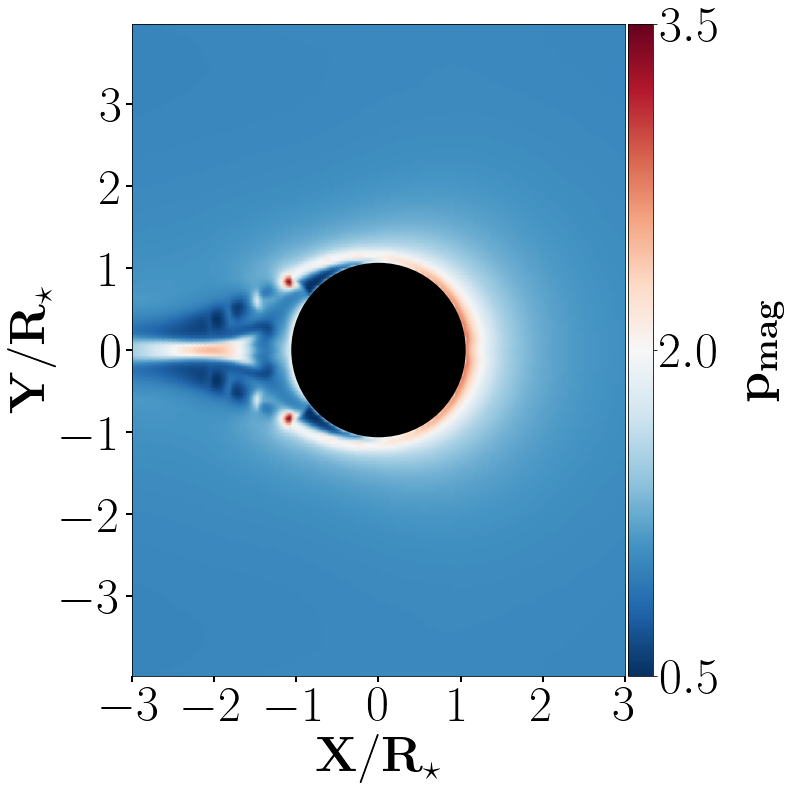}
}
\subfloat{%
\includegraphics[height=0.19\textwidth]{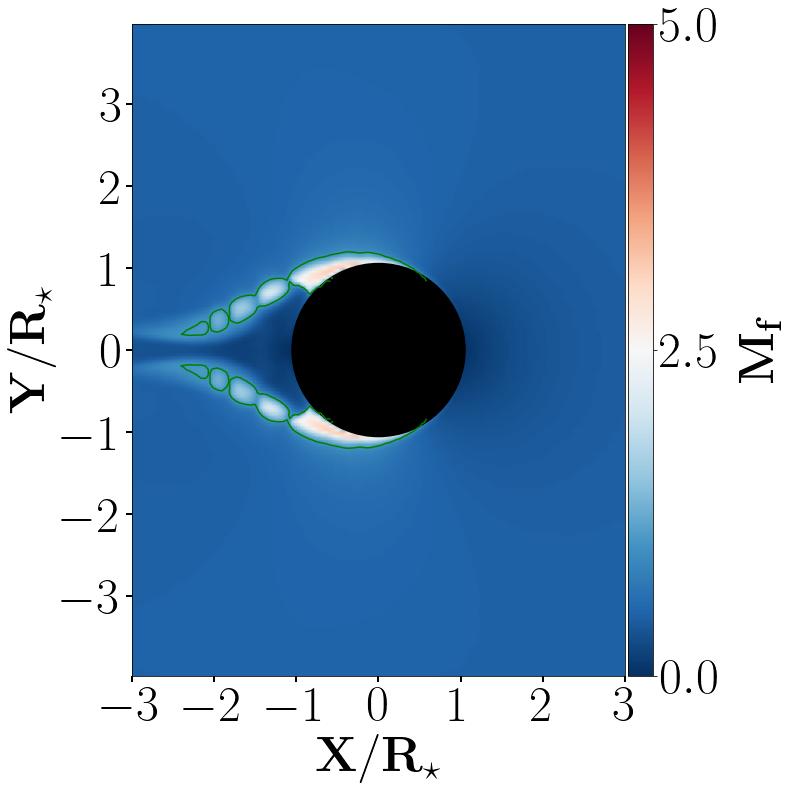}
}\\
\setcounter{subfigure}{0}

\subfloat[Velocity $v$]{%
\includegraphics[height=0.19\textwidth]{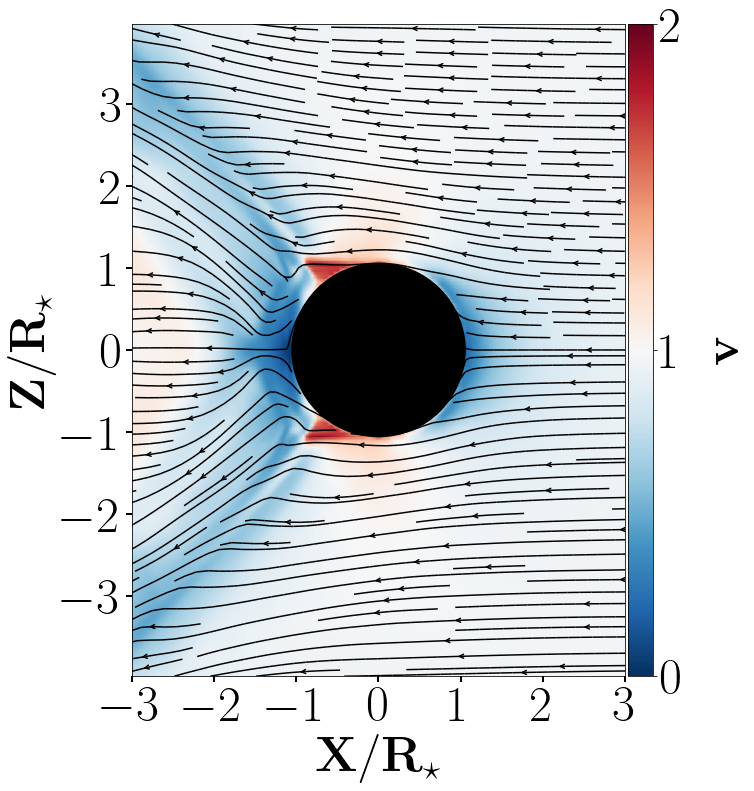}
}
\subfloat[Density $\rho$]{%
\includegraphics[height=0.19\textwidth]{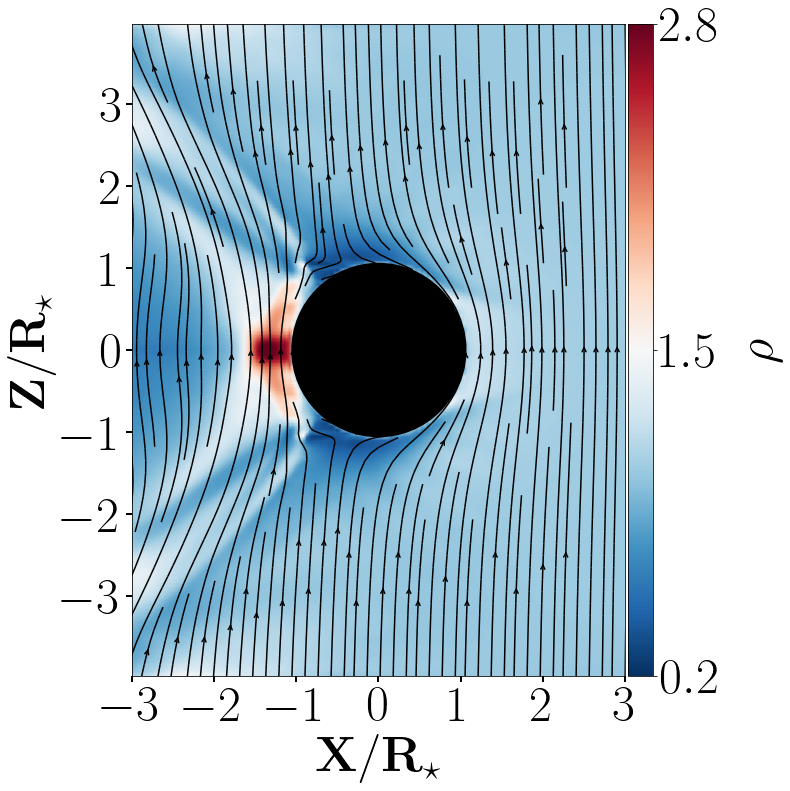}
}
\subfloat[$p_{ram}$]{%
\includegraphics[height=0.19\textwidth]{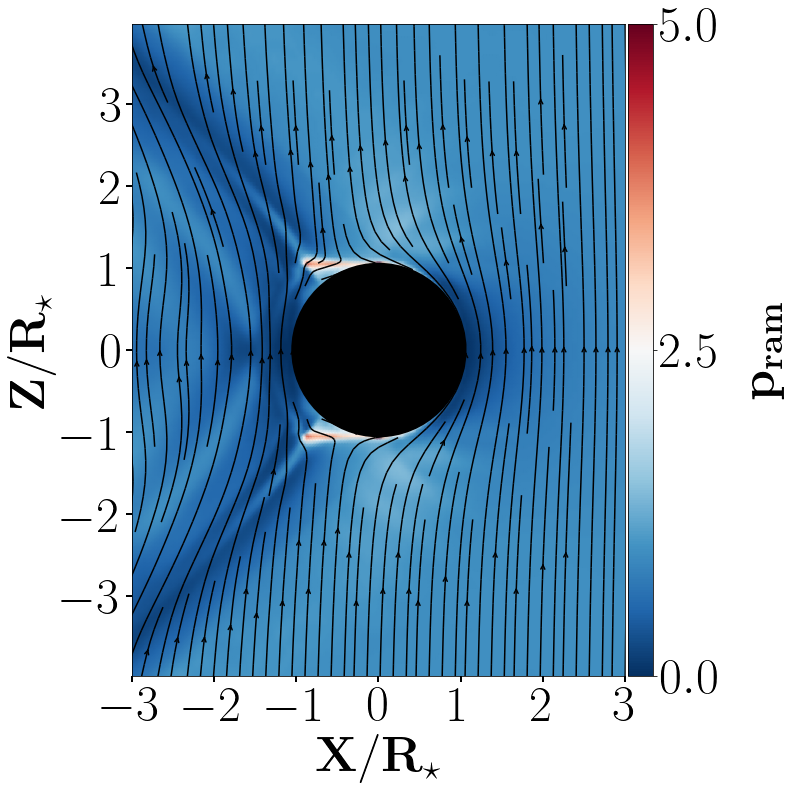}
}
\subfloat[$p_{mag}$]{%
\includegraphics[height=0.19\textwidth]{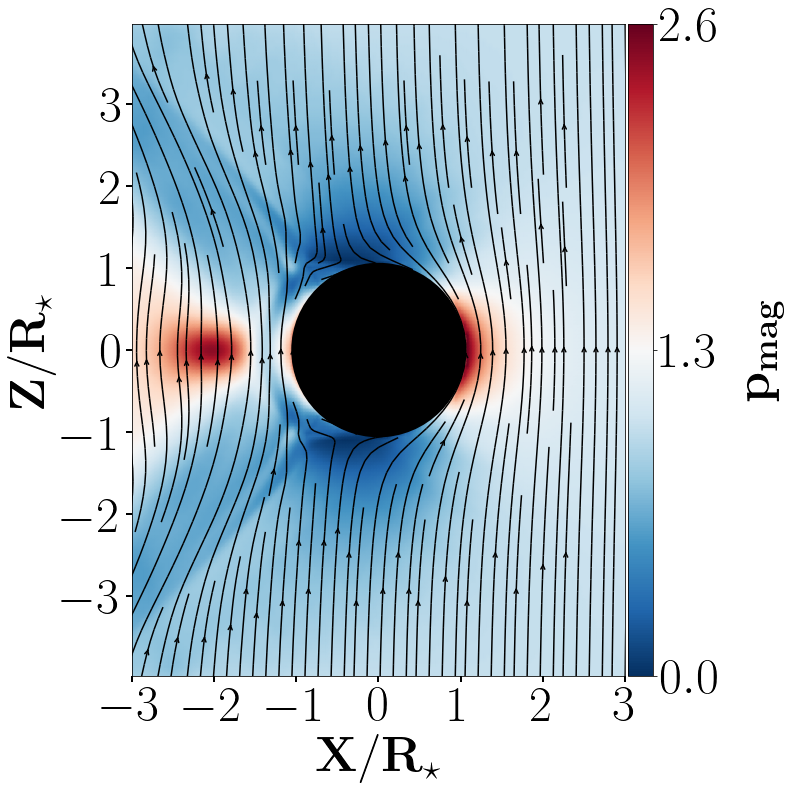}
}
\subfloat[Fast Mach $\mathcal{M}_{f}$]{%
\includegraphics[height=0.19\textwidth]{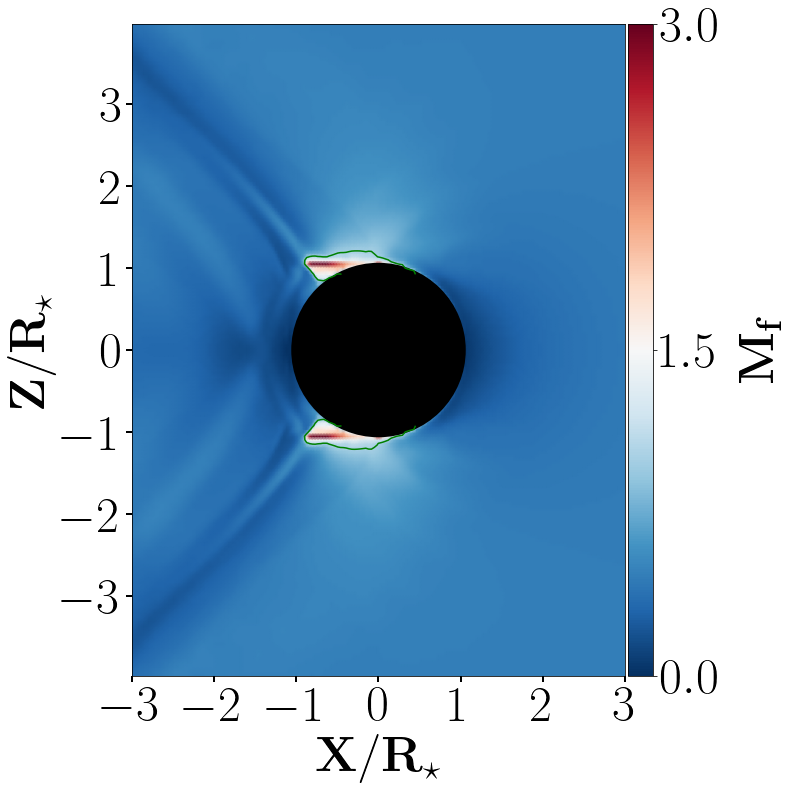}
}
\caption{Relativistic moderate sub-\Alfvenic run B2. Details are similar to Fig.~\ref{relativistic_colorplots_B1_1S}.}
\label{relativistic_colorplots_B2_1S}
\end{figure}

\begin{figure}[!htb]
\centering
\subfloat{%
\includegraphics[width=0.40\textwidth]{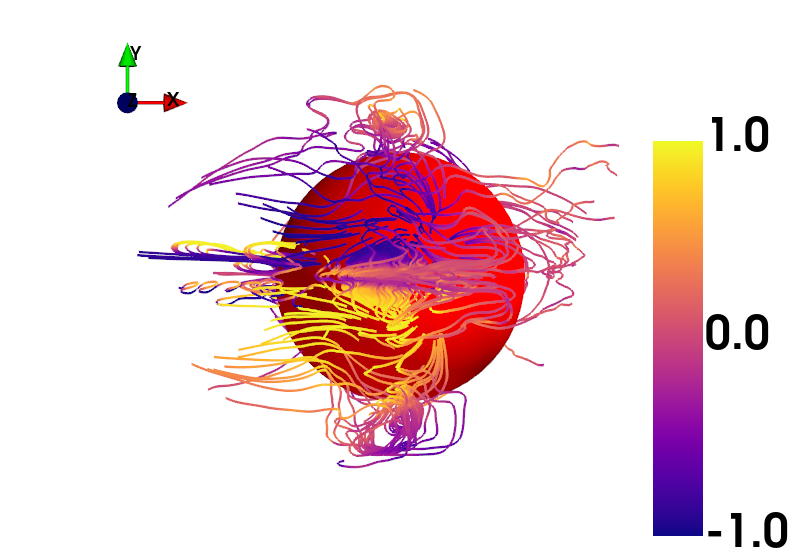}
}
\subfloat{%
\includegraphics[width=0.40\textwidth]{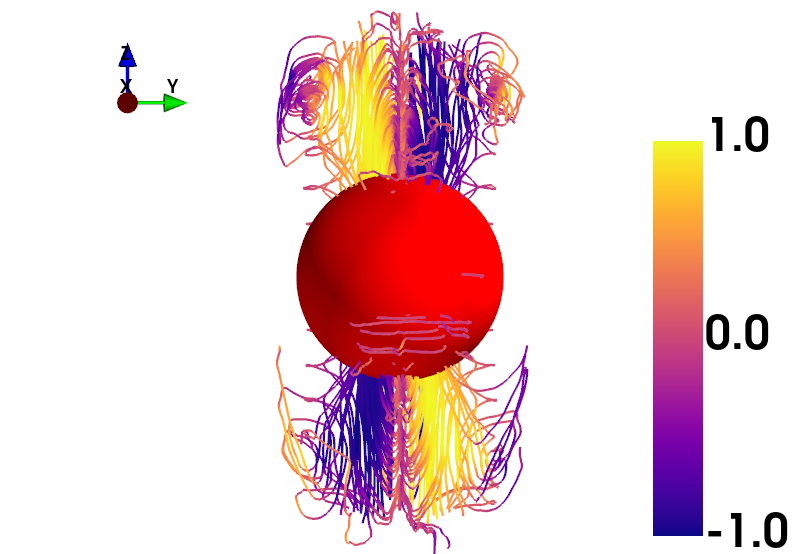}
}
\caption{Moderately relativistic sub-\Alfvenic run B2. 3D streamlines of current density at the steady state, viewed across various planes, with the viewing direction labeled in the plots. The rest of the details are the same as in Fig. \ref{J_3dstreamlines_B1}.} 
\label{relativistic_J_3dstreamlines_B2_1S}
\end{figure}

Fig.~\ref{relativistic_J_3dstreamlines_B2_1S} shows the current structure in this run. Field-aligned channels along $\pm z$ still form a bipolar \Alfven wing system, but the streamlines are significantly disrupted as the coherent columns are replaced by twisted, tangled structures indicative of current-driven instabilities. The equatorial plane reveals a fragmented wake with multiple current channels rather than a single organized sheet, reflecting asymmetry and partial breakdown of the wing geometry. Overall, the topology demonstrates that while \Alfven wing patterns persist, the increasing ram pressure induces instabilities and localized turbulence, signaling the transition from purely magnetically guided to more complex flow.

In the relativistic sub-\Alfvenic regime, the total \Alfven wing current is expected to scale as
\begin{equation}
I \sim 4 R_\star E_0 \Sigma_A,
\end{equation}
where $E_0 \sim v_0 B_0$ is the motional electric field and $\Sigma_A \sim \Gamma^2 / v_A$ is the relativistic \Alfven conductance \citep{2011A&A...532A..21M}. Since $v_A \propto B_0$ in these runs, the dependence on $B_0$ largely cancels out, yielding
\begin{equation}
I \propto v_0 \Gamma^2.
\end{equation}
Thus, even modest increases in $\Gamma$ should lead to significantly \Alfven wing currents, consistent with the enhanced current channels observed in run B2 compared to B1.

To quantify this behavior, we computed the flux of field–aligned current using the steps discussed in the \S\ \ref{results_nonrel_moderate_sub_alfvenic} using Eq. \ref{I_parallel_z0}. For comparison with analytic expectations, we extracted the values of $I_{ wings}(z_0)$ across several slices. An inspection of the computed values supports the qualitative scaling predicted above. The ratio of the wing current averaged over $z_0 \in [1.5, 3.5] $ in run B2 to sub-\Alfvenic run B1 is $\approx 1.8$, which is broadly consistent with the $I \propto v_0\Gamma^2$ relationship (theoretical ratio $\sim 2.0$) predicted by relativistic \Alfven wing theory.

\subsection{Super-\Alfvenic Relativistic Flow}
\label{results_rel_super_alfvenic}

Lastly, we examine the trans-\Alfvenic run B3, with $\mathcal{M}_A \approx 1.9$. Achieving such a high \Alfvenic Mach number in a stable configuration requires a weaker magnetic field, and we therefore adopt $\beta = 0.8$, compared to $\beta = 0.4, 0.16$ in runs B1 and B2. Attempts to reach similar $\mathcal{M}_A$ at lower $\beta$ led to numerical instabilities. This setup places the system in a regime of weak electromagnetic coupling, allowing us to test whether organized magnetic draping persists when magnetic forces are subdominant.

\begin{figure}[htbp]
\centering
\subfloat{%
\includegraphics[height=0.19\textwidth]{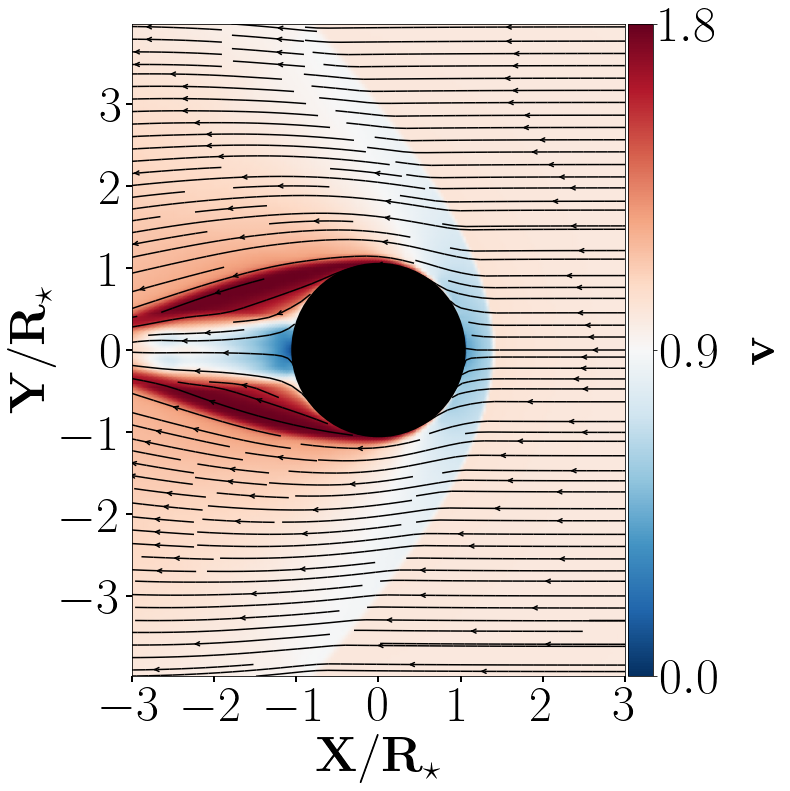}
}
\subfloat{%
\includegraphics[height=0.19\textwidth]{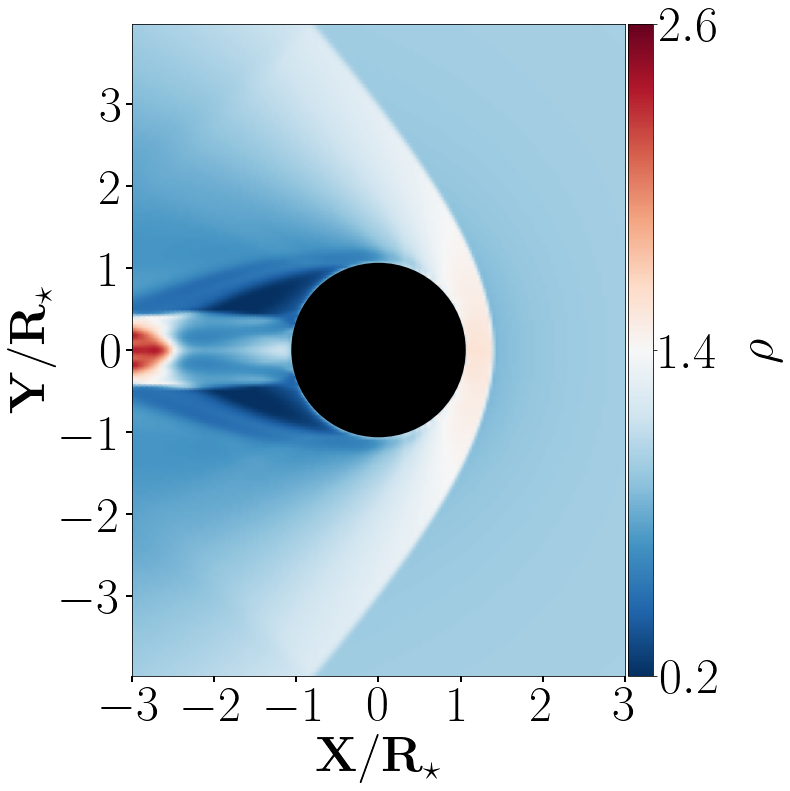}
}
\subfloat{%
\includegraphics[height=0.19\textwidth]{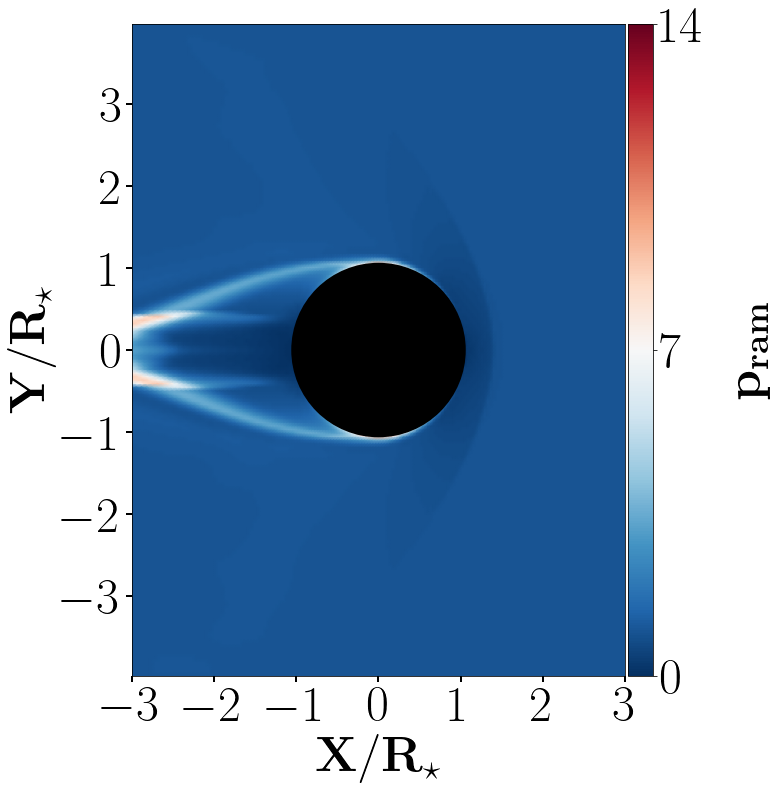}
}
\subfloat{%
\includegraphics[height=0.19\textwidth]{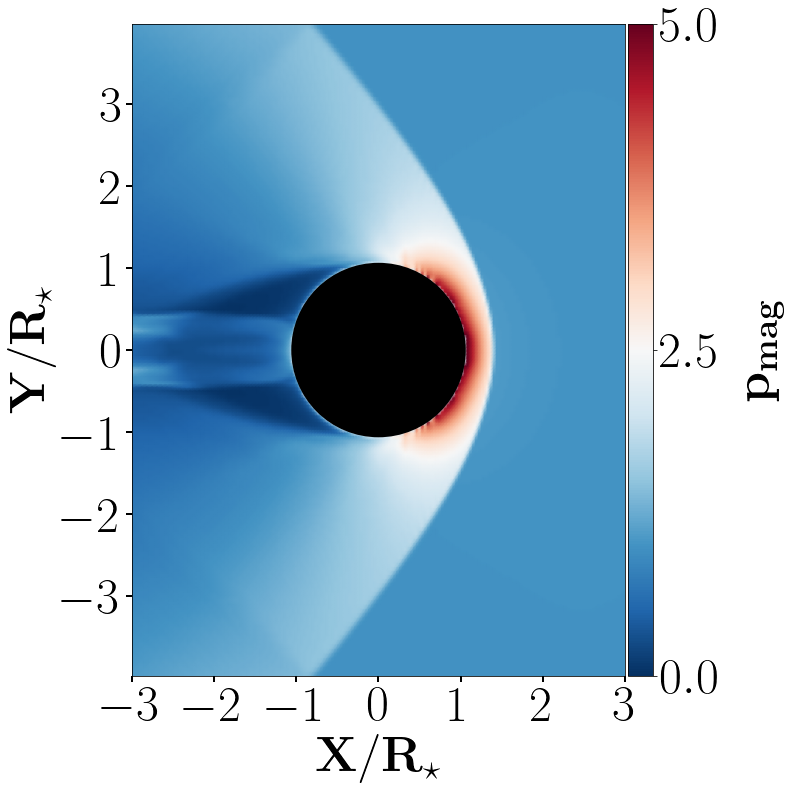}
}
\subfloat{%
\includegraphics[height=0.19\textwidth]{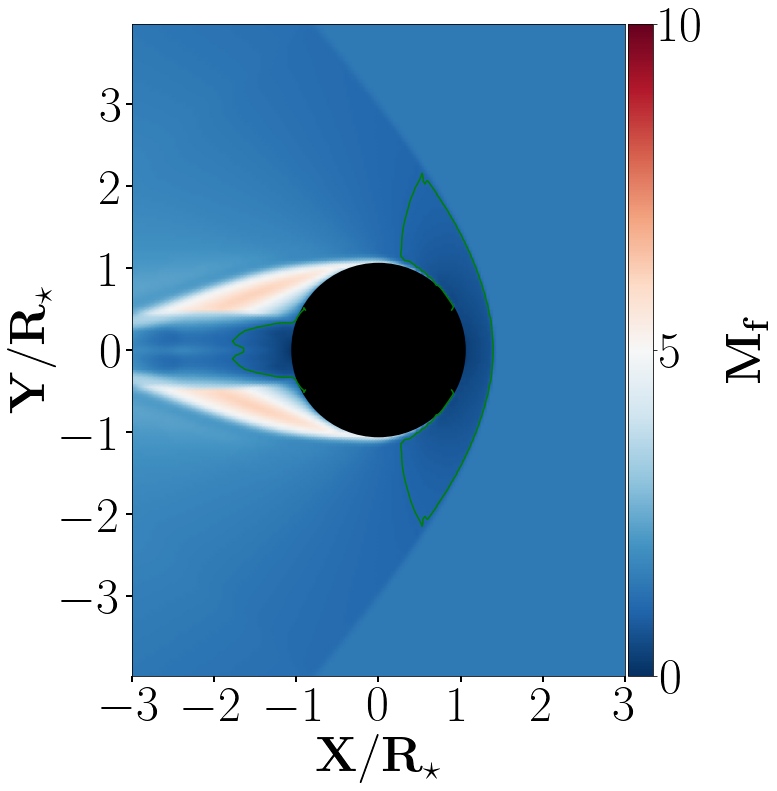}
}\\
\setcounter{subfigure}{0}

\subfloat[Velocity $v$]{%
\includegraphics[height=0.19\textwidth]{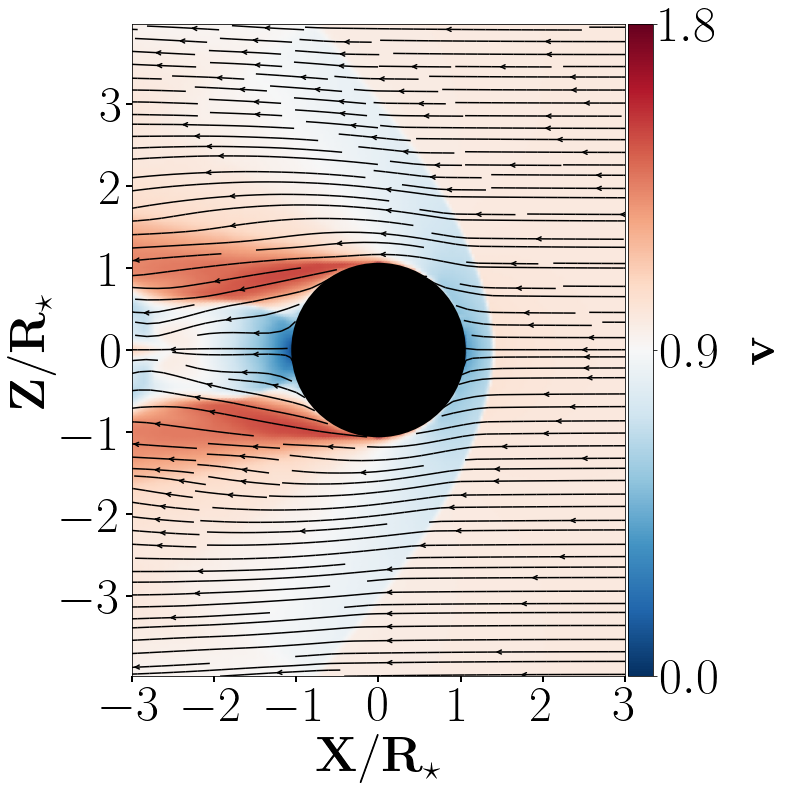}
}
\subfloat[Density $\rho$]{%
\includegraphics[height=0.19\textwidth]{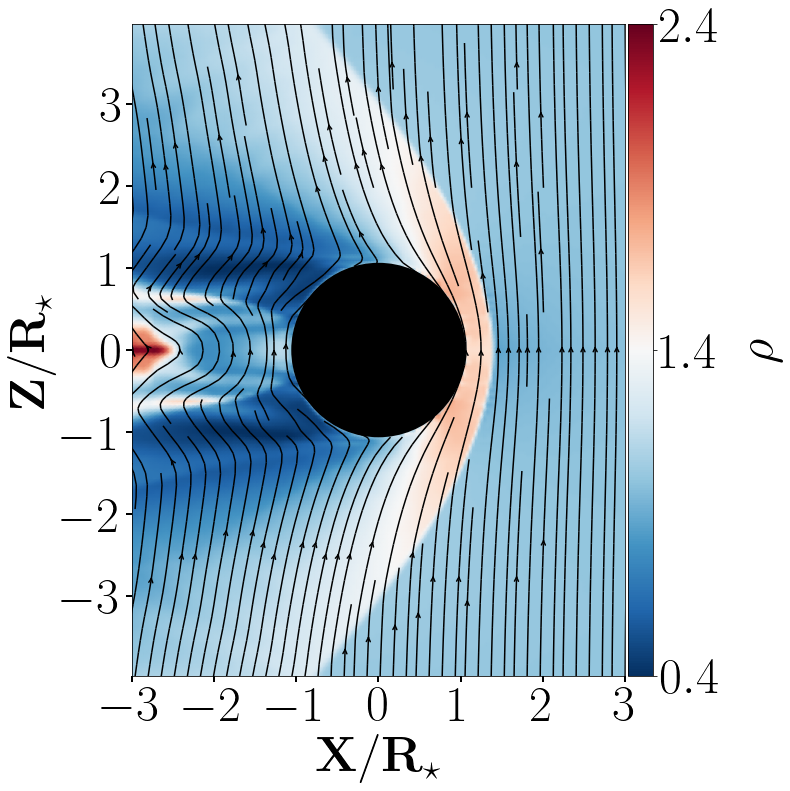}
}
\subfloat[$p_{ram}$]{%
\includegraphics[height=0.19\textwidth]{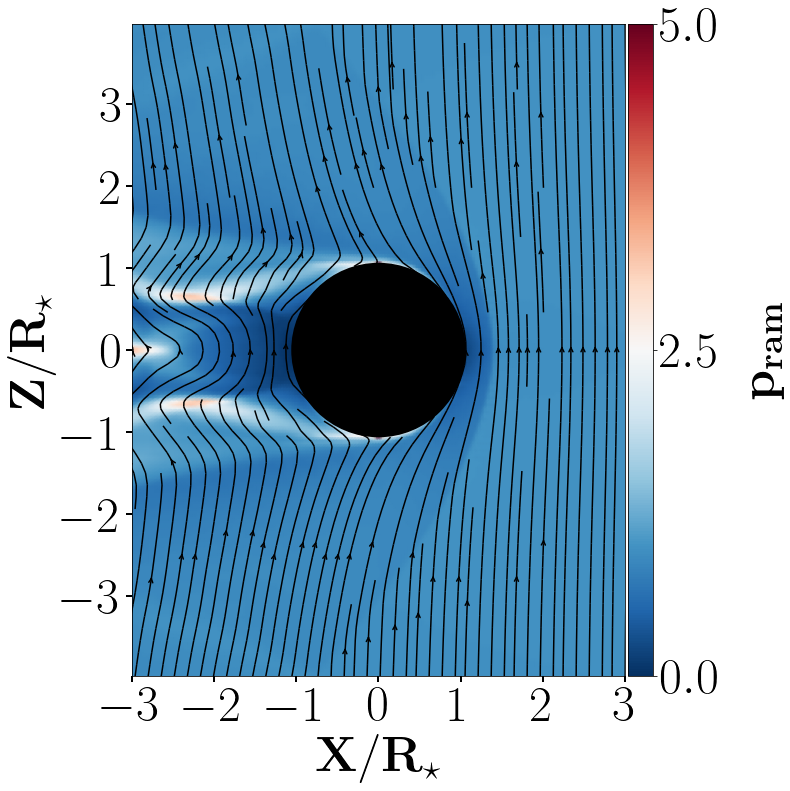}
}
\subfloat[$p_{mag}$]{%
\includegraphics[height=0.19\textwidth]{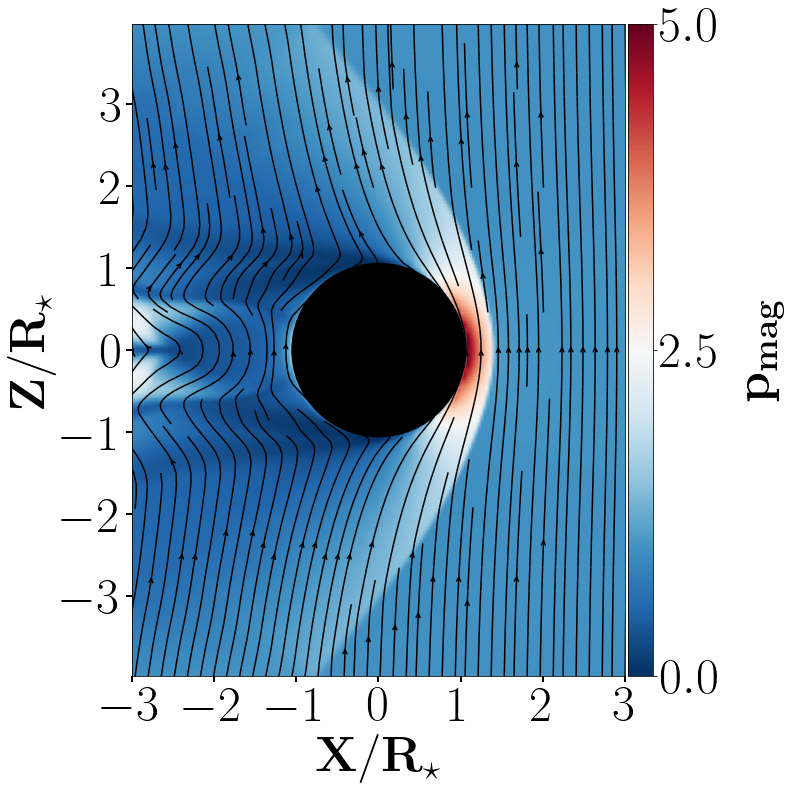}
}
\subfloat[Fast Mach $\mathcal{M}_{f}$]{%
\includegraphics[height=0.19\textwidth]{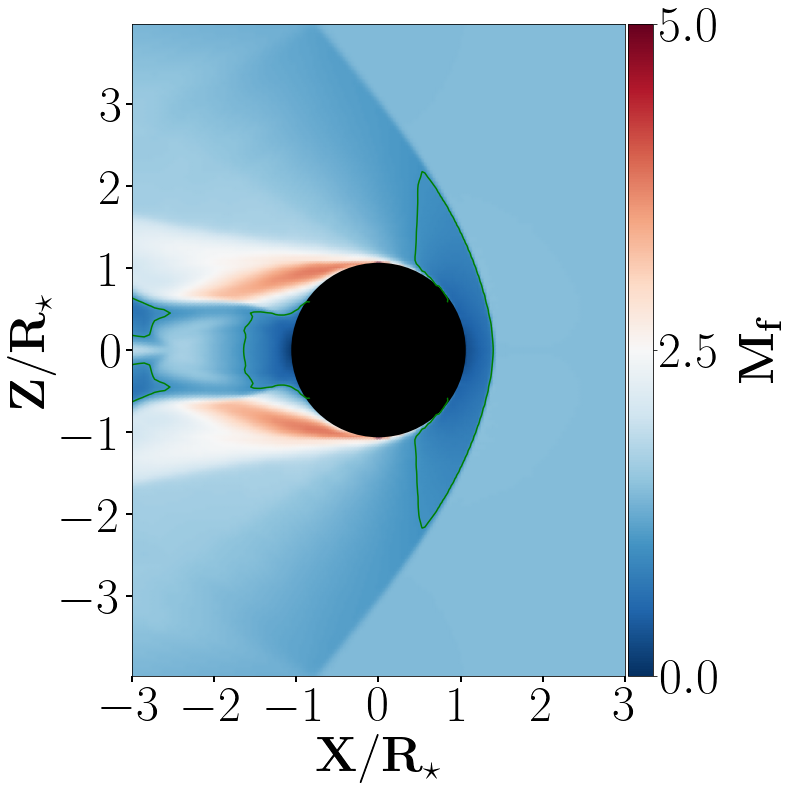}
}
\caption{Relativistic trans-\Alfvenic run B3. Details are similar to Fig.~\ref{relativistic_colorplots_B1_1S}.}
\label{relativistic_colorplots_B3_1S}
\end{figure}

Fig.~\ref{relativistic_colorplots_B3_1S} shows the relevant plots. The flow deflects smoothly around the obstacle in a largely hydrodynamic manner, without the sharp channeling or well-defined \Alfven wings seen in B1 and B2. The density distribution shows a clear bow shock with a well-defined shock cone and a broad upstream compression region. The compression is the weakest among the three cases, with $\rho_{\max} \sim 2.6$, compared to 3.54 in B2 and 10.0 in B1. Downstream, no magnetically confined bubble forms; instead, the wake is diffuse and irregular, with wavy, chaotic streamlines indicative of \Alfvenic disturbances extending far from the obstacle. The ram-pressure distribution follows a similar morphology.

Magnetic pressure amplification is modest and confined to a thin upstream boundary layer; the high-$\beta$ condition ensures that magnetic forces are subdominant. Downstream, the field is diffuse and disorganized, reflecting weak coupling between the plasma and the magnetic field at high $\mathcal{M}_A$. Fast magnetosonic Mach numbers show super-fast regions along the flanks tracing ram-pressure features. Overall, this regime is governed primarily by hydrodynamic compression and flow acceleration, consistent with a shock-dominated, magnetically weak interaction.

\begin{figure}[!htb]
\centering
\includegraphics[width=0.40\textwidth]{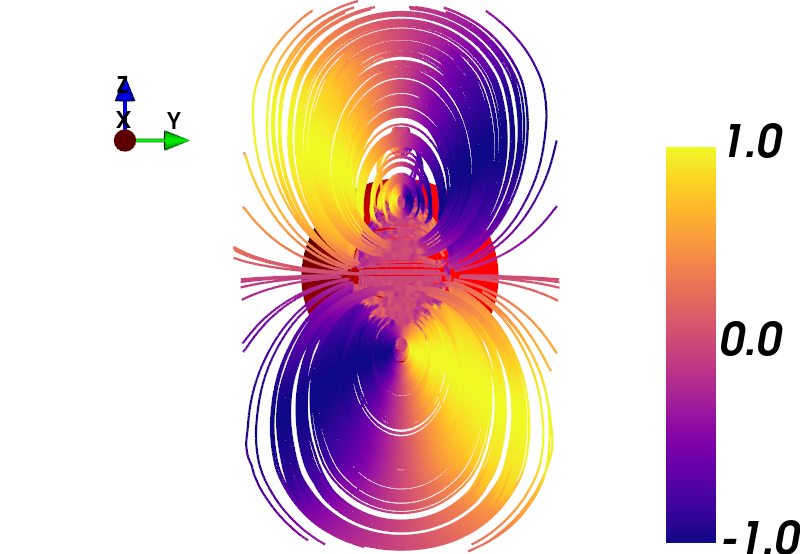}
\includegraphics[width=0.40\textwidth]{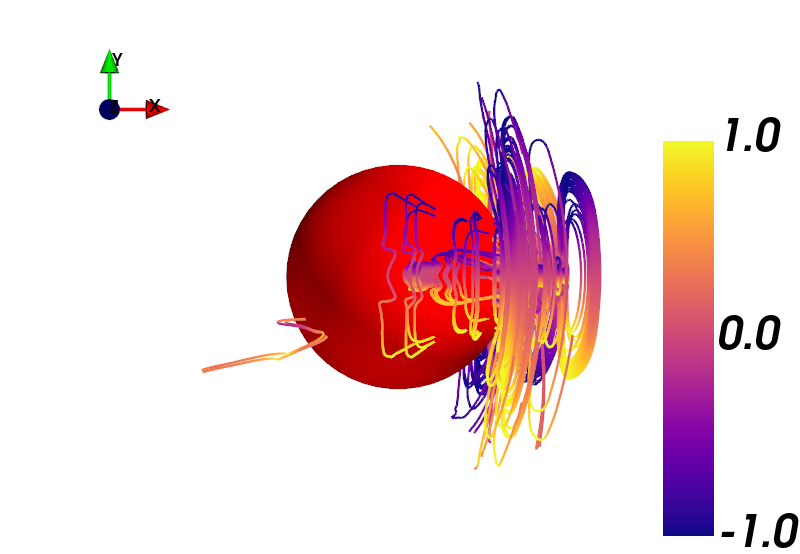}
\caption{Relativistic super-\Alfvenic run B3. 3D streamlines plot of current density at the steady state as viewed across various planes, with the viewing direction labeled in the plots. The rest of the details are the same as in Fig. \ref{J_3dstreamlines_B1}.} 
\label{relativistic_J_3dstreamlines_B3_1S}
\end {figure}

The current structure (Fig.~\ref{relativistic_J_3dstreamlines_B3_1S}) shows a topology distinctly different from runs B1 and B2. The coherent field-aligned \Alfven wings are absent. Instead, current streamlines form predominantly toroidal loops wrapping around the obstacle, with currents confined to the near-sphere region rather than extending along $\pm z$ to the far field. Viewed against the flow direction (left panel), the lack of organized vertical current channels contrasts sharply with B1's columnar structure and B2's turbulent wings. Streamlines exhibit primarily horizontal geometry ($J_z \approx 0$), indicating currents flow in planes perpendicular to the background magnetic field. The equatorial view (right panel) shows current concentration in the bow shock layer wrapping the upstream region, while the downstream wake remains largely current-depleted. The trans-\Alfvenic flow suppresses field-aligned current propagation, confining dissipation to the immediate upstream interaction region.

\section{Discussion}
\label{discussion}


In this work, we model the interaction of a mildly relativistic conducting body with highly magnetized plasma. We demonstrate that such velocity-driven \EM\ interaction provides a natural and physically motivated framework for the generation of beamed electromagnetic emission before the \NS\ merger. The \EM\ can proceed gravitational-waves by seconds to minutes, and may be detectable \citep{2024arXiv240216504L}.

Our results show that the Alfvénic Mach number $\mathcal{M}_A$ controls the transition between coherent Alfvén wing structures and more dissipative, compressed interactions. The realistic inspiral of a neutron star binary is best represented by the relativistic B-series simulations, which suggest a clear evolutionary sequence. During the early inspiral, when the flow is sub-\Alfvenic ($\mathcal{M}_A \lesssim 0.2$), coherent wings dominate the interaction. As the system evolves into the mid-to-late inspiral and approaches trans-\Alfvenic conditions ($\mathcal{M}_A \sim 0.6$--$2$), a complex regime emerges in which distorted \Alfven wings coexist with enhanced compression and turbulence. In the final stages of inspiral, should the system reach strongly trans-\Alfvenic or super-\Alfvenic velocities, \Alfven wings may collapse entirely, leaving a shock-dominated interaction with strongly suppressed wing formation.

We do not address the emission mechanism {\it per se}. Instead, we invoke a pulsar-like paradigm of ``emission follows the current'': large-scale currents would become dissipative (unstable) to the generation of both coherent radio emission and high-energy emission. A particular example of the concept ``emission follows the current'' is the horseshoe-like shape of the emission in PSR J0737-3039B \citep[Figs. 5 and 7 in][]{2014MNRAS.441..690L}, that matches the polar cap current structure  \citep[Fig. 4 in][]{2010ApJ...715.1282B}.

\subsection{Implications for neutron star merger precursors}

As neutron stars inspiral and orbital velocities increase, the interaction naturally evolves through the B1 $\rightarrow$ B2 $\rightarrow$ B3 sequence. Early inspiral exhibits stable Alfvén-wing-driven emission with strong magnetic confinement. The transitional mid-inspiral phase (B2-like, $\mathcal{M}_A \sim 0.6$), where wings become distorted and turbulent, may produce enhanced dissipation and complexity. Late inspiral (B3-like, $\mathcal{M}_A \sim 2$) becomes increasingly shock-dominated with partial suppression of large-scale electromagnetic circuits, potentially yielding different emission characteristics as the flow becomes less magnetically organized.

Finally, we note that \BH--\NS\ mergers should produce \Alfven wings as well \citep{2011PhRvD..83f4001L}. Schwarzschild black holes moving through magnetized plasma act as unipolar inductors, similar to \NSs. Though the mathematics of the corresponding effects are different in the case of \NSs\ and \BH\ (\BH\ do not have a conducting surface), heuristically, the analogy can be understood in terms of the Membrane Paradigm \citep{ThornMembrane}.

\subsection{Limitations and future directions}

Several simplifying assumptions limit the scope of this study. We model the neutron star as a perfectly conducting, unmagnetized sphere embedded in a uniform background magnetic field. While this setup captures the essential coupling between a conducting obstacle and an external magnetic field, it neglects the star’s intrinsic dipole field, internal structure, and spin. In addition, we focus exclusively on the 1M-DNS configuration. Systems in which both neutron stars are strongly magnetized (the 2M-DNS case) are expected to exhibit even stronger interactions and more complex magnetic topologies.

All simulations use ideal classical or relativistic MHD, so dissipation arises only from numerical resistivity at grid scales. We do not include explicit resistivity, radiation reaction, or kinetic plasma physics, each of which is likely essential for resolving magnetic reconnection and converting Poynting flux into observable electromagnetic emission. Finally, the simulated flows are only mildly relativistic ($\Gamma \lesssim 1.5$) and moderately magnetized. Exploring more extreme, pulsar-wind-like regimes with higher magnetization $\sigma$ will require dedicated RMHD simulations with increased resolution and tighter numerical control.

Future work should extend these calculations in several directions. Resistive RMHD or kinetic (PIC) simulations would help resolve the structure of current sheets and quantify particle acceleration more accurately, moving beyond the numerical dissipation present in ideal codes. Future observational modeling should account for both localized dissipation and extended wing transport when predicting precursor light curves and spectra.

\section{Acknowledgments}

We would like to thank Vasily Beskin for pointing out to us the original paper on the concept of \Alfven wings \citep{1962P&SS....9..321G}.


\bibliographystyle{aasjournal}

\bibliography{references}

\appendix

\section{Simulation Setup}
\label{simulation_setup}

We model the obstacle as a dense, non-rotating, perfectly conducting sphere of radius $R_{\star}$ embedded in an ambient magnetized plasma. The sphere is assumed to be unmagnetized and highly conducting, and interacts with the magnetic field of a companion star.

\subsection{Initial Conditions}
\label{initial_conditions}
Because the interaction occurs primarily at the interface between the star and the ambient plasma, it is sufficient to model the object as a uniform conducting sphere without internal structure. The density profile is initialized as
\begin{equation}
\rho =
\begin{cases}
\rho_\star, & r \le R_\star, \\
\rho_{\mathrm{amb}}, & r > R_\star .
\end{cases}
\end{equation}
and the system is initialized in pressure equilibrium with $p = p_0$.

In code units, $\rho_{amb} = p_{amb} = 1$ and $p_{\star}=p_{amb}$. The obstacle density is $\rho_{\star}=500$ and the stellar radius $R_{\star}=1$.

We work in the star’s rest frame so that the neutron star remains fixed in space. Far from the sphere ($r \gg R_{\star}$), the plasma flows with velocity $\mathbf{v} = -v_0 \hat{\mathbf{x}}$ from right (upstream) to left (downstream). We initialize the velocity field following Eq.~B4 of \citet{2023PhRvE.107b5205L}:
\begin{equation}
\begin{aligned}
\mathbf{v} &= 0 && r \le R_{\star}\\
v_r &= -v_0\!\left(1 - \frac{R_{\star}^3}{r^3}\right)\!\sin\theta\,\cos\varphi && r > R_{\star}\\
v_{\theta} &= -v_0\!\left(1 + \frac{R_{\star}^3}{2r^3}\right)\!\cos\theta\,\cos\varphi && r > R_{\star}\\
v_{\varphi} &= v_0\!\left(1 + \frac{R_{\star}^3}{r^3}\right)\!\sin\varphi && r > R_{\star}
\end{aligned}
\label{ic_v_nonrel}
\end{equation}

The external magnetic field of the companion is approximated as a uniform field of magnitude $B_0$ along the $z$-axis, valid when the stellar radius is much smaller than the orbital separation. The neutron star, being perfectly conducting, excludes magnetic flux through its surface. Thus, the total field is the sum of the uniform and induced components (see Eq.~16 of \citet{2023PhRvE.107b5205L})
\begin{equation}
\begin{aligned}
\mathbf{B} &= 0 && r \le R_{\star}\\
B_r &= B_0\!\left(1 - \frac{R_{\star}^3}{r^3}\right)\!\cos\theta && r > R_{\star}\\
B_{\theta} &= -B_0\!\left(1 + \frac{R_{\star}^3}{2r^3}\right)\!\sin\theta && r > R_{\star}\\
B_{\varphi} &= 0 && r > R_{\star}
\end{aligned}
\label{ic_B_nonrel}
\end{equation}

%

\subsection{Boundary Conditions}

In Cartesian runs, we fix the upstream ($x=+4 R_\star$) boundary to maintain the inflow velocity $\mathbf{v} = -v_0 \hat{\mathbf{x}}$ and uniform ambient state. The downstream ($x=-4 R_\star$) boundary adopts outflow conditions to allow plasma and magnetic structures to exit the domain without reflection. Lateral boundaries in $y$ ($y = \pm 4 R_\star$) and along the magnetic-field direction $z$ ($z = \pm 4 R_\star$) are also set to fixed inflow to preserve the ambient medium, preventing artificial depletion or accumulation of plasma. A subset of simulations was repeated in spherical geometry to verify coordinate independence.

Because \Alfven wings extend along the magnetic-field direction ($\pm z$), we verified that our results are not significantly affected by the finite domain size. To test for boundary effects, we repeated the relativistic sub-\Alfvenic run B1 with an extended vertical domain $z \in [-8,8] R_{\star}$ while keeping all other parameters identical. The integrated current flux through the transverse plane at $z = 1.5 R_{\star}$ differed by approximately 10\% between the standard and extended domains. Since the extended domain has twice the grid spacing in the comparison region (mean $\Delta x \approx 0.09$ versus $0.05$), we attribute this difference to resolution effects rather than boundary truncation. The qualitative current topology and wing structure remained consistent between both runs, confirming that the standard domain $[-4,4]^3$ adequately captures the relevant wing dynamics without significant boundary artifacts.


\subsection{Numerical Convergence}
\label{numerical_convergence}

To verify the robustness of our results, we performed a convergence study for run B1 across four resolutions: $N = 113^3$, $140^3$, $170^3$ (fiducial), and $190^3$. As shown in Table~\ref{convergence_table}, the maximum magnetic field amplification ($B_{max}/B_0$) shows excellent stability, varying by less than 1\% across all resolutions, indicating that the global magnetic topology is well-resolved even at the coarsest grid. The volume-integrated wing current ($I_{wings}$) and peak 
velocity ($v_{max}/v_0$) exhibit stronger resolution dependence at coarse grids but show rapid convergence: relative changes compared to the fiducial resolution decrease from $-3.2\%$ (N=113) to only $+1.6\%$ (N=190) for velocity, signaling that the $170^3$ grid has reached the asymptotic regime and effectively captures the essential physics.

\begin{table}[h]
\centering
\begin{tabular}{lccc}
\hline\hline
Resolution ($N^3$) & $v_{max}/v_0$ ($\Delta\%$) & $B_{max}/B_0$ ($\Delta\%$) & $I_{wings}$ ($\Delta\%$) \\
\hline
113   & $-3.2\%$ & $-0.7\%$  & $-10.7\%$ \\
140   & $-7.1\%$  & $-0.5\%$  & $-5.1\%$ \\
\textbf{170 (Fiducial)} & \textbf{---} & \textbf{---} & \textbf{---} \\
190   & $+1.6\%$  & $+0.7\%$  & $+3.4\%$ \\
\hline\hline
\end{tabular}
\caption{Convergence study for run B1. Values show relative percentage 
change compared to the fiducial $170^3$ resolution at $t=60$. The stabilization of all quantities at high resolution indicates the solution 
has converged.}
\label{convergence_table}
\end{table}

\subsection{Divergence Control}
\label{div_control}

The divergence-free constraint $\nabla\!\cdot\vec{B}=0$ is enforced differently for relativistic and non-relativistic runs. For the relativistic B-series, we employ the hyperbolic divergence cleaning method \citep{2002JCoPh.175..645D, 2010JCoPh.229.5896M}, which introduces 
a generalized Lagrange multiplier (GLM) coupled to the induction equation.  This approach treats divergence errors as waves that propagate out of the computational domain at a characteristic speed.

For the non-relativistic A-series, we use \textsc{pluto}'s eight-wave formulation \citep{1994arsm.rept.....P, 1999JCoPh.154..284P, 2004ApJ...602.1079B}, which adds source terms to the standard MHD equations that locally damp divergence errors. In this method, $\nabla\!\cdot\vec{B}$ is treated as an eighth non-physical wave that advects through the grid.

We monitor the volume-averaged normalized divergence error $\Delta x \langle|\nabla \cdot\vec{B}|\rangle / \langle|\vec{B}|\rangle$, where $\Delta x$ is the characteristic cell spacing. As shown in Table~\ref{div_control_table}, divergence errors remain below $sim 10^{-3}$ for all runs, demonstrating good constraint enforcement with negligible impact on the magnetic field evolution.

\begin{table}[h]
\centering
\begin{tabular}{lccc}
\hline\hline
Run & Resolution ($N^3$) & Method & Normalized Div. Error  \\
\hline
A1 & 120 & Eight-Wave & $1.8 \times 10^{-3}$ \\
A2 & 120 & Eight-Wave & $2.6 \times 10^{-3}$ \\
A3 & 120 & Eight-Wave & $3.3 \times 10^{-3}$ \\
B1 & 170 & Hyp. Div. Clean. & $4.4 \times 10^{-4}$ \\
B2 & 170 & Hyp. Div. Clean. & $6.4 \times 10^{-4}$ \\
B3 & 170 & Hyp. Div. Clean. & $1.2 \times 10^{-3}$ \\
\hline\hline
\end{tabular}
\caption{Volume-averaged normalized divergence errors for all runs, 
demonstrating robust constraint enforcement.}
\label{div_control_table}
\end{table}

\section{PLUTO Code}
\label{pluto_code}

We perform both non-relativistic and relativistic MHD simulations using the \textsc{pluto} code \citep{2007ApJS..170..228M}. The system evolves an ideal, perfectly conducting plasma, with numerical dissipation arising solely from finite resolution.

\subsection{Non-relativistic MHD Module}

The non-relativistic simulations solve the standard ideal-MHD equations with adiabatic index $\gamma=5/3$:
\begin{align}
\pd{\rho}{t} + \nabla\!\cdot(\rho\vec{v}) &= 0, \\
\pd{(\rho\vec{v})}{t} + \nabla\!\cdot\!\Big[\rho\vec{v}\vec{v} 
+ \left(p+\frac{B^2}{2}\right)\mathbf{I} - \vec{B}\vec{B}\Big] &= 0, \\
\pd{e}{t} + \nabla\!\cdot\!\Big[(e+p_{\rm t})\vec{v} - (\vec{v}\cdot\vec{B})\vec{B}\Big] &= 0,
\end{align}
with total pressure $p_{\rm t} = p + B^2/2$ and total energy
\begin{equation*}
e = \frac{p}{\gamma-1} + \frac{1}{2}\rho v^2 + \frac{1}{2}B^2.
\end{equation*}

Numerical integration employs a second-order Runge-Kutta (RK2) time-stepping scheme with CFL number 0.4, HLL Riemann solver, and the eight-wave method to enforce $\nabla\cdot\mathbf{B} = 0$. The computational domain is Cartesian, $[-4,4]^3$, with a stretched grid of $N_x = N_y = N_z = 120$ cells, providing enhanced resolution near the stellar surface.

\subsection{Relativistic MHD Module}

Relativistic simulations use the special-relativistic MHD (RMHD) module, evolving
\begin{equation*}
(D, \mathbf{m}, E_t, \mathbf{B}), \qquad D = \Gamma \rho
\end{equation*}
according to
\begin{equation}
\frac{\partial}{\partial t}
\begin{pmatrix} D \\ \mathbf{m} \\ E_t \\ \mathbf{B} \end{pmatrix} +
\nabla \cdot
\begin{pmatrix} D\mathbf{v} \\ w_t\Gamma^2\mathbf{v}\mathbf{v}-\mathbf{b}\mathbf{b}+p_t\mathbf{I} \\ \mathbf{m} \\ \mathbf{v}\mathbf{B}-\mathbf{B}\mathbf{v} \end{pmatrix}^T
= 0.
\end{equation}

Here $\Gamma = (1-v^2/c^2)^{-1/2}$ is the Lorentz factor, and
\begin{align*}
b^0 &= \Gamma\,\mathbf{v}\cdot\mathbf{B}, &
\mathbf{b} &= \frac{\mathbf{B}}{\Gamma} + \Gamma (\mathbf{v}\cdot\mathbf{B}) \mathbf{v}, \\
w_t &= \rho h + \frac{B^2}{\Gamma^2} + (\mathbf{v}\cdot\mathbf{B})^2, &
p_t &= p + \frac{1}{2}\Big(\frac{B^2}{\Gamma^2} + (\mathbf{v}\cdot\mathbf{B})^2\Big), \\
E_t &= w_t \Gamma^2 - (b^0)^2 - p_t, &
\mathbf{m} &= w_t \Gamma^2 \mathbf{v} - b^0 \mathbf{b}.
\end{align*}

Integration uses RK2 time-stepping, HLLD Riemann solver, CFL = 0.25, and hyperbolic divergence cleaning to maintain $\nabla\cdot\mathbf{B} \lesssim 10^{-3}$. The computational grid spans $[-4,4]^3$ with $N_x = N_y = N_z = 170$, and $\gamma = 4/3$.

\subsection{Unit Conversion}

Most simulations were performed in Cartesian coordinates, centered on the star. The upstream flow is along the $-x$ direction and the magnetic field along $+z$. Simulation quantities are expressed in dimensionless code units, with characteristic scales $\tilde{L}$, $\tilde{\rho}$, and $\tilde{v}$ defining the derived units $\tilde{t} = \tilde{L}/\tilde{v}$, $\tilde{p} = \tilde{\rho} \tilde{v}^2$, and $\tilde{B} = \sqrt{4\pi \tilde{\rho} \tilde{v}^2}$. The reference values in physical (cgs) units are $\tilde{\rho} = 1.67\times10^{-23}\,\text{g\,cm}^{-3}$, $\tilde{L} = 3.1\times10^{16}\,\text{cm}$, and $\tilde{v} = 10^{5}\,\text{cm\,s}^{-1}$ (or $\tilde{v} = 3 \times 10^{10}\,\text{cm\,s}^{-1}$ for relativistic runs).

\end{document}